\documentclass[10pt]{article}

\usepackage[utf8]{inputenc}
\usepackage[backend=biber,style=numeric,sorting=none]{biblatex}
\usepackage[normalem]{ulem}

\usepackage{amstext,amsmath,amssymb,amsfonts,bbm,tikz-cd}
\usepackage{dsfont}
\usepackage{hyperref}
\usepackage{amsthm}
\usepackage{color}
\usepackage{multirow}
\usepackage{psfrag}
\usepackage{graphicx}
\usepackage{extarrows}
\usepackage{enumitem}
\usepackage[colored]{shadethm} 
\usepackage{tikz}
\usepackage[compat=1.1.0]{tikz-feynman} 
\usepackage{cancel}
\usepackage{float}
\usepackage{caption}
\hypersetup{
  colorlinks=true,
  urlcolor=blue,
  citecolor=blue,
  linkcolor=red
}

\newcommand{\tetrasymbol}{%
  \begin{tikzpicture}[scale=0.22, baseline=-0.5ex]
    \draw (0,1) -- (-0.8,-0.5) -- (0.8,-0.5) -- cycle;
    \draw (0,1) -- (0,-0.2);
    \draw[dashed] (-0.8,-0.5) -- (0,-0.2);
    \draw (0.8,-0.5) -- (0,-0.2);
  \end{tikzpicture}%
}

\makeatletter
\def\ps@plain{%
  \def\@oddhead{}%
  \def\@oddfoot{\hfil--\ \thepage\ --\hfil}%
  \let\@evenhead\@oddhead
  \let\@evenfoot\@oddfoot}
\makeatother
\title{{\sc A twisted analogue of Kitaev's quantum double model from strings in a higher-form magnetic field}}

\begin{document}

\author{T.~Krajewski\thanks{\href{mailto:thomas.krajewski@cpt.univ-mrs.fr}{\texttt{thomas.krajewski@cpt.univ-mrs.fr}}}\; and M.-B.~Ziane\thanks{\href{mailto:mziane@cpt.univ-mrs.fr}{\texttt{mziane@cpt.univ-mrs.fr}}}\\
\emph{Aix Marseille Univ, Universit\'e de Toulon, CNRS, CPT, Marseille, France}}

\maketitle
\abstract{
Kitaev's quantum double model provides a natural framework for describing topological excitations in condensed matter as well as in quantum computing.
In this article, we propose to extend this theoretical framework to strings coupled to a 2-form gauge potential $B$, whose field strength is the 3-form $C=\mathrm{d}B$, invariant under the action of the finite group $G$.
This proposal then leads to successive gauge transformations which naturally give a cohomological structure encoded by a 3-cocycle.
Consequently, we obtain an extension of the Kitaev model where the strings evolve in a higher-degree magnetic field, and whose magnetic translations naturally generate a quasi-Hopf structure described by the twisted Drinfel'd double $\mathcal{D}^\omega[G]$.
We will then present the Hamiltonian formulation of this construction, detailing its formal foundations, the structure of the Hilbert space, as well as a Kitaev-type Hamiltonian built from projectors and involving two group elements of $G$, associated with the gauge fields and twisted sectors.
}

\section*{Introduction}

Quantum computing is at the interface between theoretical physics and computational science.
Now, among the types of computational approaches, topological quantum computation is distinguished since it resists noise by encoding the information in global degrees of freedom through the braiding of anyons.
The toric code of Kitaev then appears as a key model. This model is defined on a lattice where the degrees of freedom are associated with the edges and characterized by Abelian anyonic excitations whose fusion rules and braiding statistics are non-trivial.
But this is a particular case of the quantum double model $\mathcal{D}[\mathbb{Z}_2]$ \cite{Kitaev_2003}, whose generalization $\mathcal{D}[G]$ is constructed from a finite group $G$, allowing us to introduce non-Abelian excitations when $G$ is not Abelian.
Beyond its exact solvability, the quantum double model has a major interest: the \emph{universal quantum computation}, as demonstrated in the case of the symmetric group $G=S_3$ \cite{Chen_2025,lo2025universalquantumcomputations3}.

\medskip

This algebraic construction has been extended by integrating topological phases giving rise to twisted models in which the group of cocycles appear in the amplitudes associated with the configurations of a given lattice.
Two examples may be cited. The twisted quantum double model of topological phases in two dimensions \cite{Hu:2012wx} is constructed from a finite group $G$ and a 3-cocycle.
In the same spirit, the tube algebra approach, proposed by A. Bullivant and C. Delcamp in \cite{Bullivant:2019fmk} for different dimensions ($1{+}1$, $2{+}1$ and $3{+}1$), allows an algebraic characterization of the excitations and their fusion and braiding properties.
These twisted models are related to the topological gauge theories of Dijkgraaf--Witten \cite{DW_1990} since non-trivial group cocycles yield topological weights for constructing discrete actions (of the Dijkgraaf--Witten type).
In addition, the group cohomology provides a classification framework for symmetry-protected topological phases, as shown in \cite{Chen_2025}.

\medskip

In this article, we develop this same framework but, this time, based on twisted sectors coupled to a magnetic field characterized by a 2-form gauge potential $B$, with field strength $C=\mathrm{d}B$, a Kalb--Ramond field.
We consider an ribbon graph equipped with different variables: the variables $g_{\text{e}} \in G$ are the gauge fields (or longitudinal group elements), $w_{\text{e}} \in G$ are the windings (or transverse defects group elements), and by introducing a finite set $X$ equipped with a right action $\rhd_{\text{v}}$ of $G$, we consider ``matter-type'' variables $x_{\text{v}} \in X$ at the vertices.
The twisted sectors are carried by the edges of this graph and interact at the vertices.
Their dynamics (propagation and fusion, in particular) are studied in the presence of a magnetic background described by the 3-form $C$.
We then construct the Kitaev Hamiltonian analog based on a sum of three families of mutually compatible local projectors.
The first, $\Pi_{\text{v}}$, imposes the magnetic amplitudes associated with the propagation of twisted sectors, while fully fixing the $x$ dependence.
The second, $P_{\text{v}}$, imposes gauge invariance at the vertices, guaranteeing the absence of any electric and magnetic charges at these sites.
The last one, $Q_{\text{f}}$, implements face dynamics by imposing a flatness condition for the gauge fields around the face and coherently translating the transversal group elements $w_{\text{e}}$.
This construction then makes possible the joint study of gauge symmetries, non-invertible symmetries, and the quasi-triangular quasi-Hopf algebraic structure $\mathcal{D}^{\omega}[G]$ that naturally arises from this formalism.

\medskip

The article is organized into three sections.
\textbf{Section~\ref{sec1}} recalls the basic principles of topological quantum computing, before presenting the essential elements of Kitaev's quantum double model and introducing the excitations and their description in terms of representations of the Drinfel'd double.
\textbf{Section~\ref{sec2}} establishes the formalism describing particles and strings in a magnetic field, starting from a continuous description to construct a discrete analog of magnetic translations and twisted sectors.
\textbf{Section~\ref{sec3}} details the twisted model based on strings evolving in a magnetic field: we present the precise structure of the Hamiltonian built around a family of three mutually commuting local projectors, and concludes with the study of the symmetries and the resulting algebraic structure.

\section{Kitaev's quantum double model}\label{sec1}

\paragraph{Topological quantum computing.}

Quantum computing \cite{cui_2018, PR2004, SI2017} is based, in particular, on a simple idea: to use the power of the superposition principle,
    \begin{align*}
    \sum_{x}c_{x}|x\rangle
    \mathop{\longrightarrow}\limits^{U_{f}}
    \sum_{x}c_{x}|f(x)\rangle ,
    \end{align*}
where $f$ is a permutation of the computational-basis labels, so that the operator $U_f$ is unitary.

However, the conventional approach to quantum computing fundamentally faces several major challenges, such as decoherence and the need for error correction.
These problems can be mitigated by exploiting the robustness of topological states against local perturbations.

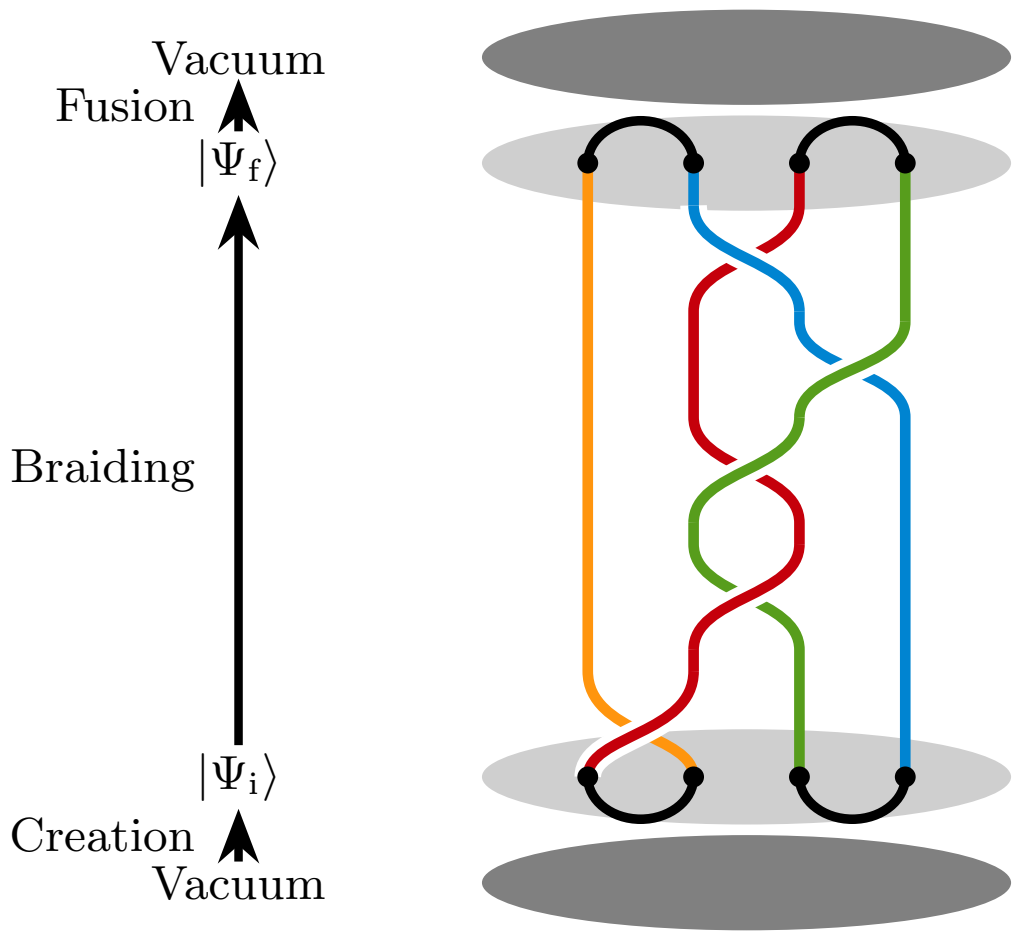
\begin{figure}[htbp]
\centering
\begin{tikzpicture}[scale=0.7, >=Stealth]
\definecolor{tqcRed}{RGB}{197, 0, 11}
\definecolor{tqcGreen}{RGB}{87, 157, 28}
\definecolor{tqcBlue}{RGB}{0, 132, 209}
\definecolor{tqcOrange}{RGB}{255, 149, 14}
\definecolor{tqcPink}{RGB}{255, 185, 190}
    \begin{scope}[font=\small]
        \node (vac_bot) at (0, 0) {Vacuum};
        \node (psi0) at (0, 1) {$|\Psi_{\textrm{i}}\rangle$};
        
        \draw[->, ultra thick] (0, 0.2) -- (0, 0.7) node[midway, left=4pt] {Creation};
        
        \node (upsi) at (0, 6.8) {$|\Psi_{\textrm{f}}\rangle$};
        \draw[->, ultra thick] (0, 1.3) -- (0, 6.5) node[midway, left=4pt] {Braiding};
        
        \node (vac_top) at (0, 7.8) {Vacuum};
        
        \draw[->, ultra thick] (0, 7.1) -- (0, 7.6) node[midway, left=4pt] {Fusion};
    \end{scope}

    \coordinate (C) at (4.8, 0); 
    
    \fill[gray] (C |- 0, 0) ellipse (2.5cm and 0.45cm);
    
    \fill[lightgray, opacity=0.75] (C |- 0, 1) ellipse (2.5cm and 0.45cm);
    
    \fill[lightgray, opacity=0.75] (C |- 0, 6.8) ellipse (2.5cm and 0.45cm);
    
    \fill[gray] (C |- 0, 7.8) ellipse (2.5cm and 0.45cm);

    \coordinate (P1) at (3.3, 1.0);
    \coordinate (P2) at (4.3, 1.0);
    \coordinate (P3) at (5.3, 1.0);
    \coordinate (P4) at (6.3, 1.0);

    \coordinate (Q1) at (3.3, 6.8);
    \coordinate (Q2) at (4.3, 6.8);
    \coordinate (Q3) at (5.3, 6.8);
    \coordinate (Q4) at (6.3, 6.8);

    \draw[line width=1.8pt] (P1) arc (180:360:0.5cm and 0.4cm);
    \draw[line width=1.8pt] (P3) arc (180:360:0.5cm and 0.4cm);

    \draw[line width=1.8pt] (Q1) arc (180:0:0.5cm and 0.4cm);
    \draw[line width=1.8pt] (Q3) arc (180:0:0.5cm and 0.4cm);

    
    \draw[tqcOrange, line width=2pt] 
        (P2) .. controls (4.3, 1.4) and (3.3, 1.4) .. (3.3, 2.0) -- (Q1);

    \draw[tqcGreen, line width=2pt] 
        (P3) -- (5.3, 2.2)
        .. controls (5.3, 2.7) and (4.3, 2.7) .. (4.3, 3.2)
        -- (4.3, 3.4)
        .. controls (4.3, 3.9) and (5.3, 3.9) .. (5.3, 4.4)
        .. controls (5.3, 4.85) and (6.3, 4.85) .. (6.3, 5.3)
        -- (Q4);

    \draw[tqcBlue, line width=2pt] 
        (P4) -- (6.3, 4.4)
        .. controls (6.3, 4.85) and (5.3, 4.85) .. (5.3, 5.3)
        -- (5.3, 5.4)
        .. controls (5.3, 5.9) and (4.3, 5.9) .. (4.3, 6.4)
        -- (Q2);

    \draw[tqcRed, line width=2pt] 
        (P1) .. controls (3.3, 1.4) and (4.3, 1.4) .. (4.3, 2.0)
        -- (4.3, 2.2)
        .. controls (4.3, 2.7) and (5.3, 2.7) .. (5.3, 3.2)
        -- (5.3, 3.4)
        .. controls (5.3, 3.9) and (4.3, 3.9) .. (4.3, 4.4)
        -- (4.3, 5.4)
        .. controls (4.3, 5.9) and (5.3, 5.9) .. (5.3, 6.4)
        -- (Q3);

    \draw[white, line width=5pt] (P1) .. controls (3.3, 1.4) and (4.3, 1.4) .. (4.3, 2.0);
    \draw[tqcRed, line width=2pt] (P1) .. controls (3.3, 1.4) and (4.3, 1.4) .. (4.3, 2.0);

    \draw[white, line width=5pt] (4.3, 2.2) .. controls (4.3, 2.7) and (5.3, 2.7) .. (5.3, 3.2);
    \draw[tqcRed, line width=2pt] (4.3, 2.2) .. controls (4.3, 2.7) and (5.3, 2.7) .. (5.3, 3.2);

    \draw[white, line width=5pt] (4.3, 3.4) .. controls (4.3, 3.9) and (5.3, 3.9) .. (5.3, 4.4);
    \draw[tqcGreen, line width=2pt] (4.3, 3.4) .. controls (4.3, 3.9) and (5.3, 3.9) .. (5.3, 4.4);

    \draw[white, line width=5pt] (5.3, 4.4) .. controls (5.3, 4.85) and (6.3, 4.85) .. (6.3, 5.3);
    \draw[tqcGreen, line width=2pt] (5.3, 4.4) .. controls (5.3, 4.85) and (6.3, 4.85) .. (6.3, 5.3);

    \draw[white, line width=5pt] (5.3, 5.4) .. controls (5.3, 5.9) and (4.3, 5.9) .. (4.3, 6.4);
    \draw[tqcBlue, line width=2pt] (5.3, 5.4) .. controls (5.3, 5.9) and (4.3, 5.9) .. (4.3, 6.4);

    \foreach \pt in {P1, P2, P3, P4, Q1, Q2, Q3, Q4} {
        \fill[black] (\pt) circle (2.8pt);
    }

\end{tikzpicture}
\caption{Process of Topological Quantum Computing (adapted from \cite{cui_2018}).}
\label{fig:tqc_process}
\end{figure}

\medskip

A topological quantum computer is a system based on anyons whose computational process is divided into three steps:
\begin{itemize}
\item \textbf{Initialization:} anyon--anti-anyon pairs, with labels $a,\bar{a}$, $b,\bar{b}$, $c,\bar{c},\ldots$, are created from the vacuum in dimension $2{+}1$.  
\item \textbf{Processing:} the braiding operations act on the fusion space of the anyons and performs the computation (inducing a generally non-Abelian representation of the braid group).
\item \textbf{Readout:} the measurement corresponds to the fusion of adjacent pairs,
\begin{equation}
a \otimes b \cong \bigoplus_c N_{ab}^c \, c,
\end{equation}
where $N_{ab}^c \in \mathbb{N}_0$, called \emph{fusion multiplicity} (or \emph{fusion coefficient}), is the number of times $c$ appears in the decomposition, or equivalently, the number of independent channels through which $a$ and $b$ can fuse into $c$.
\end{itemize}

\paragraph{Kitaev's quantum double model.}
Kitaev's quantum double model is a simple yet rich model displaying non abelian anyonic excitations \cite{Kitaev_2003}.
It is based on an oriented  ribbon graph $\Gamma=(E,V,F)$, which consists of a set of vertices $V$, connected by oriented edges $E$, together with a cyclic order of the incident half-edges at every vertex, so that there are well-defined faces $F$.
It can be realized as a graph embedded on a surface of Euler characteristic  $\chi=|V|-|E|+|F|$ or as a fat graph, whose edges are ribbons. 

The Hilbert space is $L^{2}(G^{|E|})$ whose elements are functions of $|E|$ group elements assigned to the edges
    \begin{align}
    {\cal H}=\bigg\{|\Psi\rangle=\sum_{\left\{g_{\text{e}}\right\}\in G^{E}}\Psi\big(\left\{g_{\text{e}}\right\}\big)\,\big|\left\{g_{\text{e}}\right\}\rangle\bigg\}
    \end{align}
where $\left\{g_{\text{e}}\right\}$ is an assignment of a group element to every edge,  a map from the edge set $E$ to $G$.
In the sequel, we use only the wave function representation, so that we define the operators directly on the wave function $\Psi\big(\left\{g_{\text{e}}\right\}\big)$

The Hamiltonian is defined as a combination of mutually commuting Hermitian projectors $P_{\text{v}}$ and $Q_{\text{f}}$   
    \begin{align}
    H=\sum_{\text{v}}\mu_{\text{v}}\big(1-P_{\text{v}}\big) +\sum_{\text{f}}\nu_{\text{f}}\big(1-Q_{\text{f}}\big)\qquad \mu_{\text{v}},\nu_{\text{f}} > 0
    \end{align}
with $\mu_{\text{v}}$ and $\nu_{\text{f}}$ positive coupling constants. 

The first projector $P_{\text{v}}$ implements gauge transformations at vertex $\text{v}$
    \begin{align}
    P_{\text{v}}=\frac{1}{|G|}\sum_{g\in G}U_{\text{v},g}
    \end{align}
with
    \begin{align}
    U_{\text{v},g}\Psi\big(\left\{g_{\text{e}}\right\}\big)=\Psi\big(\left\{g\rhd_{\text{v}}g_{\text{e}}\right\}\big)
    \end{align}
where we use the shorthand
    \begin{align}
    g\rhd_{\text{v}}g_{\text{e}}=
    \begin{cases}
    g^{-1}g_{\text{e}}&\text{ if  e is leaving v}\\
    g_{\text{e}}g&\text{ if e is entering v}\\
    g^{-1}g_{\text{e}}g&\text{ if both ends of e are connected to v}\\ 
    g_{\text{e}}&\text{ if  e is not  connected to v} 
    \end{cases}
    \end{align}
Alternatively, $U_{\text{v},g}$ can be defined on the basis by $U_{\text{v},g}\big|\left\{g_{\text{e}}\right\}\big)=\big|\left\{g^{-1}\rhd_{\text{v}}g_{\text{e}}\right\}\rangle$.
These operators are unitary and obey $U_{\text{v},g}U_{\text{v},h}=U_{\text{v},gh}$, so that $P_{\text{v}}$ is a Hermitian projector.
Moreover, $U_{\text{v},g}$ commutes with $U_{\text{v'},g'}$ for $\text{v} \neq \text{v'}$, so that $P_{\text{v}}$ and $P_{\text{v}'}$ commute.
These projectors enforce gauge invariance, in the sense that we can translate that the operators $U_{\text{v},g_{\text{v}}}$ implement independent shifts of the wave function at every vertex.

The second projector $Q_{\text{f}}$ implements the flatness constraint around the face $\text{f}$
    \begin{align} Q_{\text{f}}\Psi\big(\left\{g_{\text{e}}\right\}\big)= \delta\Big(\prod\limits_{\text{e}\in\partial\text{f}}g_{\text{e}}^{\pm}\Big)
    \Psi\big(\left\{g_{\text{e}}\right\}\big) 
    \qquad\text{with}\qquad\delta(h)=
    \begin{cases}
    1& \text{if $h=e$ (unit element)}\\
    0& \text{if $h\neq e$ }
    \end{cases}
    \end{align}
where the product is taken around the face, following its orientation from right to left.
We take $+$ if the orientations of $\text{e}$ and $\text{f}$ agree, and $-$ otherwise when travelling around the face.
Note that a starting point has to chosen on the face but the projector does not depend on this choice.
Once such a starting point is chosen on every face, it will be convenient to denote by $h_{\text{f}} = h_{\text{f}}(\{g_{\text{e}}\})$ the holonomy around the face, expressed in terms of edge variables.

By construction, the operators $Q_{\text{f}}$ define mutually commuting Hermitian projectors for all faces $\text{f}$ that commute with gauge transformations $U_{\text{v},g}$, so that they also commute with the projectors $P_{\text{v}}$.
This is the key to a complete description of all eigenstates of the Hamiltonian.
In particular, the ground states have to be flat and gauge invariant
    \begin{align}
    H\Psi_{0}=0\qquad\Leftrightarrow\qquad 
    \begin{cases}
    P_{\text{v}}\Psi_{0}=\Psi_{0}\text{ for all vertices v}\\
    Q_{\text{f}}\Psi_{0}=\Psi_{0}\text{ for all faces f}
    \end{cases}
    \end{align}
For a planar graph, i.e., one that can be embedded on a genus $0$ surface, gauge invariance allows writing a general flat configuration as a pure gauge, using for instance gauge transformations along a spanning tree. 
The ground state is therefore unique up to a normalization and can be written as
    \begin{align}
    \Psi\big(\left\{g_{\text{e}}\right\}\big)=
    \frac{1}{|G|^{|V|}}\sum_{\left\{u_{\text{v}}\right\}}
    \prod_{\text{e}\in E}\delta\big(u_{s(e)}g_{e} u^{-1}_{t(e)}\big)
    \end{align}   
where $s(\text{e})$ and $t(\text{e})$ are the source and target vertices of $\text{e}$.
On a higher genus surface $\Sigma$, gauge holonomies around the non contractible cycles cannot be gauged away and ground states are parametrized by morphisms of $\pi_{1}(\Sigma)$ into $G$, up to gauge transformations.
For example, on a torus with only two edges, ground states are $\Psi_{0}(g_{1},g_{2})=\delta(g^{}_{1}g^{}_{2}g^{-1}_{1}g^{-1}_{2})F(g_{1},g_{2})$, with $F$ an arbitrary function of two commuting elements, invariant under the adjoint action $F(g^{-1}g_{1}g,g^{-1}g_{2}g)=F(g_{1},g_{2})$ for all $g\in G$.

\bigskip

Before briefly discussing the excited states, it is interesting to give a second description in the representation basis.
First, let us express $Q_{\text{f}}$ in terms of the characters $\chi_{\rho}$ of the unitary irreducible representations of $G$.
Using
    \begin{align}
    \delta(h)=\frac{1}{|G|}\sum_{\rho}d_{\rho}\chi_{\rho}(h)
    \end{align}
with $d_{\rho}$ the dimension of $\rho$, one can write
    \begin{align}
    Q_{\text{f}}=\frac{1}{|G|}\sum_{\rho} d_{\rho}V_{\text{f},\rho}
    \end{align}
where $V_{\text{f},\rho}$ is a multiplication by $\chi_{\rho}(h_{\text{f}})$
    \begin{align}
    V_{\text{f},\rho}\Psi\big(\left\{g_{\text{e}}\right\}\big)=\chi_{\rho}(h_{\text{f}})\,\Psi\big(\left\{g_{\text{e}}\right\}\big).
    \end{align}
These operators commute with the gauge transformations, but are not necessarily unitary or invertible, and obey the algebra
    \begin{align}
    V_{\text{f},\rho}V_{\text{f},\rho'}=\sum_{\rho''}N^{\rho''}_{\rho\rho'}V_{\text{f},\rho''}
    \end{align}
where $N^{\rho''}_{\rho\rho'}$ are the fusion coefficients of the representations of $G$, $\rho\otimes\rho'=\bigoplus_{\rho''}N^{\rho''}_{\rho\rho'}\rho''$.
Furthermore, these operators, like the unitary ones $U_{\text{v},g}$, commute with the Hamiltonian 
    \begin{align}
    U_{\text{v},g}H=HU_{\text{v},g}\qquad V_{\text{f},\rho}H=HV_{\text{f},\rho}.
    \end{align}
Second, recall the Schur orthogonality relation of the matrix elements of irreducible representations of $G$
    \begin{align}
    \sum_{g} \overline{D_{ij}^{\rho}}(g)D_{kl}^{\rho'}(g)=\frac{|G|}{d_{\rho}}\delta^{\rho\rho'}\delta_{ik}\delta_{jl}.
    \end{align}
Thus, an orthonormal basis of $L^{2}(G^{|E|})$ can be constructed by multiplying normalized matrix elements on all edges.
Furthermore, these matrix elements can be contracted with intertwiners at the vertices.
If edges carrying representations $\rho_{1},\dots,\rho_{n}$ are incoming at vertex $v$, we insert an intertwiner ${\cal I}^{\rho_{1},\dots,\rho_{n}\rightarrow\rho}_{i_{1},\dots,i_{n}\rightarrow i}$  from  $\rho_{1}\otimes\cdots\otimes\rho_{n}$ to $\rho$.
For those edges oriented away from $\text{v}$, we replace $\rho$ by its dual $\rho^{\ast}$, defined as $D^{\rho\ast}_{ij}(g)=D^{\rho}_{ji}(g^{-1})$ and consider it as incoming.
Then, contracting all edge indices between matrix elements and intertwiners, we obtain a spin network basis of  $L^{2}(G^{|E|})$.
Then, gauge invariance amounts to imposing that all vertex representations be trivial.

\paragraph{Excitations and representations of the Drinfel'd quantum.}

Elementary excitations are created by ribbon operators ${\cal L}^{u}_{g}(\text{v}_{1},\text{f}_{1}|\text{v}_{2},\text{f}_{2})$ that create excitations on  two sites \cite{Chen:2021gfu}, i.e., a pair of a face and a vertex belonging to that face,  $(\text{v}_{1},\text{f}_{1}$, $\text{v}_{2},\text{f}_{2})$. They are connected by a path in the graph and a parallel one in the dual graph. $g=g_{1}^{\pm}\cdots g_{n}^{\pm}$ is the product of group elements encountered along the path with signs chosen according to orientation. $u\in G$ is parallel transported along the path 
    \begin{align}
    u_{0}=u\rightarrow u_{1}=g^{-1}_{1}ug^{}_{1}\rightarrow u_{2}=(g_{1}g_{2})^{-1}ug_{1}g_{2}\rightarrow\cdots\rightarrow  u_{n}=(g_{1}\cdots g_{n})^{-1}ug^{}_{1}\cdots g^{}_{n}.
    \label{paralleltransport}
    \end{align}
Moreover, every edge met transversally by the dual path is transformed as $h_{m}\rightarrow u_{m}^{-1}h_{m}$ if it is outgoing and  $h_{m}\rightarrow h_{m}u_{m}$ if incoming. 
The parallel transport of $u$ ensures that the faces remain flat, except at the boundary.
To explicitly construct the wave function and control its gauge dependence, one has to use the representation of the quantum double of $G$, which we briefly describe.
  
First, recall that the group algebra is the set of all linear combination of elements of $G$  
    \begin{align}
    C[G]=\bigg\{\sum_{g\in G} a_{g}\mathop{\square}\limits_{g}\bigg\},    \end{align}
where are $a_{g}$ are complex numbers and $\mathop{\square}\limits_{g}$ denote the group elements, which form a basis of $C[G]$, with multiplication 
    \begin{align}  \mathop{\square}\limits_{g}\mathop{\square}\limits_{h}=\mathop{\square}\limits_{gh}.
    \end{align}
Representations of $C[G]$ as an algebra are in one to one correspondence with representations of the group $G$.

The Drinfel'd  double \cite{Drinfeld:1989st, DPR_1990, Altschuler_1992, Kassel_1995} is obtained by letting the coefficients $a_{g}$ also depend on a second group element $w$
    \begin{align}
    \mathcal{D}[G]=\bigg\{\sum_{g,w\in G} a_{g,w}\,{  w}\mathop{\square}\limits_{g}\bigg\}
    \end{align}
with a semi-direct product law
    \begin{align}
    w\mathop{\square}\limits_{g}\quad g^{-1}wg\mathop{\square}\limits_{h}=w\mathop{\square}\limits_{gh} .
    \end{align}
From the point of view of the Kitaev model, $g$ represents the gauge fields and $w$ the variable to be transported along the ribbon.

Furthermore, $\mathcal{D}[G]$ turns out to be a quasi-triangular Hopf algebra.
This means that there is a coproduct
$\Delta:\; \mathcal{D}[G]\rightarrow \mathcal{D}[G]\otimes \mathcal{D}[G] $ defined by
    \begin{align}
    \Delta\!\left(w\mathop{\square}\limits_{g}\right)=\sum_{uv=w}u\mathop{\square}\limits_{g}\otimes v\mathop{\square}\limits_{g}
    \end{align}
a counit $\epsilon( w\mathop{\square}\limits_{g})=\delta(w)$ and an antipode 
    \begin{align}
    S\!\left(w\mathop{\square}\limits_{g}\right)= g^{-1}w^{-1}g\mathop{\square}\limits_{g^{-1}}.
    \end{align}
These operations have to obey a certain number of rules like the coassociativity  $(\Delta\otimes\text{id})\circ \Delta=(\text{id}\otimes\Delta)\circ \Delta$ which can be found in many texts on the subject.
Moreover, denoting by $\tau$ the flip of two factors, $\tau(a\otimes b)=b\otimes a$, the opposite coproduct is related to the coproduct through the $R$-matrix
$\tau\circ\Delta=R\Delta R^{-1}$, with $R\in \mathcal{D}[G]\otimes \mathcal{D}[G]$.
If we denote by $R_{ij}$ the matrix $R$ acting non-trivially only on the factors $i$ and $j$ of a tensor product and  $\tau_{ij}$ the transposition of the factors $i$ and $j$, then $\sigma_{i}=\tau_{i,i+1}\circ R_{i,i+1}$, then $\sigma_{i}$ provides a representation of the braid group, $\sigma_{i}\sigma_{i+1}\sigma_{i}=\sigma_{i+1}\sigma_{i}\sigma_{i+1}$.
       
We end this very short overview of $\mathcal{D}[G]$ by sketching its representations, as an associative algebra.
Irreducible representations are labelled by pairs $\rho=(\kappa,r)$, where $\kappa$ is a conjugacy class of $G$ and $r$ an irreducible representation of the centralizer of an arbitrary element of $\kappa$, i.e., of the subgroup that commutes with this element.
We denote the matrix elements $D^{\rho}_{ij}$, where $i=(u,\alpha)$ is a pair of a group element $u\in\kappa$ and an index $\alpha$ in the corresponding representation of the centralizer.
These form a representation of 
    \begin{align}
    \sum_{j}D^{\rho}_{ij}\big(u\mathop{\square}\limits_{g}\big)D^{\rho}_{jk}\big(v\mathop{\square}\limits_{h}\big)=
    \delta_{v,g^{-1}ug}D^{\rho}_{ik}\big(u\mathop{\square}\limits_{gh}\big) .
    \end{align}
Moreover, composition with the coproduct allows us to take tensor products of representations, which is  defined by,
    \begin{align}
    \big[D^{\rho}_{ij}\otimes D^{\rho'}_{kl}\big]\big(w\mathop{\square}\limits_{g}\big)=
    \sum_{uv=w}D^{\rho}_{ij}\big(u\mathop{\square}\limits_{g}\big)D^{\rho'}_{kl}\big(v\mathop{\square}\limits_{g}\big) .
    \end{align}
The counit corresponds to the trivial representation $\rho=1$ and composition with the antipode to the dual representation
    \begin{align}
    D^{\rho^{\ast}}_{ij}\big(u\mathop{\square}\limits_{g}\big)=
    D^{\rho}_{ji}\big(S\,\big[ u\mathop{\square}\limits_{g}\big]\big).
    \end{align} 
Finally, the representations can be braided using $\sigma: \rho\otimes\rho'\rightarrow \rho'\otimes\rho$ and fused $\rho\otimes \rho'=\sum_{\rho''}N^{\rho''}_{\rho\rho'}\rho^{''}$.

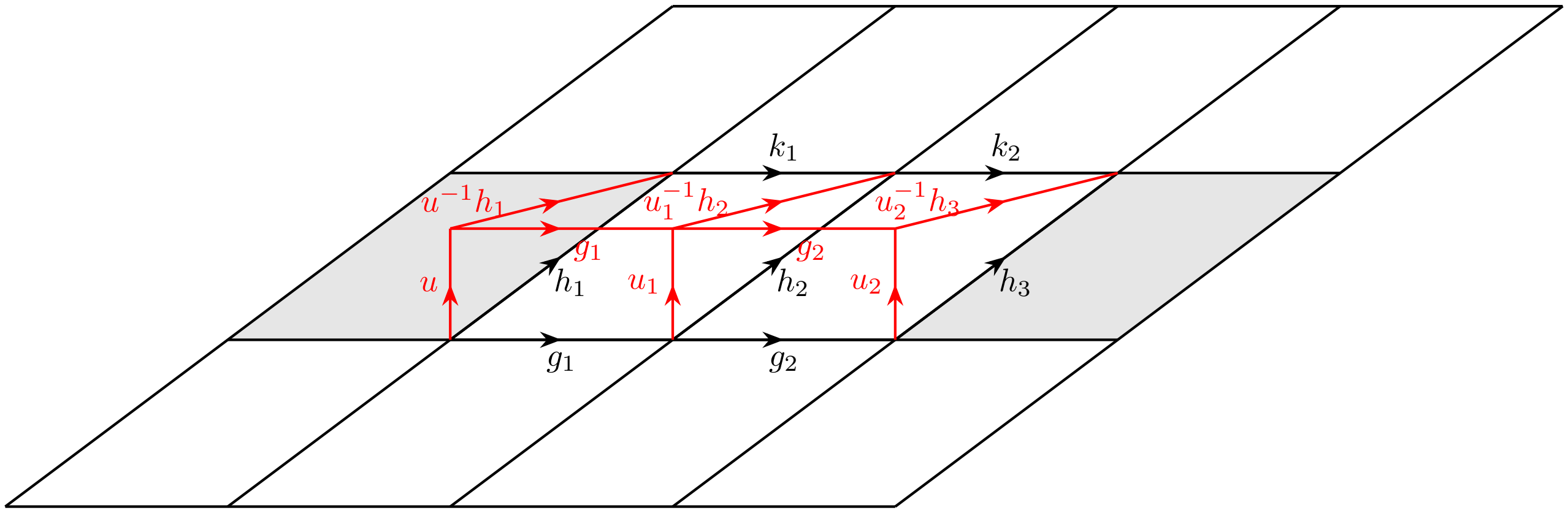
\begin{figure}[H]
\centering


\begin{tikzpicture}[scale=6,
  midarrow/.style={
    postaction={
      decorate,
      decoration={
        markings,
        mark=at position 0.5 with {\arrow{Stealth}}}}},scale=0.5]

 \pgfmathsetmacro{\a}{0.8}
 \pgfmathsetmacro{\b}{0.8}
 \pgfmathsetmacro{\c}{0.6}
\pgfmathsetmacro{\d}{0.4}

\coordinate (A0) at (0,0);
\coordinate (A1) at (0+\a,0);
\coordinate (A2) at (0+2*\a,0);
\coordinate (A3) at (0+3*\a,0);
\coordinate (A4) at (0+4*\a,0);

\coordinate (B0) at (0+\b,0+\c);
\coordinate (B1) at (0+\a+\b,0+\c);
\coordinate (B1') at (0+\a+\b,0+\c+\d);
\coordinate (B2) at (0+2*\a+\b,0+\c);
\coordinate (B2') at (0+2*\a+\b,0+\c+\d);
\coordinate (B3) at (0+3*\a+\b,0+\c);
\coordinate (B3') at (0+3*\a+\b,0+\c+\d);
\coordinate (B4) at (0+4*\a+\b,0+\c);

\coordinate (C0) at (0+2*\b,0+2*\c);
\coordinate (C1) at (0+1*\a+2*\b,0+2*\c);
\coordinate (C2) at (0+2*\a+2*\b,0+2*\c);
\coordinate (C3) at (0+3*\a+2*\b,0+2*\c);
\coordinate (C4) at (0+4*\a+2*\b,0+2*\c);

\coordinate (D0) at (0+3*\b,0+3*\c);
\coordinate (D1) at (0+1*\a+3*\b,0+3*\c);
\coordinate (D2) at (0+2*\a+3*\b,0+3*\c);
\coordinate (D3) at (0+3*\a+3*\b,0+3*\c);
\coordinate (D4) at (0+4*\a+3*\b,0+3*\c);

\fill[black!10] (B0) -- (B1) -- (C1) -- (C0) -- (B0);
\fill[black!10] (B3) -- (B4) -- (C4) -- (C3) -- (B3);


\draw[thick] (A0) -- (A4) ;
\draw[thick] (B0) -- (B4) ;
\draw[thick] (C0) -- (C4) ;
\draw[thick] (D0) -- (D4) ;

\draw[thick] (A0) -- (D0) ;
\draw[thick] (A1) -- (D1) ;
\draw[thick] (A2) -- (D2) ;
\draw[thick] (A3) -- (D3) ;
\draw[thick] (A4) -- (D4) ;

\draw[midarrow, thick] (B1) -- (B2) node[midway, below ] {$g_{1}$};
\draw[midarrow, thick] (B2) -- (B3) node[midway, below ] {$g_{2}$};
\draw[midarrow, thick] (C1) -- (C2) node[midway, above ] {$k_{1}$};
\draw[midarrow, thick] (C2) -- (C3) node[midway, above] {$k_{2}$};
\draw[midarrow, thick] (B1) -- (C1) node[midway, below ] {$\;\;h_{1}$};
\draw[midarrow, thick] (B2) -- (C2) node[midway, below ] {$\;\;h_{2}$};
\draw[midarrow, thick] (B3) -- (C3) node[midway, below ] {$\;\;h_{3}$};

\draw[midarrow, thick,red] (B1') -- (B2') node[midway, below right] {$g_{1}$};
\draw[midarrow, thick,red] (B2') -- (B3') node[midway, below right] {$g_{2}$};

\draw[midarrow, thick,red] (B1) -- (B1') node[midway, left] {$u$};
\draw[midarrow, thick,red] (B2) -- (B2') node[midway, left] {$u_{1}$};
\draw[midarrow, thick,red] (B3) -- (B3') node[midway, left] {$u_{2}$};

\draw[midarrow, thick,red] (B1') -- (C1) node[midway,  left ] {$u^{-1}h_{1}\;\quad$};
\draw[midarrow, thick,red] (B2') -- (C2) node[midway,  left] {$u_{1}^{-1}h_{2}\;\quad$};
\draw[midarrow, thick,red] (B3') -- (C3) node[midway,  left] {$u_{2}^{-1}h_{3}\quad$};

\end{tikzpicture}
\caption{Lift by $u$ parallel transported $u_{1}=g_{1}^{-1}ug_{1},u_{2}=(g_{1}g_{2})^{-1}ug_{1}g_{2}$ to ensure flatness (except on boundary).}
\label{ribbon:fig}
\end{figure}

Consider now a ribbon operator acting along a path $\ell$ from site $(v_{1},f_{1})$ to $(v_{2},f_{2})$.
Its representation expression reads
    \begin{align}  
    {\cal L}^{\rho}_{ij}\big[\ell_{(\text{v}_{1},\text{f}_{1})\rightarrow(\text{v}_{2},\text{f}_{2})}\big]=\sum_{u\in G}
    D^{\rho}_{ij}\big(u\mathop{\square}\limits_{g_{\ell}}\big)
    \Psi\big(\left\{u\rhd_{\ell}g_{e}\right\}\big)
    \label{ribbonrepresentation}
    \end{align} 
with $g_{\ell}$ the product of the group elements encountered along the path.
For example, for the ribbon operator of figure \ref{ribbon:fig}, one has
    \begin{multline}  
    {\cal L}^{\rho}_{ij}\big[\ell_{(\text{v}_{1},\text{f}_{1})\rightarrow(\text{v}_{2},\text{f}_{2})}\big]
    \Psi(h_{1},h_{2},g_{1},g_{2},\dots)
    =\\\sum_{u\in G}
    D^{\rho}_{ij}\big(u\mathop{\square}\limits_{g_{1}g_{2}}\big)
    \Psi\big(u^{-1}h_{1},\;g_{1}^{-1}u^{-1}g_{1}\,h_{2},\;(g_{1}g_{2})^{-1}u^{-1}g_{1}g_{2}\,h_{3},\; g_{1},\; g_{2},\;\dots\big)
    \label{ribbonrepresentation2}
    \end{multline}
These operators have the following interesting properties.

\begin{itemize}
\item Ribbon operators are homotopically invariant
    \begin{align}
    {\cal L}^{\rho}_{ij}\big[\ell_{(\text{v}_{1},\text{f}_{1})\rightarrow(\text{v}_{2},\text{f}_{2})}\big]\Psi=
    {\cal L}^{\rho}_{ij}\big[\ell'_{(\text{v}_{1},\text{f}_{1})\rightarrow(\text{v}_{2},\text{f}_{2})}\big]\Psi
    \end{align}
provided $\ell$ can be deformed on $\Sigma$ (the surface on which the graph is embedded), with fixed extremities and without encountering any defect, i.e., $P_{\text{v}}\Psi=Q_{\text{f}}\Psi=\Psi$ on the surface swept by the deformation.
\item Composition of ribbons on the same path corresponds to tensor product of representations
    \begin{align}
    {\cal L}^{\rho}_{ij}\big[\ell_{(\text{v}_{1},\text{f}_{1})\rightarrow(\text{v}_{2},\text{f}_{2})}\big]
    \circ{\cal L}^{\rho'}_{i'j'}\big[\ell_{(\text{v}_{1},\text{f}_{1})\rightarrow(\text{v}_{2},\text{f}_{2})}\big]\Psi=
    {\cal L}^{\rho\otimes\rho'}_{ii'jj'}\big[\ell_{(\text{v}_{1},\text{f}_{1})\rightarrow(\text{v}_{2},\text{f}_{2})}\big]
    \end{align}
with the tensor product defined using the coproduct. 

\item Concatenation of ribbons  with the same representation is obtained by contracting the operators
    \begin{align}
    \sum_{j}{\cal L}^{\rho}_{ij}\big[\ell_{(\text{v}_{1},\text{f}_{1})\rightarrow(\text{v}_{2},\text{f}_{2})}\big]
    \circ {\cal L}^{\rho}_{jk}\big[{\ell'}_{(\text{v}_{2},\text{f}_{2})\rightarrow(\text{v}_{3},\text{f}_{3})}\big]=
    {\cal L}^{\rho}_{ik}\big[{\ell''}_{(\text{v}_{1},\text{f}_{1})\rightarrow(\text{v}_{3},\text{f}_{3})}\big].
    \end{align}
    \end{itemize}
Then, braiding using $\sigma=\tau\circ R$
    \begin{align}
    {\cal L}^{\rho\otimes\rho'}_{ii'jj'}\big[\ell_{(\text{v}_{1},\text{f}_{1})\rightarrow(\text{v}_{2},\text{f}_{2})}\big]=\sum_{klk'l'}
    \sigma_{ii'k'k}
    {\cal L}^{\rho'\otimes\rho}_{k'kl'l}\big[\ell_{(\text{v}_{1},\text{f}_{1})\rightarrow(\text{v}_{2},\text{f}_{2})}\big]
    (\sigma^{-1})_{jj'l'l}
    \end{align}
and fusing (${\cal I}$ denotes suitable intertwiners associated with the fusion channel $\alpha$)
    \begin{align}
    {\cal L}^{\rho\otimes\rho'}_{ii'jj'}\big[\ell_{(\text{v}_{1},\text{f}_{1})\rightarrow(\text{v}_{2},\text{f}_{2})}\big]=\sum_{\rho''} \sum_{\alpha=1}^{N^{\rho''}_{\rho\rho'}} \sum_{i'',j''}
    {\cal I}_{i'',ii'}^{\rho\otimes\rho'\rightarrow \rho'', \alpha}
    {\cal L}^{\rho''}_{i''j''}\big[\ell_{(\text{v}_{1},\text{f}_{1})\rightarrow(\text{v}_{2},\text{f}_{2})}\big]
    {\cal I}_{jj', j''}^{\rho''\rightarrow \rho\otimes\rho', \alpha}
    \end{align}
form the basis of a model of topological quantum computation.
In general, the ribbon operators, when acting on the vacuum, create excitations on both ends that are labelled by a pair of dual representations $\rho$ and $\rho^*$.
The representations are the label of the anyons $\rho=(\kappa,r)$ with a magnetic charge corresponding to the conjugacy class $\kappa$ and an electric charge labelled by the representation of the centralizer.
The index $i$ is an internal state, corresponding to a gauge degree of freedom.

\section{Particles and strings in magnetic fields}\label{sec2}

\paragraph{Particle in a magnetic field.}

Consider first a particle moving on a manifold ${\cal M}$, in an external magnetic field.
The latter is a closed 2-form $B$, which we assume to be exact, $B=\mathrm{d}A$.
In general, one only assumes that $\mathrm{d}B=0$ so that $B=\mathrm{d}A$ holds locally and one can patch all potentials to form a connection on a line bundle.
In the sequel, we are going to work with a discretization, so that these global considerations play no role.\footnote{
Since we will work in a discrete case afterwards, we have not insisted on an additional condition that must nevertheless be imposed in the continuous setting: the \emph{Weil integrality condition}.
Indeed, for the line bundle to be well-defined, the magnetic flux must be quantized:
\[
\frac{1}{2\pi} \int_{\Sigma} B \in \mathbb{Z}
\]
for every closed integral 2-cycle $\Sigma$.
}
Consider now a particle moving on ${\cal M}$ in this magnetic field and express the propagator \begin{align}
\psi(x,t)=\int_{{\cal M}} \mathrm{d}x_{0} K(x,t|x_{0},t_{0})\psi(x_{0},t_{0})
\end{align}
as a path integral
\begin{align}
K(x,t|x_{0},t_{0})=\int_{\substack{\gamma(t_{0})=x_{0}\\ \gamma(t)=x}} [D\gamma]
\exp\text{i}S[\gamma]\exp\text{i}\int_{\gamma}A,
\end{align} 
with $S[\gamma]$ the non-magnetic part of the action.
Note that this formulation is gauge invariant, provided we simultaneously change $A\rightarrow A+\mathrm{d}\Lambda$ and $\psi\rightarrow \exp\text{i}\Lambda\times \psi$, with $\Lambda$ a function.

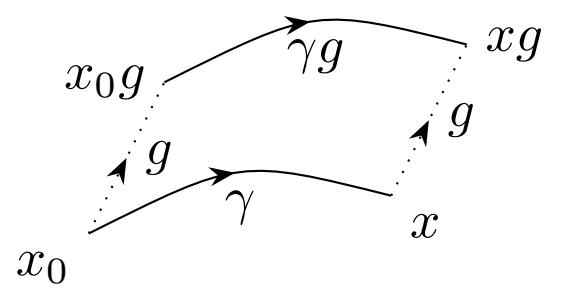
\begin{figure}[htbp]
\centering
\begin{tikzpicture}[scale=0.50,
  midarrow/.style={
    postaction={
      decorate,
      decoration={
        markings,
        mark=at position 0.5 with {\arrow{Stealth}}}}},scale=1]
  \draw[midarrow] (0,0) .. controls (2,1) .. (4,0.5) node[midway, below ] {$\gamma$};
      \draw (0,0) node[below left]{$x_{0}$};
       \draw (4,0.5) node[below right]{$x$};
         \draw[midarrow] (0+1,0+2) .. controls (2+1,1+2) .. (4+1,0.5+2) node[midway, below ] {$\gamma g$};
      \draw (0+1,0+2) node[left]{$x_{0}g$};
       \draw (4+1,0.5+2) node[right]{$xg$};
       \draw[midarrow, dotted]  (0,0) --  (0+1,0+2)node[midway, right ] {$g$};
             \draw[midarrow, dotted]  (4,0.5) --  (4+1,0.5+2)node[midway, right ] {$g$};
       \end{tikzpicture}
\caption{Translated path under the action of $g$.}
\label{translatedpath:fig}
\end{figure}

Consider now a group $G$ acting on ${\cal M}$ on the right, denoted by $x\mapsto xg$ while preserving $B$, $g^{\ast}B=B$.
However, the potential is invariant only up to a gauge transformation, assuming that every closed 1-form is also exact,
\begin{align}
\mathrm{d}(g^{\ast}A-A)=g^{\ast}B-B=0\qquad\Rightarrow\qquad g^{\ast}A-A=\mathrm{d}\phi_{g}.
\end{align}
Then, translating the path by $g\in G$ leads to (see figure \ref{translatedpath:fig})
\begin{align}
\exp\text{i}\int_{\gamma g}A= 
\exp\text{i}\int_{\gamma}g^{\ast}A=
\exp-\text{i}\phi_{g}(x_{0})\quad\times\quad\exp\text{i}\int_{\gamma t}A\quad\times\quad\exp\text{i}\phi_{g}(x)
\end{align}
which is compensated by the following magnetic translation acting on the wave function  
\begin{align}
T_{g}\psi(x)=\exp\big\{-\text{i}\phi_{g}(x)\big\}\,\psi(xg).
\end{align}
They commute with propagation\footnote{
The non-magnetic dynamics is assumed to be invariant under $G$ since the action is invariant under the translation:
\[
S[\gamma g] = S[\gamma],
\]
unlike the magnetic part.
So by including $\phi$, the magnetic translations commute with the time-evolution operator.
}, but they form only a projective representation of the group $G$
\begin{align*}
T_{g}T_{h}=\omega_{g,h}T_{gh} \qquad\text{with}\qquad
\omega_{g,h}=\exp-\text{i}\big\{g^{\ast}\phi_{h}-\phi_{gh}+\phi_{g}\big\}
\end{align*}
where $\omega$ is a constant 2-cocycle in group cohomology
    \begin{align}
    \delta\omega^{}_{g,h,k}=\omega^{}_{h,k}\;\omega^{-1}_{gh,k}\;\omega^{}_{g,hk}\;\omega^{-1}_{g,h}=1.
    \end{align}
This cocycle condition ensures associativity of the multiplication $(T_{g}T_{h})T_{k}=T_{g}(T_{h}T_{k})$.

\paragraph{Strings in an invariant magnetic field.}

For strings in a magnetic field,  which is a closed 3-form $C$ on the target space ${\cal M}$, let us assume the existence of a 2-form potential $B$ such that  $C=\mathrm{d}B$.
In general, the potential only exists locally and defines a gerbe with connection, but as soon as we move to the finite setting this does not matter.
\begin{figure}[H]
\begin{align*}
\parbox{2.2cm}{\includegraphics[width=2.2cm]{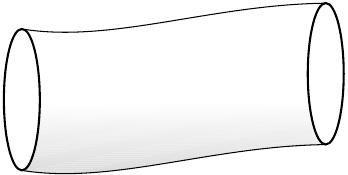}}\qquad\parbox{2cm}{\includegraphics[width=2cm]{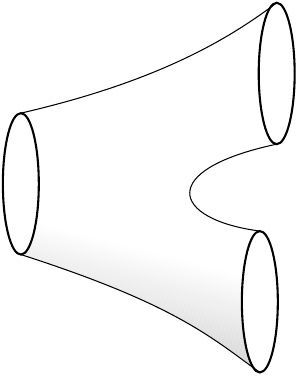}}\qquad\rightarrow\qquad \exp\text{i}\int_{\Sigma}B 
\end{align*} 
\caption{Strings in magnetic fields.}
\label{stringsB}
\end{figure}
The 2-form potential is integrated on the world sheet ${\Sigma}$ (see figure \ref{stringsB}) and the string states correspond to connected components of its boundary $\partial\Sigma$, outgoing for strings that are oriented in accordance with the orientation of $\Sigma$ and incoming if the orientations disagree. 

Consider a finite group $G$ acting on ${\cal M}$.
If $C$ is invariant, $B$ need not necessarily be invariant but there exist a 1-form $A_{g}$ and a 0-form $\Phi_{g,h}$, defined up to gauge transformations $\Lambda$ and $\Theta_{g}$ such that
    \begin{align}
    \begin{cases}
    C&=\mathrm{d}B\\
    g^{\ast}B-B&=\mathrm{d}A_{g}\\
    g^{\ast}A_{h}-A_{gh}+A_{g}&=\mathrm{d}\Phi_{g,h}
    \end{cases}
    \label{descent}
    \end{align}
defined up to gauge transformations $\Lambda$ (1-form) and $\Theta_{g}$ (0-form) such that
    \begin{align}
    \begin{cases}
    B&\rightarrow B+\mathrm{d}\Lambda\\
    A_{g}&\rightarrow A_{g}+g^{\ast}\Lambda-\Lambda+\mathrm{d}\Theta_{g}\\
    \Phi_{g,h}&\rightarrow \Phi_{g,h}+g^{\ast}\Theta_{h}-\Theta_{gh}+\Theta_{g}
    \end{cases}
    \label{descentgauge}
    \end{align}
These equations lead to a constant 3-cocycle 
    \begin{align}
    \omega_{g,h,k}=\exp-\text{i}\big\{g^{\ast}\Phi_{h,k}-\Phi_{gh,k}+\Phi_{g,hk}-\Phi_{g,h}\big\}\label{3cocycle}
    \end{align}
obeying the cocycle condition in group cohomology
    \begin{align}
    \delta\omega^{}_{g,h,k,l}=\omega^{}_{h,k,l}\;\omega^{-1}_{gh,k,l}\;\omega^{}_{g,hk,l}\;\omega^{-1}_{g,h,kl}\;\omega^{}_{g,h,k}=1.
    \end{align}
In this case, the objects of interest are twisted sectors, i.e., strings that close up to a group element: if the initial point is $x\in X$, then the end point is $xw$, with $w\in G$ labelling the twisted sector.
Then, the operators that translate the twisted sectors obey the multiplication law of the twisted Drinfel'd double, as follows from the commutation with propagation along a cylinder.

\begin{figure}[H]
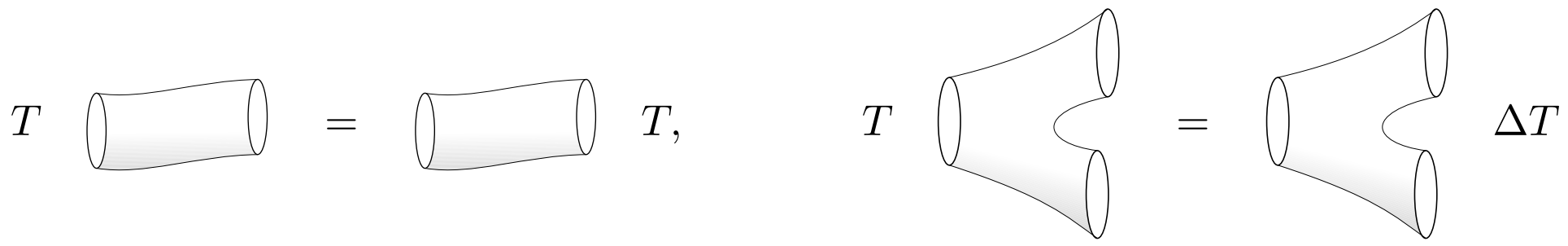

\begin{align*}
T\quad\parbox{1.5cm}{\mbox{\includegraphics[width=1.5cm]{minicy.pdf}}}
\quad=\quad
\parbox{1.5cm}{\mbox{\includegraphics[width=1.5cm]{minicy.pdf}}}
\quad T,\qquad\qquad
T\quad
\parbox{1.5cm}{\mbox{\includegraphics[width=1.5cm]{minipant.pdf}}}
\quad=\quad
\parbox{1.5cm}{\mbox{\includegraphics[width=1.5cm]{minipant.pdf}}}
\quad\Delta T
\end{align*}
\caption{Commutation with propagation and interaction.}
\end{figure}

\paragraph{Discretized analogues of differential forms.}

Since we aim at constructing a finite model on a graph, we are going to replace the target manifold ${\cal M}$ by a finite set $X$ and differential forms by their finite analogues.
In this setting, we define discrete $n$-forms $\Omega$ to be completely antisymmetric real functions of $n+1$ elements in $X$, $(x_{0},\dots,x_{n})\in X^{n+1}\mapsto \Omega(x_{0},\dots,x_{n})\in{\mathbb R}$.
Intuitively, $\Omega(x_{0},\dots,x_{n})$ can be considered as the integral of an actual $n$-form over a simplex with vertices $x_{0}$,\dots,$x_{n}$ in the manifold ${\cal M}$.
In analogy with Stokes' theorem $\int_{\Sigma}d\Omega=\int_{\partial\Sigma}\Omega$, we define the discrete differential as the alternating sum over the faces of the simplex,
    \begin{multline}
    \mathrm{d}\Omega(x_{0},\dots,x_{n},x_{n+1})=
    \Omega(x_{1},x_{2},\dots,x_{n},x_{n+1})-\Omega(x_{0},x_{2},\dots,x_{n},x_{n+1})\\
    +\cdots+(-1)^{i}\Omega(x_{0},x_{1},\dots,x_{i-1},x_{i+1},\dots,x_{n},x_{n+1})+\cdots+(-1)^{n+1}\Omega(x_{0},x_{1},\dots,x_{n})
    \label{discreted:eq}
    \end{multline}
$\mathrm{d}$ takes an $n$-form into an  $(n{+}1)$-form and is nilpotent, $\mathrm{d}^{2}=0$, so that an exact form $\Omega=\mathrm{d}\Theta$ is also closed, $\mathrm{d}\Omega=\mathrm{d}^{2}\Theta=0$.
In this very simple setting, it turns out that the converse (Poincar\'e lemma) also holds.
Indeed, suppose that $\Omega$ is a closed $(n{+}1)$-form and define an $n$-form $\Theta(x_{0},x_{1},\dots,x_{n})=\Omega(y,x_{0},x_{1},\dots,x_{n})$ with $y\in X$ a fixed constant.
Then,
    \begin{align}
    \mathrm{d}\Theta(x_{0},\dots,x_{n},x_{n+1})&=
    \Omega(y_{0},x_{1},x_{2},\dots,x_{n},x_{n+1})-\Omega(y_{0},x_{0},x_{2},\dots,x_{n},x_{n+1})\nonumber\\
    &\qquad+\cdots+(-1)^{n+1}\Omega(y_{0},x_{0},x_{1},\dots,x_{n})\\
    &=\Omega(x_{0},x_{1},x_{2},\dots,x_{n},x_{n+1})
    \end{align}
where the last relation follows from $\mathrm{d}\Omega=0$.
Moreover, the right action of a group $G$ on $X$, denoted by a simple juxtaposition $(x,g)\in X\times G\mapsto xg\in X$, induces a left action on differential forms by pull-back, $g^{\ast}\Omega(x_{0},\dots,x_{n})=\Omega(x_{0}g,\dots,x_{n}g)$.
This is indeed a left action, $(gh)^{\ast}\Omega=g^{\ast}(h^{\ast}\Omega)$, and commutes with the differential, $\mathrm{d}g^{\ast}\Omega=g^{\ast}\mathrm{d}\Omega$.
Consequently, we can use the same descent equation as in the continuous case.

\paragraph{Magnetic translation of twisted sectors.}

In this finite framework, the wave function of a particle is nothing but a function $\psi:\; X\rightarrow{\mathbb C}$, and the magnetic holonomy along a single edge from $x$ to $y$ is $\exp\text{i}A(x,y)$.
It commutes with the magnetic translation $T_{g}\psi(x)$ in the sense that 
    \begin{align}
    \chi(y)=\sum_{x\in X}\exp\text{i}A(x,y)\;\psi(x)\qquad\Rightarrow\qquad
    T_{g}\chi(y)=\sum_{x\in X}\exp\text{i}A(x,y)\; T_{g}\psi(x).
    \end{align}  
In the case of strings, we consider elementary strings given by a starting point $x$ and a winding $w$, so that their endpoint is $xw$.
Note that we retain from the string only its endpoints and parametrize them by the pair $(x,xw)$, even if $xw=x$.
We decompose the string Hilbert space as a direct sum over twisted sectors 
    \begin{align}
    {\cal H}=\mathop{\bigoplus}\limits_{w\in G} {\cal H}_{w},
    \qquad
    {\cal H}_{w}=\big\{\psi_{w}:\; X\rightarrow \mathbb{C},\; x\mapsto \psi_{w}(x)\big\}.
    \end{align}
The magnetic amplitude for a finite string twisted sector $(x,xw)$ propagating to  $(y,yw)$ with $x,y\in X$ and winding $w\in G$ reads
    \begin{align}
    {\cal K}_{w}(x,y)= \exp\text{i}\big\{B(x,y,yw)+B(yw,xw,x)+A_{w}(x,y)\big\}.
    \label{stringpropagator}
    \end{align}
This describes the propagation of the twisted sector around an elementary cylinder, triangulated by 2 triangles.
Note that there is a 2-form contribution similar to $\int_{\Sigma} B$ as well as a 1-form $\int A_{w}$ along the cut connecting the starting point of each string.
This amplitude is compatible with a gauge transformation of the string wave function as
    \begin{align}
    \psi_{w}(x)\rightarrow \exp\text{i}\big\{\Lambda(x,xw)+\Theta_{w}(x)\big\}\; \psi_{w}(x)\label{stringgauge}
    \end{align}
in the sense that a gauge transformation of the fields can be compensated by a gauge transformation on the initial and final strings. 
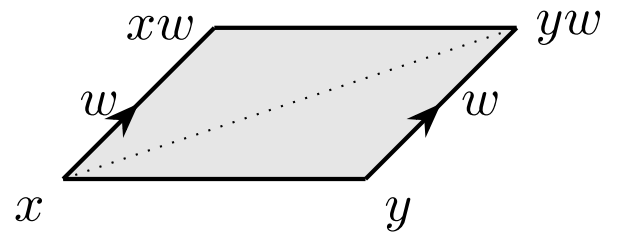
\begin{figure}[H]
\begin{center}
\begin{tikzpicture}[scale=0.50,
  midarrow/.style={
    postaction={
      decorate,
      decoration={
        markings,
        mark=at position 0.5 with {\arrow{Stealth}}}}},scale=1]
\fill[black!10] (0,0) -- (4,0) -- (6,2) -- (2,2);

 \draw[thick] (0,0) -- (4,0);
     \draw[midarrow, thick] (4,0) -- (6,2) node[midway, right] {$w$};
     \draw[thick] (6,2) -- (2,2) ;
      \draw[midarrow, thick] (0,0) -- (2,2) node[midway, left] {$w$};
           \draw[dotted] (0,0) -- (6,2) ;
      \draw (0,0) node[below left]{$x$};
       \draw (4,0) node[below right]{$y$};
      \draw (6,2) node[right]{$yw$};
       \draw (2,2) node[left]{$xw$};
\end{tikzpicture}
\end{center}
\caption{Propagation along a triangulated cylinder with a cut.}
\end{figure}

Propagation along a cylinder commutes with stringy magnetic translations $T_{g}^{w}:\;{\cal H}_{g^{-1}wg}\rightarrow {\cal H}_{w}$ defined by
    \begin{align}
    \big[T^{w}_{g}\psi\big]_{w}(x)=\exp\text{i}\big\{-A_{g}(x,xw)+\phi_{w,g}(x)-\phi_{g,g^{-1}wg}(x)\big\}\; \psi_{g^{-1}wg}(xg).
    \end{align}
Indeed,
    \begin{align}
    \chi_{g^{-1}wg}(y)=\sum_{x\in X}{\cal K}_{g^{-1}wg}(x,y)\psi_{g^{-1}wg}(x)
    \qquad\Leftrightarrow\qquad
    \big[T_{g}^{w}\chi\big]_{w}(y)=\sum_{x\in X}{\cal K}_{w}(x,y)\big[T_{g}^{w}\psi\big]_{w}(x).
    \end{align}
The latter obey the associative multiplication law
    \begin{align}
    T_{g}^{w}\;T_{h}^{v}=\delta_{g^{-1}wg,v}\;\frac{\omega_{w,g,h}\;\omega_{g,h,(gh)^{-1}wgh}}{\omega_{g,g^{-1}wg,h}}\;T^{w}_{gh}.
    \label{Tgwmultiplication}
    \end{align}
This is the multiplication in the twisted Drinfel'd double $\mathcal{D}^{\omega}[G]$ \cite{DPR_1990}, as derived in \cite{JureitKrajewski}.

Furthermore, compatibility with gauge transformations fixes the elementary amplitude for the merging of two strings ${\cal H}_{u}\otimes {\cal H}_{v}\rightarrow {\cal H}_{uv}$ up to a constant to be 
    \begin{align}
    {\cal V}_{(u,v)\rightarrow uv}(x)
    =\exp-\text{i}\big\{B(x,xu,xuv)+\Phi_{u,v}(x)\big\}\label{vertexstring}
    \end{align}
so that the wave function of the outgoing string is expressed in terms of the ingoing ones as
    \begin{align}
    \psi_{uv}(x)= {\cal V}_{(u,v)\rightarrow uv}(x)\,\eta_{u}(x)\,\zeta_{v}(xu).
    \end{align}
Commutation with stringy magnetic translations
    \begin{align}
    T^{g}_{uv}\psi_{uv}(x)&= {\cal V}_{(u,v)\rightarrow uv}(x)\,
    T^{g}_{u}\,\eta_{u}(x)\,T^{g}_{v}\,\zeta_{v}(xu)
    \qquad\Leftrightarrow\qquad\\
    \psi_{g^{-1}uvg}(xg)&= {\cal V}_{(g^{-1}ug,g^{-1}vg)\rightarrow g^{-1}uvg}(xg)\,\eta_{g^{-1}ug}(xg)\,\zeta_{g^{-1}vg}(xug)
    \end{align}
forces the action on the tensor product ${\cal H}_{u}\otimes {\cal H}_{v}$ to be
    \begin{align}
    \frac{\omega_{u,v,g} \, \omega_{g,g^{-1}ug,g^{-1}vg}}{\omega_{u,g,g^{-1}vg}}\,T^{u}_{g}\otimes T^{v}_{g}
    \end{align}
which is the coproduct in the twisted Drinfel'd double
    \begin{align}
    \Delta\!\left(T^{w}_{g} \right)=\sum_{\substack{u,v \in G \\ uv=w}} \delta_{uv,w}\,\frac{\omega_{u,v,g} \, \omega_{g,g^{-1}ug,g^{-1}vg}}{\omega_{u,g,g^{-1}vg}}\,T^{u}_{g}\otimes T^{v}_{g}.
    \label{coproduct} 
    \end{align}

\begin{figure}[H]
\begin{align*}
\parbox{3.5cm}{\includegraphics[width=3.5cm]{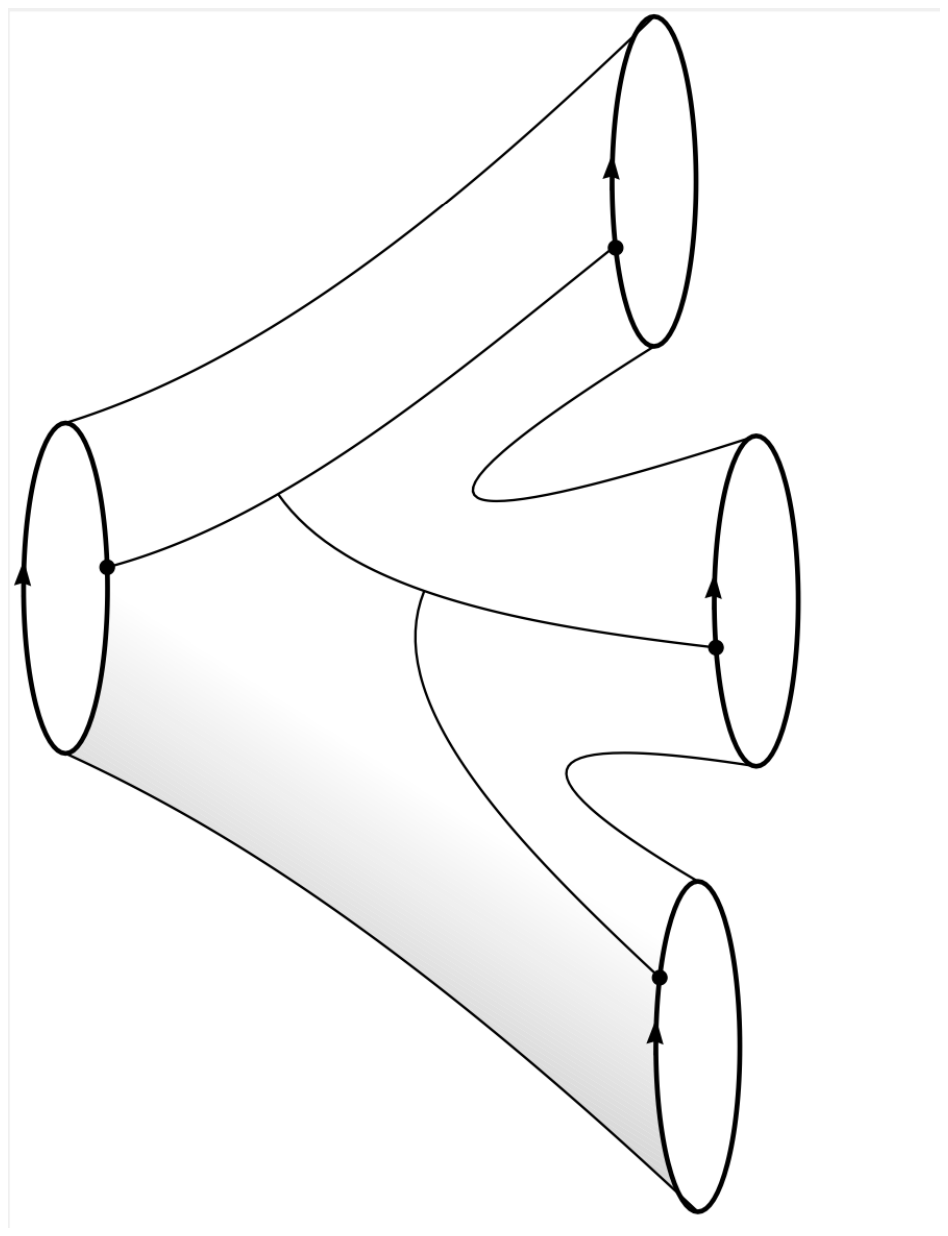}}\qquad\rightarrow\qquad
\parbox{3.5cm}{\includegraphics[width=3.5cm]{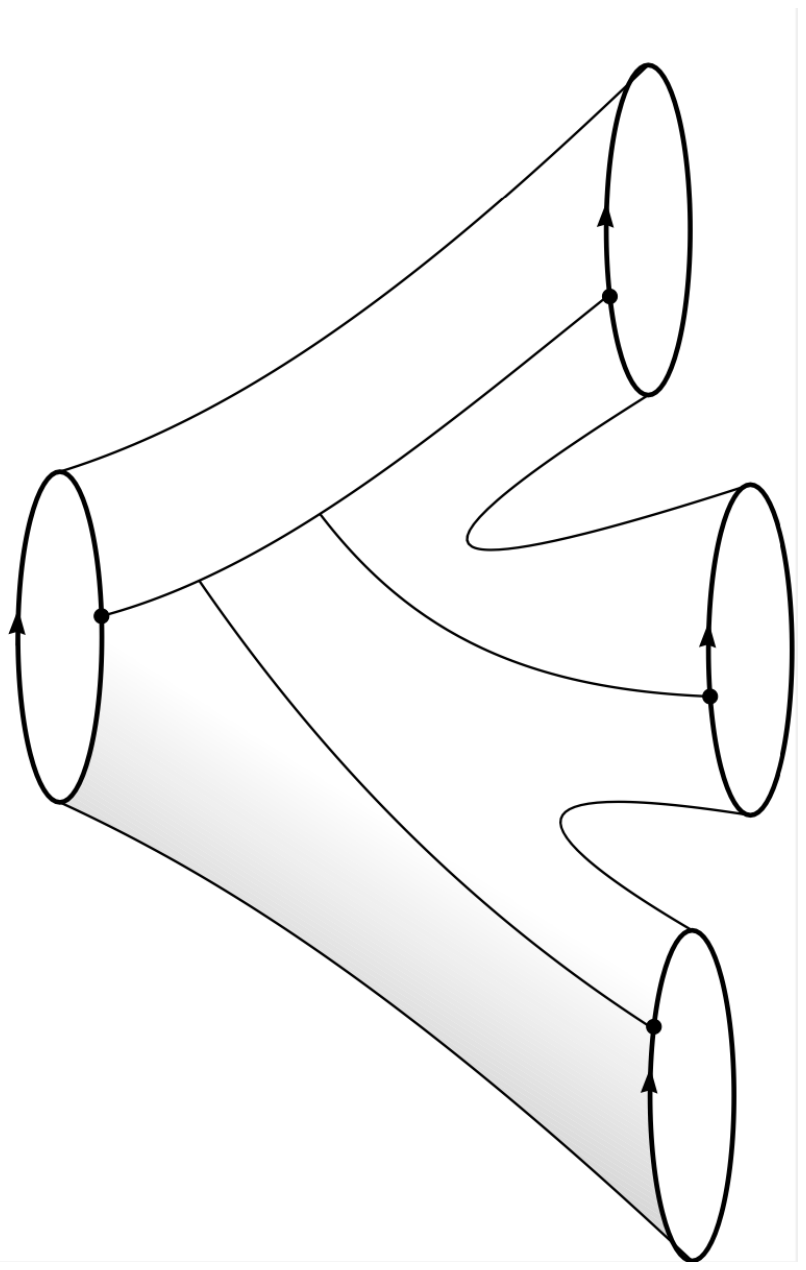}}
\end{align*}
\caption{Associator on the worldsheet.}
\label{associatorstring}
\end{figure}

\paragraph{Quasi-triangular quasi-Hopf structure from strings.}
The previous coproduct is coassociative only up to a Drinfel'd associator
    \begin{align} 
    (\mathrm{id}\otimes\Delta)\circ\Delta=\Omega\left[
    (\Delta\otimes\mathrm{id})\circ\Delta\right]\Omega^{-1}
    \qquad
    \text{with} 
    \qquad
    \Omega=\sum_{u,v,w\in G}\frac{1}{\omega_{u,v,w}}\, T_{e}^{u}\otimes T_{e}^{v}\otimes T_{e}^{w}.
    \end{align}
At the level of string states, this induces a non-trivial associator $({\cal H}_{u}\otimes {\cal H}_{v}) \otimes {\cal H}_{w}\rightarrow {\cal H}_{u}\otimes ({\cal H}_{v}\otimes {\cal H}_{w})$, materialized by the tree defining the cut on the string worldsheet, see figure \ref{associatorstring}.

\begin{figure}[H]
\begin{align*}
\parbox{2.5cm}{\includegraphics[width=2.5cm]{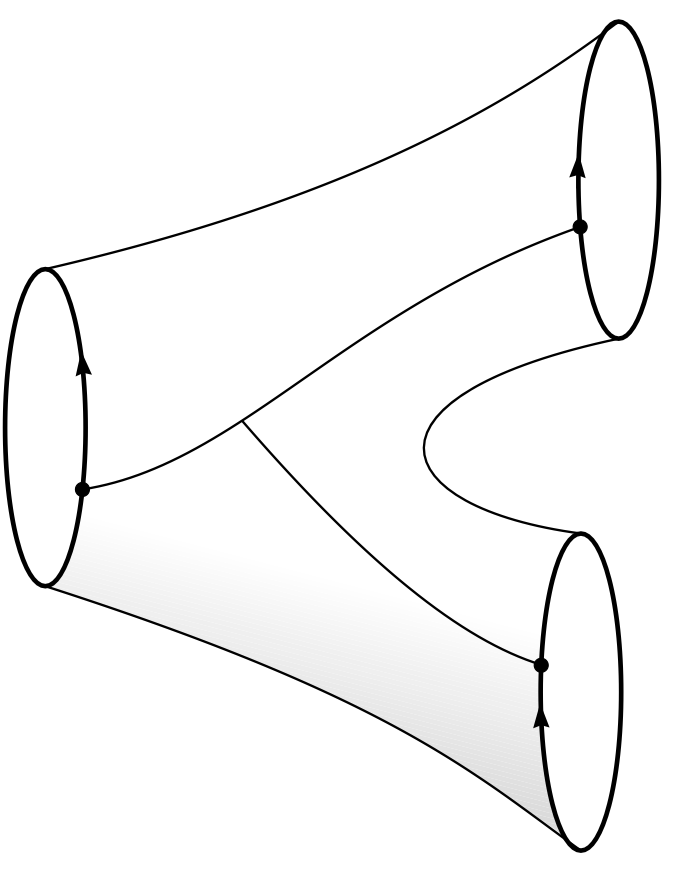}}\qquad\rightarrow\qquad
\parbox{2.7cm}{\includegraphics[width=2.7cm]{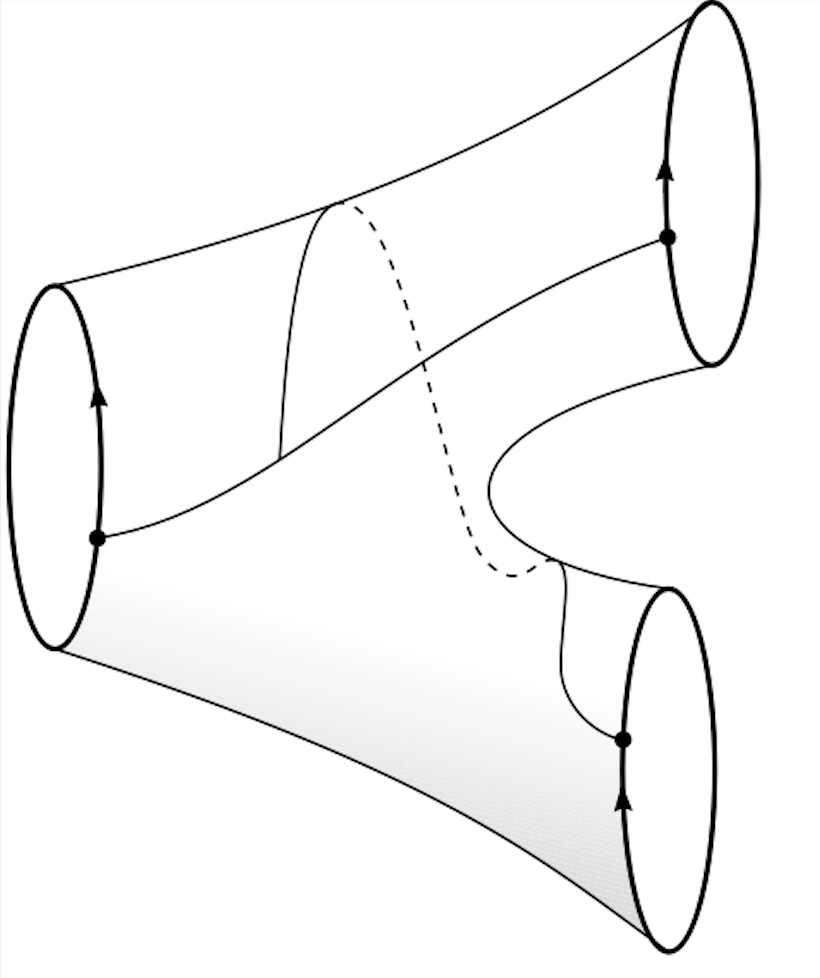}}
\end{align*}
\caption{Braiding on the string worldsheet.}
\label{stringbraiding}
\end{figure}

By the same way, there is a non trivial braiding due to the quasi-triangular structure of $\mathcal{D}^{\omega}[G]$ relating the coproduct with its opposite 
    \begin{align}
    \tau\circ \Delta=R\Delta R^{-1}
    \end{align}
with $\tau$ the flip $\tau(a\otimes b)=b\otimes a$ and the $R$-matrix
    \begin{align}
    R=\sum_{u,v}T_{\text{e}}^{u}\otimes T_{u}^{v}.
    \end{align}
This further defines a representation of the braid group by $\sigma=\tau \circ R$ acting on two factors\footnote{
Note that when more than two factors are involved, the Drinfel'd associator $\Omega$ must be used to re-associate the tensor products before braiding \cite{JureitKrajewski}.
}.
At the string theory level, it is materialized by a braiding  of the cut on a pair of pants, see figure \ref{stringbraiding}.

\paragraph{Antipode.}
Finally, there is an antipode defined by
    \begin{align}
    S\!\left(T^{w}_{g}\right)&=
    \frac{\omega_{w,g,g^{-1}w^{-1}g}}{\omega_{w,w^{-1},g}\;\omega_{g,g^{-1}wg,g^{-1}w^{-1}g}}\;
    \frac{\omega_{g,g^{-1}w^{-1}g,g^{-1}}}{\omega_{w^{-1},g,g^{-1}}\;\omega_{g,g^{-1},w^{-1}}}\;
    T^{g^{-1}w^{-1}g}_{g^{-1}}
    \\
    &=
    \frac{\omega_{w,g,g^{-1}w^{-1}g}}{\omega_{w,w^{-1},g}\;\omega_{g,g^{-1}wg,g^{-1}w^{-1}g}}
    \left[T^{w^{-1}}_{g}\right]^{-1}.
    \label{antipodeinverse}
    \end{align}
The antipode is related to the orientation reversing operation. Following the general spirit of the categorical formulation of quantum field theory, the dual is naturally identified with incoming strings, whose orientations disagree.
Accordingly, the dual space  ${\cal H}_{w}^{\ast}$ can be identified with the incoming strings through the orientation reversing operation, taking a string starting at $x$ and ending at $xw$, into one starting at $xw$ and ending at $x$   
    \begin{align}
    \psi_{w}(x)\rightarrow \psi^{\ast}_{w}(x)=\psi_{w^{-1}}(xw).
    \end{align}
This is made more precise using the following bilinear pairing between $\psi_{w}^{\ast}\in {\cal H}_{w}^{\ast}$ and $\zeta_{w}\in{\cal H}_{w}$,
    \begin{align}
    \big(\psi_{w}^{\ast}(x),\zeta_{w}(x)\big)=\exp-\text{i}\Phi_{w,w^{-1}(x)}\;\psi_{w}(x)\zeta_{w}(x).
    \end{align}
Then the antipode renders the following diagram commutative
\begin{center}
\begin{tikzcd}
  {\cal H}_{g^{-1}wg} \arrow[r, "\ast"] \arrow[d, "T_{g}^{w}"]
    & {\cal H}_{g^{-1}w^{-1}g}\simeq {\cal H}_{g^{-1}wg}^{\ast} \\
  {\cal H}_{w} \arrow[r, "\ast" ]
&  {\cal H}_{w^{-1}}\simeq {\cal H}_{w}^{\ast}\arrow[u, "S\big(T_{g}^{w}\big)"]
\end{tikzcd}
\end{center}
where $\ast$ stands for the orientation reversal operation.

\section{A  twisted model based on strings in  a magnetic field}\label{sec3}

\paragraph{Hilbert space and Hamiltonian.}
The Hilbert space of the model we present is associated to an oriented ribbon graph $\Gamma=(E,V,F)$ made of edges $E$, vertices $V$ and faces $F$. 
This is an ordinary graph, with edges connecting the vertices and a cyclic orientation of the half-edges at each vertex, so that faces are well defined.
Alternatively, it can be constructed using edges which are oriented ribbons, whose sides alternate between short sides and long sides.
The short sides connect vertices that are oriented polygons, and the faces are the connected components of its boundary.
It can be embedded in a surface $\Sigma$ of genus $2-2g(\Sigma)=V-E+F$, in such a way that each faces of $\Gamma$ bounds a disc in $\Sigma$. 

Given a finite group $G$ and  a finite set $X$, we assume that there is a right action of $G$ on $X$, denoted by $(g,x)\in G\times X\mapsto xg\in X$, such that $(xg)h=x(gh)$.
To each vertex v, we associate a variable $x^{\text{v}}\in X$ and to each edge e a pair of group elements $(g_{\text{e}},w_{\text{e}})$, respectively called longitudinal  and transversal.
Then, the Hilbert space is ${\cal H}=L^{2}(G^{2|E|}\times X^{|V|})$, with $|A|$ denoting the cardinality of the set $|A|$.
It has dimension $|G|^{2|E|}X^{|V|}$ and admits a canonical basis $\big|\left\{g_{\text{e}},w_{\text{e}},x_{v}\right\}\big\rangle$, so that every state vector is determined by its wave function $\Psi$ from $G^{2|E|}\times X^{|V|}$
    \begin{align}
    |\Psi\rangle=
    \sum_{\left\{g_{\text{e}},w_{\text{e}},x_{v}\right\}\in G^{2|E|}\times X^{|V|} }\Psi\big(\left\{g_{\text{e}},w_{\text{e}},x_{v}\right\}\big) \big|\left\{g_{\text{e}},w_{\text{e}},x_{v}\right\}\big\rangle.
    \end{align}
To construct the Hamiltonian, we need to specify the assignment of the edge and vertex variables.

\begin{itemize}[label=\textbullet]
\item Every edge is a triangulated rectangle with a longitudinal variable $g_{\text{e}}$ associated to the long sides and transversal variables $g_{\text{e}}^{-1}w_{\text{e}}g_{\text{e}}$ and outgoing $w_{\text{e}}^{-1}$ to the short edges.
Note that the short sides inherit their orientation from that of the rectangle defining the edge, see figure \ref{ribbonedge}.

\begin{figure}[H]
\begin{center}
\begin{tikzpicture}[scale=0.75,
  midarrow/.style={
    postaction={
      decorate,
      decoration={
        markings,
        mark=at position 0.5 with {\arrow{Stealth}}}}},scale=1]
        
    \coordinate (A) at (0,0);
     \coordinate (B) at (6,0);
    \coordinate (C) at (7,1);
        \coordinate (D) at (1,1);

\fill[black!10] (A) -- (B) -- (C) -- (D);

\draw[midarrow,thick] (A) -- (B) node[midway, below] {$g$};

\draw[midarrow,thick] (D) -- (A) node[midway, above left] {$w^{-1}$};

\draw[midarrow,thick] (B) -- (C) node[midway, right] {$g^{-1}wg$};

\draw[midarrow,thick] (D) -- (C) node[midway, above ] {$g$};

\draw[dotted] (A) -- (C) ;
\end{tikzpicture}
\caption{Ribbon edge.}
\label{ribbonedge}
\end{center}
\end{figure}
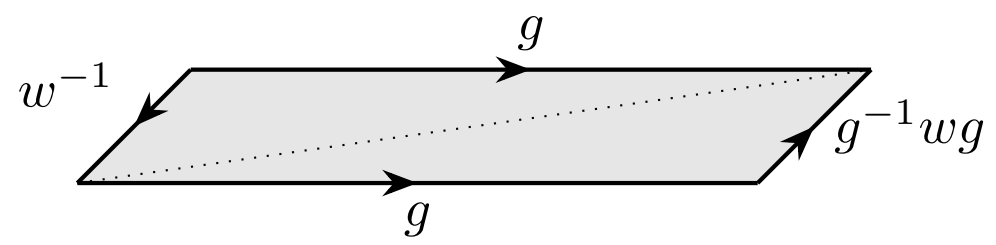

\item Every vertex with $n$ incident edges is a triangulated polygon with $n+1$ sides, a corner carrying $x_{v}\in X$, next $n$ sides attached to transversal elements $w_{1},\dots,w_{n}$ and $x_{\text{v}}w_{1}\cdots w_{i}$ at corner $i$, with defect $w_{1},\dots,w_{n}$ on last edge.
Note that the sides corresponding to the incident edges are negatively oriented with respect to that of the surface, only the side carrying the defect is positively oriented.
This vertex is triangulated and we assume that all the triangles have $x_{\text{v}}$ as a common vertex.  

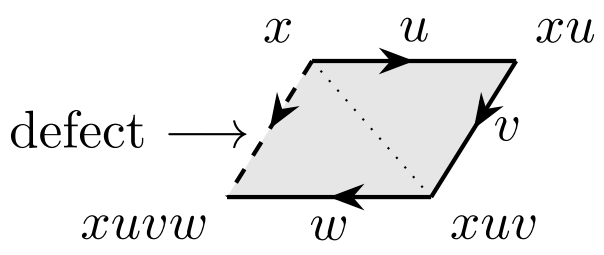
\begin{figure}[H]
\begin{center}
\begin{tikzpicture}[scale=0.75,
  midarrow/.style={
    postaction={
      decorate,
      decoration={
        markings,
        mark=at position 0.5 with {\arrow{Stealth}}}}},scale=0.3]      
 \pgfmathsetmacro{\a}{3}
 \pgfmathsetmacro{\b}{2}
 \pgfmathsetmacro{\c}{2.5}

\coordinate (A) at (-\a,-\b);
\coordinate (B) at (\a,-\b);
\coordinate (C) at (\a+\c,\b);
\coordinate (D) at (-\a+\c,\b);

\fill[black!10] (A) -- (B) -- (C) -- (D) -- (A);

\draw[thick,dashed,midarrow]  (D) -- (A) node[midway, left] {defect $\longrightarrow$};

\draw[dotted] (B) -- (D);

\draw[midarrow,thick] (B) -- (A) node[midway, below] {$w$};
\draw[midarrow,thick] (C) -- (B) node[midway, right] {$v$};

\draw[midarrow,thick] (D) -- (C) node[midway, above] {$u$};

\draw (A) node[below left]{$xuvw$};
\draw (B) node[below right]{$xuv$};
\draw (C) node[above right]{$xu$};
\draw (D) node[above left]{$x$};

\end{tikzpicture}
\end{center}
\caption{A trivalent vertex.}
\end{figure}

\end{itemize}

Note that vertices and edges have been triangulated, for later use.
For the sake of simplicity, some choices have been made: orientations of sides, location of the defect on the vertex, triangulation, ...
Different choices lead to an equivalent Hilbert space ${\cal H}'$, with a specific unitary transformation $U:\;{\cal H}'\rightarrow {\cal H}$.
The graph $\Gamma$ is obtained by gluing the short sides of the edges to the sides of the polygonal vertices matching the transversal elements, except the ones carrying the defect which are left unpaired. 

From a string theory viewpoint, the transversal variables $w$ are winding, so that $(x,xw)$ represents a twisted sector, and longitudinal variables $g$ finite group gauge transformation. 
The formalism developed in the previous section on stringy magnetic amplitudes will be instrumental in constructing the Hamiltonian, but not that the surface $\Sigma$ on which the graph is embedded is not the string worldsheet.
Rather, it is more convenient to think of the model presented here as a finite group gauge theory on the surface $\Sigma$, with twisted sectors and gauge fields as degrees of freedom.

We seek to construct a Hamiltonian similar to that of the Kitaev quantum double model, i.e., a sum of mutually commuting projectors, 
    \begin{align}
    H=\sum_{\text{v}}\lambda_{\text{v}}\big(1-\Pi_{\text{v}}\big)+\sum_{\text{v}}\mu_{\text{v}}\big(1-P_{\text{v}}\big)
    +\sum_{\text{f}}\nu_{\text{f}}\big(1-Q_{\text{f}}\big)
    \end{align}
with $\lambda_{\text{v}},\mu_{\text{v}},\nu_{\text{f}}$ positive coupling constants.
This allows to easily describe the ground state by the conditions $\Pi_{\text{v}}|\Psi_{0}\rangle=|\Psi_{0}\rangle$, $P_{\text{v}}|\Psi_{0}\rangle=|\Psi_{0}\rangle$ and $Q_{\text{f}}|\Psi_{0}\rangle=|\Psi_{0}\rangle$ for any vertex v and face f, as well as the elementary excitations above the ground state.

\begin{itemize}[label=\textbullet]
\item $\Pi_{\text{v}}$ projects onto magnetic amplitudes and fully determines the $x$ dependences;
\item $P_{\text{v}}$ projects onto gauge invariant states at vertex, i.e., vanishing electric and magnetic charges;
\item $Q_{{\text{f}}}$ imposes a flatness condition relating the gauge holonomy and the defects around face f and translates $w$.
\end{itemize}
Note that it is possible to separate the vanishing of the electric and magnetic charges, as well as the flatness condition and translation by $w$, thus leading to five sets of commuting projectors.
In the original formulation of Kitaev \cite{Kitaev_2003}, it was already pointed out that the Kitaev model can be formulated for a general Hopf algebra, with excitations constructed using the quantum double of this Hopf algebra.
This model with five projectors would then correspond to a model based on the twisted quantum double $\mathcal{D}^{\omega}[G]$, with excitations based on the quantum double of $\mathcal{D}^{\omega}[G]$.
It would need a slight extension of Kitaev's formulation since $\mathcal{D}^{\omega}[G]$ is only a quasi-Hopf algebra, but its quantum double is well defined, see \cite{majid1998quantumdoublequasihopfalgebras}.
In the sequel, we restrict our attention to this set of three projectors, as they are already sufficient to ensure uniqueness of the ground state on a genus 0 surface.
In the remainder of this section, we present the construction of these projectors.
To this aim, we assume that we are given on $X$ a discretized 3-form $C$, which is closed and invariant.
Therefore, we can also derive a sequence of forms, $B$, $A_{g}$ and $\Phi_{g,h}$ ending in a 3-cocycle $\omega_{g,h,k}$ according to the descent equations \eqref{descent}.

\paragraph{Projection onto magnetic amplitude.}

For a vertex $\text{v}$ with defect $w_{\text{v}}=w_{1}\cdots w_{n}$, we define the projector as
    \begin{multline} \Pi_{\text{v}}\Psi\big(\left\{g_{\text{e}},w_{\text{e}},x_{\text{v'}}\right\}\big)=\frac{1}{|X|}\sum_{y_{\text{v}}\in X}
    \exp\text{i}\left\{C\left(\tetrasymbol \text{ on $y_{\text{v}}, x_{\text{v}},\left\{g_{\text{e}},w_{\text{e}}\right\}_{\text{v}\ni \text{e}}$}\right)\right\} \\\underbrace{\exp\text{i}\big\{B(y_{\text{v}},x_{\text{v}},x_{\text{v}}w_{\text{v}})+B(x_{\text{v}}w_{\text{v}},y_{\text{v}}w_{\text{v}},y_{\text{v}})+A_{w_{\text{v}}}(y_{\text{v}},x_{\text{v}})\big\}}_{\text{string propagator } {\cal K}_{w_{\text{v}}}(y_{\text{v}},x_{\text{v}})} \Psi\big(\left\{g_{\text{e}},w_{\text{e}},x_{\text{v'}\neq \text{v}},y_{\text{v}}\right\}\big)
    \label{definitionPiv}
    \end{multline}
where in the last line $x_{\text{v}}$ has been replaced by $y$. 
This projector only acts on the variable $x_{\text{v}}$, but reads off all the variables on the vertex v and on the edges related to v.
$C\big(\tetrasymbol \text{ on $y_{\text{v}}, x_{\text{v}},\left\{g_{\text{e}},w_{\text{e}}\right\}_{\text{v}\ni \text{e}}$}\big)$ is a product of tetrahedra equipped with the 3-form $C$, triangulating the volume between the initial graph and the lifted one, with all faces oriented outwards.
Finally, on the face between the initial defect $(y_{\text{v}},y_{\text{v}}w_{\text{v}})$ and the final one $(x_{\text{v}},x_{\text{v}}w_{\text{v}})$, we insert a string propagator ${\cal K}_{{\cal H}_{w_{\text{v}}}\rightarrow {\cal H}_{w_{\text{v}}}}(y_{\text{v}},x_{\text{v}})$, see equation \eqref{stringpropagator}.

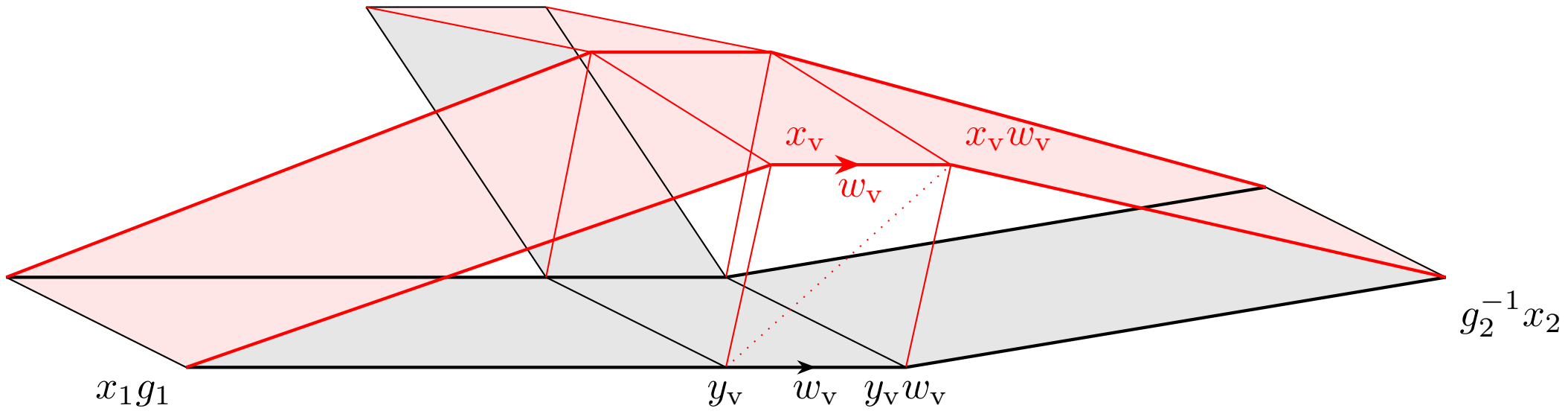
\begin{figure}[H]
\begin{center}
\begin{tikzpicture}[scale=0.8,
  midarrow/.style={
    postaction={
      decorate,
      decoration={
        markings,
        mark=at position 0.5 with {\arrow{Stealth}}}}},scale=1]

  \pgfmathsetmacro{\l}{2}
  \pgfmathsetmacro{\h}{2.5}

 \coordinate (A0) at (0-2*\l,0+2*\l);
\coordinate (A1) at (-1*\l-2*\l,1*\l+3/2*\l);
\coordinate (B0) at (\l,\l+\l);
\coordinate (B0T) at (\l+0.2*\h,\l+\l+0.9*\h);
 \coordinate (B1) at (0,3/2*\l+\l);
\coordinate (B1T) at (0+0.2*\h,3/2*\l+\l+\h);
\coordinate (B2) at (-\l,3*\l+\l);
\coordinate (C0) at (\l+\l,\l+\l);
\coordinate (C0T) at (\l+\l+0.2*\h,\l+\l+0.9*\h);
\coordinate (C1) at (0+\l,3/2*\l+\l);
\coordinate (C1T) at (0+\l+0.2*\h,3/2*\l+\l+\h);
\coordinate (C2) at (-\l+\l,3*\l+\l);
\coordinate (D0) at (5*\l,5/2*\l);
\coordinate (D1) at (4*\l,3*\l);

    \fill[black!10] (A0) -- (B0) -- (C0) -- (D0) -- (D1)-- (C1)-- (C2)-- (B2)-- (B1)-- (A1)-- (A0);

    \fill[red!10] (A0) -- (B0T) -- (C0T) -- (D0) -- (D1)-- (C1T)-- (C2)-- (B2)-- (B1T)-- (A1)-- (A0);

\draw (A0) -- (A1) ;
 \draw (B0) -- (B1) -- (B2);
\draw (C0) -- (C1) -- (C2);
\draw (D0) -- (D1) ;
\draw (B2) -- (C2) ;

\draw[thick] (A0) -- (B0) -- (C0) -- (D0);
\draw[thick] (A1) -- (B1) -- (C1) -- (D1);
\draw[] (C0) -- (D0) ;

\draw (B0) node[below ] {$y_{\text{v}}$} ;
\draw (C0) node[below ] {$y_{\text{v}}w_{\text{v}}$} ;

\draw[red] (C0) -- (C0T);
\draw[red] (B1) -- (B1T) ;
\draw[red] (B0) -- (B0T) ;
\draw[red] (C1) -- (C1T) ;

\draw[red] (B0T) -- (B1T) --(B2) ;
\draw[red] (C0T) -- (C1T) --(C2) ;

\draw[red] (B0T) node[above right] {$x_{\text{v}}$} ;
\draw[red] (C0T) node[above right] {$x_{\text{v}}w_{\text{v}}$} ;

\draw[red,dotted] (B0) -- (C0T) ;

\draw[red, thick] (A1) -- (B1T) -- (C1T) -- (D1);

\draw[] (A0) -- (B0)  ;
\draw (A0) node[below left]{$x_{1}g_{1}$};
\draw (D0) node[below right]{$g_{2}^{-1}x_{2}$};

\draw[midarrow] (B0) -- (C0) node[midway, below ] {$w_{\text{v}}$} ;

\draw[red,thick] (A0) -- (B0T)  ;

\draw[red, midarrow,thick] (B0T) -- (C0T) node[midway, below ] {$w_{\text{v}}$} ;
\draw[red, thick] (C0T) -- (D0) ;



 ;


\end{tikzpicture}
\end{center}
\caption{Construction of the projector $\Pi_{\text{v}}$, with $g_{1}$ incoming from $x_{1}$ and $g_{2}$ outgoing to $x_{2}$.}
\label{projectorPi}
\end{figure}

First, let us explain why $\Pi_{\text{v}}\Pi_{\text{v}}=\Pi_{\text{v}}$.
When composing two such operators, one piles up two series of triangulations, and aims at rewriting them as a single series of triangulation, as depicted on figure \ref{projectorPi}, moving from bottom to top.
In the process, there are missing tetrahedra on both sides of the edges, that cancel because of the invariance of $C$.
There is also a missing prism at the level of the defect, but the latter is compensated by the multiplication of string propagators (see \eqref{stringpropagator}), that obey
    \begin{align}
    {\cal K}_{w}(z,y)\,
    {\cal K}_{w}(y,x)=
    \exp\text{i}\big\{C(z,y,x,xw)+C(yw,xw,z,zw)+C(y,xw,yw,z)\big\}\,{\cal K}_{w}(z,x).
    \end{align}
Obviously, $\Pi_{\text{v}}$ and $\Pi_{\text{v'}}$ commute if the vertices $\text{v}$ and $\text{v'}$ do not share an edge.
If $\text{v}$ and $\text{v'}$ are related by an edge $\text{e}$ the triangulations constructed in  the product  $\Pi_{\text{v}}\Pi_{\text{v'}}$ are separated by a translation by the longitudinal group element $g_{\text{e}}$.
The invariance of $C$ and the commutation of the string propagator with stringy magnetic translation allow to equalize the triangulation and propagators $\Pi_{\text{v}}\Pi_{\text{v}'}$ with that in  $\Pi_{\text{v}'}\Pi_{\text{v}}$.

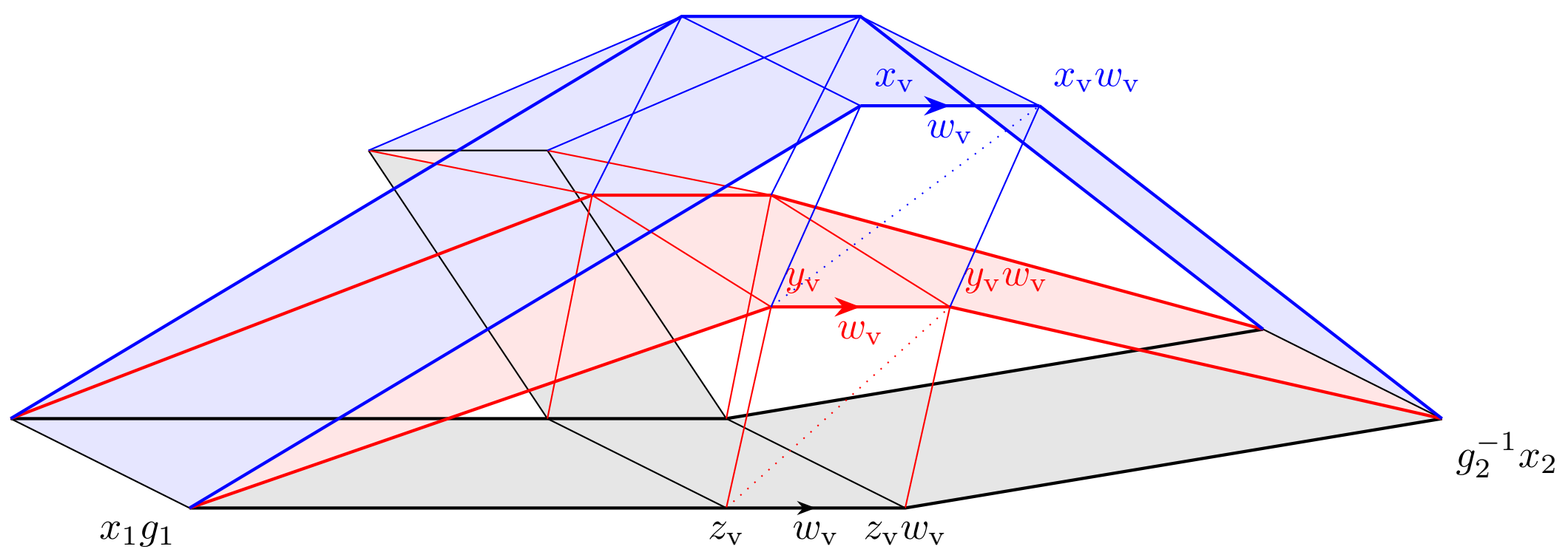
\begin{figure}[H]
\begin{center}
\begin{tikzpicture}[scale=0.8,
  midarrow/.style={
    postaction={
      decorate,
      decoration={
        markings,
        mark=at position 0.5 with {\arrow{Stealth}}}}},scale=1]

  \pgfmathsetmacro{\l}{2}
  \pgfmathsetmacro{\h}{2.5}

 \coordinate (A0) at (0-2*\l,0+2*\l);
\coordinate (A1) at (-1*\l-2*\l,1*\l+3/2*\l);
\coordinate (B0) at (\l,\l+\l);
\coordinate (B0T) at (\l+0.2*\h,\l+\l+0.9*\h);
 \coordinate (B1) at (0,3/2*\l+\l);
\coordinate (B1T) at (0+0.2*\h,3/2*\l+\l+\h);
\coordinate (B2) at (-\l,3*\l+\l);
\coordinate (C0) at (\l+\l,\l+\l);
\coordinate (C0T) at (\l+\l+0.2*\h,\l+\l+0.9*\h);
\coordinate (C1) at (0+\l,3/2*\l+\l);
\coordinate (C1T) at (0+\l+0.2*\h,3/2*\l+\l+\h);
\coordinate (C2) at (-\l+\l,3*\l+\l);
\coordinate (D0) at (5*\l,5/2*\l);
\coordinate (D1) at (4*\l,3*\l);

\coordinate (B0TT) at (\l+0.6*\h,\l+\l+1.8*\h);
\coordinate (B1TT) at (0+0.6*\h,3/2*\l+\l+1.8*\h);
\coordinate (C0TT) at (\l+\l+0.6*\h,\l+\l+1.8*\h);
\coordinate (C1TT) at (0+\l+0.6*\h,3/2*\l+\l+1.8*\h);

    \fill[black!10] (A0) -- (B0) -- (C0) -- (D0) -- (D1)-- (C1)-- (C2)-- (B2)-- (B1)-- (A1)-- (A0);

    \fill[red!10] (A0) -- (B0T) -- (C0T) -- (D0) -- (D1)-- (C1T)-- (C2)-- (B2)-- (B1T)-- (A1)-- (A0);
;

    \fill[blue!10] (A0) -- (B0TT) -- (C0TT) -- (D0) -- (D1)-- (C1TT)-- (B1TT)-- (A1)-- (A0);
    
        \fill[blue!10] (A0) -- (B0TT) -- (C0TT) -- (C1TT) -- (B1TT)-- (A1);
    
     \fill[blue!10] (C1TT)-- (B1TT)-- (B2)-- (C2);

\draw (A0) -- (A1) ;
 \draw (B0) -- (B1) -- (B2);
\draw (C0) -- (C1) -- (C2);
\draw (D0) -- (D1) ;
\draw (B2) -- (C2) ;

\draw[thick] (A0) -- (B0) -- (C0) -- (D0);
\draw[thick] (A1) -- (B1) -- (C1) -- (D1);
\draw[] (C0) -- (D0) ;

\draw (B0) node[below ] {$z_{\text{v}}$} ;
\draw (C0) node[below ] {$z_{\text{v}}w_{\text{v}}$} ;

\draw[red] (C0) -- (C0T);
\draw[red] (B1) -- (B1T) ;
\draw[red] (B0) -- (B0T) ;
\draw[red] (C1) -- (C1T) ;

\draw[red] (B0T) -- (B1T) --(B2) ;
\draw[red] (C0T) -- (C1T) --(C2) ;

\draw[red] (B0T) node[above right] {$y_{\text{v}}$} ;
\draw[red] (C0T) node[above right] {$y_{\text{v}}w_{\text{v}}$} ;

\draw[red,dotted] (B0) -- (C0T) ;

\draw[red, thick] (A1) -- (B1T) -- (C1T) -- (D1);

\draw[] (A0) -- (B0)  ;
\draw (A0) node[below left]{$x_{1}g_{1}$};
\draw (D0) node[below right]{$g_{2}^{-1}x_{2}$};

\draw[midarrow] (B0) -- (C0) node[midway, below ] {$w_{\text{v}}$} ;

\draw[red,thick] (A0) -- (B0T)  ;

\draw[red, midarrow,thick] (B0T) -- (C0T) node[midway, below ] {$w_{\text{v}}$} ;
\draw[red, thick] (C0T) -- (D0) ;

\draw[blue] (C0T) -- (C0TT);
\draw[blue] (B1T) -- (B1TT) ;
\draw[blue] (B0T) -- (B0TT) ;
\draw[blue] (C1T) -- (C1TT) ;

\draw[blue] (B0TT) -- (B1TT) --(B2) ;
\draw[blue] (C0TT) -- (C1TT) --(C2) ;

\draw[blue] (B0TT) node[above right] {$x_{\text{v}}$} ;
\draw[blue] (C0TT) node[above right] {$x_{\text{v}}w_{\text{v}}$} ;

\draw[blue,dotted] (B0T) -- (C0TT) ;

\draw[blue, thick] (A1) -- (B1TT) -- (C1TT) -- (D1);

\draw[blue,thick] (A0) -- (B0TT)  ;

\draw[blue, midarrow,thick] (B0TT) -- (C0TT) node[midway, below ] {$w_{\text{v}}$} ;
\draw[blue, thick] (C0TT) -- (D0) ;



 ;


\end{tikzpicture}
\end{center}
\caption{Composition $\Pi_{\text{v}}\Pi_{\text{v}}=\Pi_{\text{v}}$.}
\label{projectorP}
\end{figure}

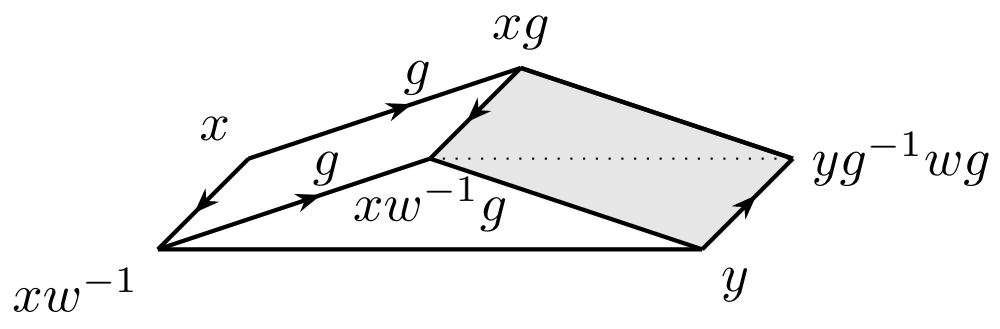
\begin{figure}[H]
\begin{center}
\begin{tikzpicture}[scale=0.60,
  midarrow/.style={
    postaction={
      decorate,
      decoration={
        markings,
        mark=at position 0.6 with {\arrow{Stealth}}}}},scale=1]
    \coordinate (A) at (0,0);
     \coordinate (B) at (6,0);
    \coordinate (C) at (7,1);
        \coordinate (D) at (1,1);
  \coordinate (B') at (3,1);
    \coordinate (C') at (4,2);
   \fill[black!10] (B') -- (B) -- (C) -- (C');
\draw[thick] (A) -- (B) ;
\draw[thick] (B) -- (C) ;
\draw[dotted] (B') -- (C) ;
\draw[thick] (C') -- (C) ;
\draw[thick] (D) -- (A) ;
\draw[thick] (B') -- (C') ;
\draw[thick] (A) -- (B') ;
\draw[thick] (B') -- (B) ;
\draw[thick] (D) -- (C') ;
\draw[thick] (C) -- (C') ;
\draw[midarrow] (C') -- (B') ;
\draw[midarrow] (D) -- (A) ;
\draw[midarrow] (B) -- (C) ;
\draw (A) node[below left]{$xw^{-1}$};
\draw (B) node[below right]{$y$};
\draw (B') node[below]{$xw^{-1}g$};
\draw (C') node[above]{$xg$};
\draw (D) node[above left]{$x$};
\draw (C) node[right]{$yg^{-1}wg$};
\draw[midarrow] (A) -- (B') node[midway, above right] {$g$};
\draw[midarrow] (D) -- (C') node[midway, above right] {$g$};
\end{tikzpicture}
\end{center}
\caption{Feynman rule for the edge.}
\begin{align*}
\Psi_{\substack{\text{\tiny string}\\\text{\tiny edge}}}(g,w,x,y)&=
\exp\text{i}\big\{B(xw^{-1}g,y,yg^{-1}wg)+B(yg^{-1}wg,xg,xw^{-1}g)+A_{g^{-1}wg}(xw^{-1}g,y)\big\}\\
&\quad\times 
\exp\text{i}\big\{\Phi_{w^{-1},w}(x)+A_{g}(xw^{-1},x)-\Phi_{w,g}(xw^{-1})+\Phi_{g,g^{-1}wg}(xw^{-1})\big\}
\end{align*}
\label{Feynmanedge}
\end{figure}

Then, the product of all the projectors is itself a projector and projects onto their common range.
It turns out that the dependence of the latter on the $x$ variable is entirely fixed, 
    \begin{align}
    \Pi_{\text{v}}\Psi=\Psi,\;\forall \text{v}
    \qquad\Leftrightarrow\qquad
    \Psi(\left\{g_{\text{e}},w_{\text{e}},x_{\text{v}}\right\})=
    \Psi_{\text{\tiny string}}(\left\{g_{\text{e}},w_{\text{e}},x_{\text{v}}\right\})\times F(\left\{g_{\text{e}},w_{\text{e}}\right\})
    \label{PivString}
    \end{align}
with $\Psi_{\text{\tiny string}}(\left\{g_{\text{e}},w_{\text{e}},x_{\text{v}}\right\})$ constructed in the sequel and $F(\left\{g_{\text{e}},w_{\text{e}}\right\})$ an arbitrary function of the group variables.
To explicitly construct $\Psi_{\text{\tiny string}}(\left\{g_{\text{e}},w_{\text{e}},x_{\text{v}}\right\})$, recall that the graph is made by gluing edges and vertices with the short sides of the vertices oriented as outgoing (same orientation as the rectangle) and the side of the vertices as incoming (opposite orientation).
Then, $\Psi_{\text{\tiny string}}(\left\{g_{\text{e}},w_{\text{e}},x_{\text{v}}\right\})$ is obtained by multiplying contributions of edges and trivalent vertices, see figures \ref{Feynmanedge} and \ref{Feynmanvertex},
    \begin{align}
    \Psi_{\text{\tiny string}}=\prod_{\text{\tiny edge}} \Psi_{\substack{\text{\tiny string}\\\text{\tiny edge}}} \prod_{\text{\tiny vertex}}\Psi_{\substack{\text{\tiny string}\\\text{\tiny vertex}}}.
    \end{align}

\begin{figure}[h]
\begin{center}
\begin{tikzpicture}[scale=1.2,
  midarrow/.style={
    postaction={
      decorate,
      decoration={
        markings,
        mark=at position 0.5 with {\arrow{Stealth}}}}},scale=1]
    \coordinate (A) at (-1,0);
     \coordinate (B) at (0.2,-0.5);
    \coordinate (C) at (0.5,0.6);
    \fill[black!10] (A) -- (B) -- (C) -- (A);
\draw[midarrow,thick] (A) -- (B) ;
\draw[midarrow, thick] (C) -- (B) ;
\draw[midarrow, thick] (A) -- (C) ;
\draw (A) node[left]{$x$};
\draw (B) node[right]{$xuv$};
\draw (C) node[right]{$xu$};
\end{tikzpicture}
\end{center}
\begin{align*}
\Psi_{\substack{\text{\tiny string}\\\text{\tiny vertex}}}(u,v,x)=
\exp-\text{i}\big\{B(x,xu,xuv)+\Phi_{u,v}(x)\big\}
\end{align*}
\caption{Feynman rule for the vertex.}
\label{Feynmanvertex}
\end{figure}
The Feynman rule for the edge is obtained by first applying the projector $\Pi_{\text{v}_{y}}$ at the vertex carrying the variable $x$ to the constant function 1,  $\Pi_{\text{v}_{y}}1$, which generate the first line in figure~\ref{Feynmanedge}.
Then, we apply the projector at the second vertex, $\Pi_{\text{v}_{x}}\Pi_{\text{v}_{y}}1$, which generates the second line.
The vertex Feynman rule is nothing but the amplitude for the merging of two strings, see \eqref{vertexstring}.

\paragraph{Finite group gauge transformations at vertices.}

Our next step consists in defining the projector $P_{\text{v}}$, that implements the finite group gauge transformation at vertex $\text{v}$, which is made of independent transformations by group elements $g_{\text{v}}$ at each vertex.
The transformation at vertex $\text{v}$ is local in the sense that it only affects the variables attached to $\text{v}$.
Let us define a right action $\rhd_{\text{v}}$ of $G$ at v by $g_{\text{v}}\rhd_{\text{v}}x_{\text{v}}=x_{\text{v}}g_{\text{v}}$ and $g_{\text{v}}\rhd_{\text{v}}x_{\text{v}'}=x_{\text{v}'}$ for $\text{v}' \neq \text{v}$ as well as the action on the group elements (see figure~\ref{ribbonedge} for the attachment of group elements to edges)
    \begin{align}
    g_{\text{v}}\rhd_{\text{v}} (g_{\text{e}},w_{\text{e}})=
    \begin{cases}
    (g_{\text{v}}^{-1}g_{\text{e}}, g_{\text{v}}w_{\text{e}}g_{\text{v}}^{-1} )&\text{ if e is leaving v}\\
    (g_{\text{e}}g_{\text{v}}, w_{\text{e}})&\text{ if e is entering  v}\\
    (g_{\text{v}}^{-1}g_{\text{e}}g_{\text{v}}, g_{\text{v}}^{-1}w_{\text{e}}g_{\text{v}})&\text{ if e is attached on both sides to v}\\
    (g_{\text{e}},w_{\text{e}})&\text{ if e is not attached to v}\\
    \end{cases}
    \end{align}
This action is lifted on ${\cal H}$ by the operator $U_{\text{v},g}$ defined by
    \begin{multline}U_{\text{v},g}\Psi\big(\left\{g_{\text{e}},w_{\text{e}},x_{\text{v}'}\right\}\big)=\Psi\big(\underbrace{\left\{g\rhd_{\text{v}}g_{\text{e}},g\rhd_{\text{v}}w_{\text{e}},g\rhd_{\text{v}}x_{\text{v}'}\right\}}_{\text{action on variable at vertex $\text{v}$}}\big)
    \times \underbrace{\exp-\text{i}A_{g}(x_{\text{v}},x_{\text{v}}w_{\text{v}})
    \times\frac{\exp\text{i}\phi_{w,g}(x_{\text{v}})}{\exp\text{i}\phi_{g,g^{-1}wg}(x_{\text{v}})}}_{\text{string magnetic translation}}.
    \end{multline}
This insertion of the string magnetic translation on the defect ensure the commutation with $\Pi_{v}$, since the latter involves a string propagator ${\cal K}_{w_{\text{v}}}(y_{\text{v}},x_{\text{v}})$. 
The other contribution to $\Pi_{v}$ involve the 3-form $C$, which is invariant under $G$, so that $U_{\text{v},g}$ and $\Pi_{\text{v}}$ commute.
Commutation of $U_{\text{v},g}$ and $\Pi_{\text{v'}}$ is clear if they are not on both ends of the same edge.
If it is the case, assume that the edge $\text{e}$ points from $\text{v}$ to $\text{v}'$. 
The definition of $\Pi_{\text{v}}$ only involves  variables that are invariant under the action of $g_{\text{v}}$.

\begin{figure}[H]
\begin{center}
\begin{tikzpicture}[scale=0.8,
  midarrow/.style={
    postaction={
      decorate,
      decoration={
        markings,
        mark=at position 0.5 with {\arrow{Stealth}}}}},scale=1]

  \pgfmathsetmacro{\l}{2}
  \pgfmathsetmacro{\h}{2.5}

    \fill[black!10] (A0) -- (B0) -- (C0) -- (D0) -- (D1)-- (C1)-- (C2)-- (B2)-- (B1)-- (A1)-- (A0);

    \fill[red!10] (A0) -- (B0T) -- (C0T) -- (D0) -- (D1)-- (C1T)-- (C2)-- (B2)-- (B1T)-- (A1)-- (A0);
    
 \coordinate (A0) at (0-2*\l,0+2*\l);
\coordinate (A1) at (-1*\l-2*\l,1*\l+3/2*\l);
\coordinate (B0) at (\l,\l+\l);
\coordinate (B0T) at (\l+0.2*\h,\l+\l+\h);
 \coordinate (B1) at (0,3/2*\l+\l);
\coordinate (B1T) at (0+0.2*\h,3/2*\l+\l+\h);
\coordinate (B2) at (-\l,3*\l+\l);
\coordinate (C0) at (\l+\l,\l+\l);
\coordinate (C0T) at (\l+\l+0.2*\h,\l+\l+\h);
\coordinate (C1) at (0+\l,3/2*\l+\l);
\coordinate (C1T) at (0+\l+0.2*\h,3/2*\l+\l+\h);
\coordinate (C2) at (-\l+\l,3*\l+\l);
\coordinate (D0) at (5*\l,5/2*\l);
\coordinate (D1) at (4*\l,3*\l);

\vspace{-0.5cm}
\draw (A0) -- (A1) ;
 \draw (B0) -- (B1) -- (B2);
\draw (C0) -- (C1) -- (C2);
\draw (D0) -- (D1) ;
\draw (B2) -- (C2) ;

\draw[thick] (A0) -- (B0) -- (C0) -- (D0);
\draw[thick] (A1) -- (B1) -- (C1) -- (D1);
\draw[midarrow] (C0) -- (D0) node[midway, below left] {$g_{2}$} ;

\draw (B0) node[below ] {$x$} ;
\draw (C0) node[below ] {$xw$} ;

\draw[red,midarrow] (C0) -- (C0T) node[midway, below left ] {$g$};
\draw[red] (B1) -- (B1T) ;
\draw[red] (B0) -- (B0T) ;
\draw[red] (C1) -- (C1T) ;

\draw[red] (B0T) -- (B1T) --(B2) ;
\draw[red] (C0T) -- (C1T) --(C2) ;

\draw[red] (B0T) node[above] {$xg$} ;

\draw[thick,red] (A1) -- (B1T) -- (C1T) -- (D1);

\draw[midarrow] (A0) -- (B0) node[midway, below left] {$g_{1}$} ;
\draw[midarrow] (B0) -- (C0) node[midway, below ] {$w$} ;

\draw[midarrow,red,thick] (A0) -- (B0T) node[midway, above left  ] {$g_{1}g$} ;

\draw[midarrow,red,thick] (B0T) -- (C0T) node[midway, below ] {$g^{-1}wg$} ;
\draw[midarrow,red,thick] (C0T) -- (D0) node[midway, below left ] {$g^{-1}g_{2}$} ;



 ;


\end{tikzpicture}
\end{center}
\caption{Action of  the vertex gauge transformation $U_{\text{v},g}$.}
\end{figure}
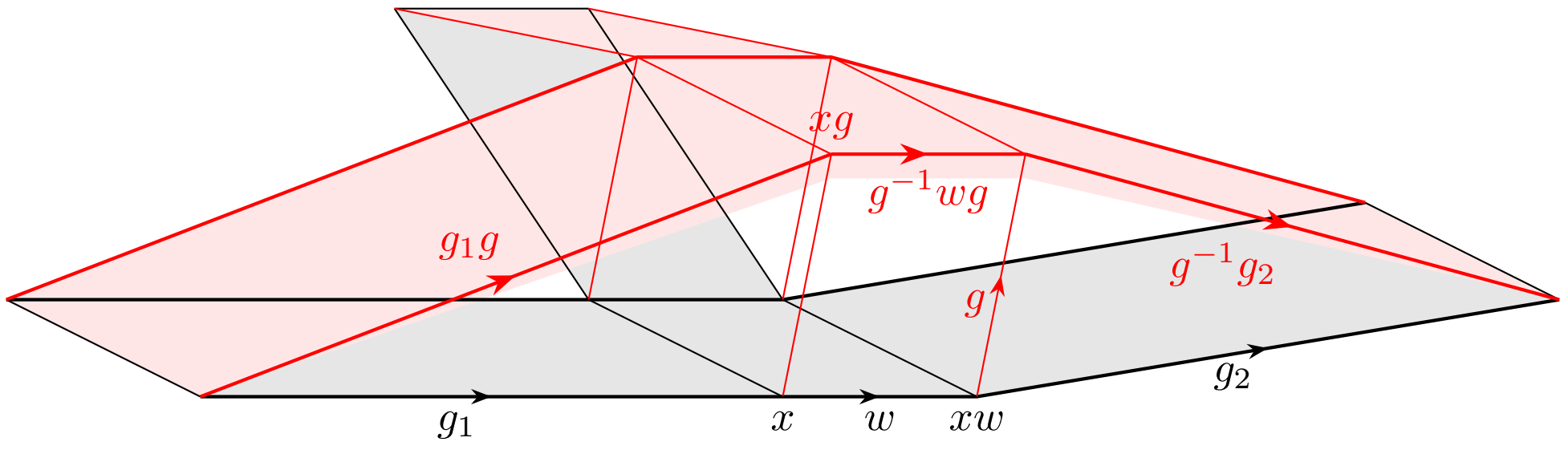

Moreover, $U_{\text{v},g}$ commutes with $U_{\text{v},g'}$ for different vertices v and v'.
Because of the phase in the composition of stringy magnetic translation (see \eqref{Tgwmultiplication}) 
    \begin{align}
    U_{\text{v},g}U_{\text{v},h}\Psi\big(\left\{g_{\text{e}},w_{\text{e}},x_{\text{v}'}\right\}\big)=
    \;\frac{\omega_{w_{\text{v}},g,h}\;\omega_{g,h,(gh)^{-1}w_{\text{v}gh}}}{\omega_{g,g^{-1}w_{\text{v}g,h}}}\;U_{\text{v},gh}\Psi\big(\left\{g_{\text{e}},w_{\text{e}},x_{\text{v}'}\right\}\big).
    \end{align}
Because of the extra phase, $U_{\text{v},g}$ only define a projective representation of $G$ and one imposes the condition $w_{\text{v}}=e$ (unit of the group), in such a way that $U_{\text{v},g}U_{\text{v},h}=U_{\text{v},gh}$.
Then, 
    \begin{align}
    P_{\text{v}}=\frac{\delta_{w_{{\text v}},e}}{|G|}\sum_{g\in G}U_{\text{v},g}
    \end{align}
is a projector since
    \begin{align}
    P_{\text{v}}^{2}=\frac{\delta_{w_{{\text v}},e}}{|G|^{2}}\sum_{g,h\in G}U_{\text{v},g}U_{\text{v},h}
    =\frac{\delta_{w_{{\text v}},e}}{|G|^{2}}\sum_{g\in G}\sum_{gh\in G}U_{\text{v},gh}
    =P_{\text{v}}.
    \end{align}
Moreover, projector at different vertices commute, and they commute with finite gauge transformations, summarized by the relations
    \begin{align}
    P_{\text{v}}^{2}=P_{\text{v}},\quad
    P_{\text{v}}P_{\text{v}'}=P_{\text{v}'}P_{\text{v}},\quad
    P_{\text{v}}U_{\text{v}',g'}=U_{\text{v}',g'}P_{\text{v}},\quad
    P_{\text{v}}\Pi_{\text{v}'}=\Pi_{\text{v}'}P_{\text{v}}.
    \end{align}
In particular $U_{\text{v},g}$ commutes with $\Pi_{\text{v}}$, so that it must preserve its common ground state, defined up to multiplication by a function of the group variables only 
    \begin{align}
    U_{\text{v},g}\Psi_{\text{\tiny string}}(\left\{g_{\text{e}},w_{\text{e}},x_{\text{v}}\right\})
    \times F(\left\{g_{\text{e}},w_{\text{e}}\right\})
    =
    F(\left\{g\rhd_{\text{v}}g_{\text{e}},g\rhd_{\text{v}}w_{\text{e}}\right\})
    \times 
    \Upsilon_{\text{v},g}(\left\{g_{\text{e}},w_{\text{e}}\right\})
    \Psi_{\text{\tiny string}}(\left\{g_{\text{e}},w_{\text{e}},x_{\text{v}}\right\}).
    \label{groupgauge}
    \end{align}
This means that after projecting out by all $\Pi_{\text{v}}$, the action of $\Upsilon_{\text{v},g}$ on the remaining degrees of freedom is fully captured by its action on the group theoretical part of the wave function.

\paragraph{Spin network basis.}

According to the Peter--Weyl theorem, matrix elements of irreducible representations of a finite group form a basis of $L^{2}(G)$.  A similar result holds for the matrix elements of irreducible representations of the Drinfel'd double $\mathcal{D}[G]$ and its deformed version $\mathcal{D}^{\omega}[G]$, which define a basis of $L^{2}(G\times G)$.

\medskip

In analogy with the undeformed Drinfel'd double $\mathcal{D}[G]$, the irreducible representations of the twisted one $\mathcal{D}^{\omega}[G]$ are parametrized as $\rho=(\kappa,r)$ with $\kappa$ a conjugation class and $r$ a projective representation of any centralizer of $w\in C$, i.e., of the subgroup $G_{w}$ of the elements of $G$ that commute with $w$.
The precise form of these representations is not needed here, we refer to the original work \cite{DPR_1990}, as well as \cite{Altschuler:2003nn} and the more recent \cite{Bullivant:2019fmk} for an up-to-date overview.  Starting from the multiplication in $\mathcal{D}^{\omega}[G]$,
\begin{align}
w\mathop{\square}\limits_{g}\;v\mathop{\square}\limits_{h}=\delta_{gvg^{-1},w}\;\frac{\omega_{w,g,h}\;\omega_{g,h,(gh)^{-1}wgh}}{\omega_{g,g^{-1}wg,h}}\;w\mathop{\square}\limits_{gh}
\end{align}
the matrix elements in the irreducible representation $\rho$ obey
\begin{align}
\sum_{j}{\cal D}_{ij}^{\rho}\big(w\mathop{\square}\limits_{g}\big){\cal D}_{jk}^{\rho}\big(v\mathop{\square}\limits_{h}\big)=\delta_{gvg^{-1},w}\;\frac{\omega_{w,g,h}\;\omega_{g,h,(gh)^{-1}wgh}}{\omega_{g,g^{-1}wg,h}}\;{\cal D}_{ik}^{\rho}\big(w\mathop{\square}\limits_{gh}\big).
\end{align}
The unitarity condition imposes that the inverse is related to the complex conjugate by
    \begin{align}
    {\cal D}_{ij}^{\rho}\big(w\mathop{\square}\limits_{g}\big)^{-1}=\frac{\omega_{g,g^{-1}wg,g^{-1}}}{\omega_{w,g,g^{-1}}\;\omega_{g,h,w}}\;{\cal D}_{ij}^{\rho}\big(g^{-1}wg\mathop{\square}\limits_{g^{-1}}\big)= 
    \overline{\cal D}_{ji}^{\rho}\big(w\mathop{\square}\limits_{g}\big)^{-1}.
    \end{align}
Denoting by $d_{\rho}$ the dimension of $\rho$, the matrix elements define an orthonormal basis of $L^{2}(G^{2})$, analogue to the Peter--Weyl theorem for finite group representations, 
    \begin{align}
    \frac{1}{|G|}\sum_{g,w}\sqrt{d_{\rho}}{\cal D}_{ij}^{\rho}\big(w\mathop{\square}\limits_{g}\big)
    \;\sqrt{d_{\rho}}\,\overline{{\cal D}}_{kl}^{\rho'}\big(w\mathop{\square}\limits_{g}\big)
    =\delta^{\rho\rho'}\delta_{ik}\delta_{jl}.
    \label{PeterWeyl}
    \end{align}
Furthermore, $\mathcal{D}^{\omega}[G]$ is equipped with a coproduct which is a morphism of algebras $\Delta:\; \mathcal{D}^{\omega}[G]\rightarrow \mathcal{D}^{\omega}[G]\otimes \mathcal{D}^{\omega}[G]$ 
    \begin{align}
    \Delta\!\left(w\mathop{\square}\limits_{g}\right)
    =\sum_{u,v\in G}\delta_{uv,w}\;
    \frac{\omega_{u,v,g}\;\omega_{g,g^{-1}ug,g^{-1}vg}}{\omega_{u,g,g^{-1}vg}}
     u\mathop{\square}\limits_{g}\otimes w\mathop{\square}\limits_{g}.
    \end{align} 
This allows us to define the tensor product of two representations $\rho$ and $\rho'$ as
    \begin{align}
    {\cal D}_{ij}^{\rho}\otimes {\cal D}_{kl}^{\rho'}
    \big(w\mathop{\square}\limits_{g}\big)
    =\sum_{u,v\in G}\delta_{uv,w}\;
    \frac{\omega_{u,v,g}\;\omega_{g,g^{-1}ug,g^{-1}vg}}{\omega_{u,g,g^{-1}vg}}
    {\cal D}_{ij}^{\rho}\big( u\mathop{\square}\limits_{g}\big) {\cal D}_{kl}^{\rho'}\big(  w\mathop{\square}\limits_{g}\big).
    \label{tensorrepresentations}
    \end{align} 
The tensor product is associative up to the action of the Drinfel'd associator and can be braided using the same $R$-matrix as in the untwisted case.
The trivial representation is given by the counit.   

An important ingredient in the sequel is the antipode, 
    \begin{align}
    S\!\left(w\mathop{\square}\limits_{g}\right)&=
    \frac{\omega_{w,g,g^{-1}w^{-1}g}}{\omega_{w,w^{-1},g}\;\omega_{g,g^{-1}wg,g^{-1}w^{-1}g}}\;
    \frac{\omega_{g,g^{-1}w^{-1}g,g^{-1}}}{\omega_{w^{-1},g,g^{-1}}\;\omega_{g,g^{-1},w^{-1}}}
    {{\textstyle g^{-1}w^{-1}g}}\mathop{\square}\limits_{g^{-1}}
    \end{align}
that allows us to define the dual representation by
    \begin{align}
    {\cal D}_{ij}^{\rho^{\ast}}
    \big(w\mathop{\square}\limits_{g}\big)=
    {\cal D}_{ji}^{\rho}
    \big(S\big[w\mathop{\square}\limits_{g}\big]\big).
    \end{align}

\begin{figure}[H]
\begin{center}
\begin{tikzpicture}[scale=0.7,
  midarrow/.style={
    postaction={
      decorate,
      decoration={
        markings,
        mark=at position 0.5 with {\arrow{Stealth}}}}},scale=0.3]

 \pgfmathsetmacro{\a}{3}
 \pgfmathsetmacro{\b}{3}
 \pgfmathsetmacro{\c}{2}

\coordinate (A) at (-\a,-\b);
\coordinate (B) at (\a,-\b);
\coordinate (C) at (\a,\b);
\coordinate (D) at (-\a,\b);

\coordinate (E) at (-\a+1.5*\a,-\b-\c);
\coordinate (F) at (\a+\c,-\b+0.5*\b);
\coordinate (I) at (-\a+1.5*\a,-\b+0.5*\b);

\coordinate (G) at (\a-1.5*\a,\b+\c);
\coordinate (H) at (-\a-\c,0.5*\b);
\coordinate (J) at (\a-1.5*\a,0.5*\b);

\fill[black!10] (A) -- (B) -- (C) -- (D) -- (A);

\draw[thick,midarrow] (D) -- (A) ;
\draw[thick,midarrow] (D) -- (C) ;
\draw[thick,midarrow] (C) -- (B) ;
\draw[thick,midarrow] (B) -- (A) ;

\draw (A) node[below left]{$xuvw$};
\draw (B) node[below right]{$xuv$};
\draw (C) node[above right]{$xu$};
\draw (D) node[above left]{$x$};

\draw[dotted] (A) -- (C);

\draw[midarrow,red] (E) -- (I);
\draw[midarrow,red] (F) -- (I);
\draw[midarrow,red] (G) -- (J);
\draw[midarrow, red] (I) -- (J);
\draw[midarrow, red] (J) -- (H);

\draw (E) node[below]{\textcolor{red}{$\rho_{3}$}};

\draw (F) node[right]{\textcolor{red}{$\rho_{2}$}};

\draw (G) node[above]{\textcolor{red}{$\rho_{1}$}};
\draw (H) node[left]{\textcolor{red}{$\rho$}};

\end{tikzpicture}
\end{center}
\caption{An intertwiner $\rho_{1}\otimes (\rho_{2}\otimes \rho_{3})\rightarrow\rho$.}
\label{intertwinerfigure}
\end{figure}
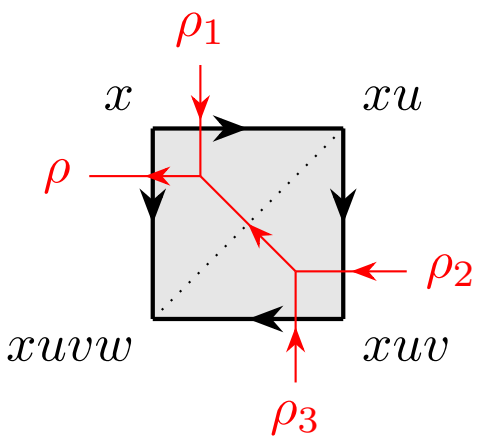

To apply this formalism to our context, consider first a graph made of two vertices $\text{v}_{x}$ and $\text{v}_{y}$, carrying respective variables $x$ and $y$ in $X$.
We orient the edge from  $\text{v}_{x}$ to $\text{v}_{y}$ and assign it a gauge field $g\in G$ and a winding $w^{-1}$ located at $\text{v}_{x}$, so that the vertex $\text{v}_{y}$ carries winding $g^{-1}wg$.
The two strings are outgoing and the resulting edge graph is that defining the Feynman rules, see figure \ref{Feynmanedge}.
After applying the conditions $\Pi_{\text{v}_{x}}\Psi=\Psi$ and $\Pi_{\text{v}_{y}}\Psi=\Psi$, the resulting wave function can be expanded, using the Peter--Weyl theorem, over the basis (orthonormal up to the insertion of the factor $\sqrt{d_{\rho}/|X|}$)
    \begin{align}
    \Psi^{\rho}_{ij}(x,y,g,w)&=\exp\text{i}\big\{B(xw^{-1}g,y,yg^{-1}wg)+B(yg^{-1}wg,xg,xw^{-1}g)+A_{g^{-1}wg}(xw^{-1}g,y)\big\}\nonumber\\
    &\quad\times 
    \exp\text{i}\big\{\Phi_{w^{-1},w}(x)+A_{g}(xw^{-1},x)-\Phi_{w,g}(xw^{-1})+\Phi_{g,g^{-1}wg}(xw^{-1})\big\}
    \;{\cal D}^{\rho}_{ij}\big(\textstyle{w}\mathop{\square}\limits_{\textstyle{g}}\big).
    \end{align}
Then, a finite group gauge transformation by $h$ at $\text{v}_{y}$ ($x\rightarrow x$, $y\rightarrow yh$, $g\rightarrow gh$, $w\rightarrow w$) reduces to a matrix multiplication on the right 
    \begin{align}
    U_{\text{v}_{y},h}\Psi^{\rho}_{ij}(x,y,g,w)&=
    \exp\text{i}\big\{-A_{h}(y,yg^{-1}wg)+\Phi_{g^{-1}wg,h}(y)-\Phi_{h,(gh)^{-1}wgh}(y)\big\}
    \,\Psi^{\rho}_{ij}(x,yh,gh,w)\\
    &=\sum_{k}\Psi^{\rho}_{ik}(x,y,g,w)\,{\cal D}^{\rho}_{kj}\big(\textstyle{w}\mathop{\square}\limits_{\textstyle{h}}\big).
    \end{align}
Similarly, a finite group  gauge transformation at $\text{v}_{x}$ ($x\rightarrow xh$, $y\rightarrow y$, $g\rightarrow h^{-1}g$, $w\rightarrow h^{-1}wh$) reduces to a matrix multiplication on the left, composed with the antipode 
    \begin{align}
    U_{\text{v}_{x},h}\,\Psi^{\rho}_{ij}(x,y,g,w)&=
    \exp\text{i}\big\{-A_{h}(x,xw^{-1})+\Phi_{w^{-1},g}(x)-\Phi_{g,g^{-1}w^{-1}g}(x)\big\}
    \,\Psi^{\rho}_{ij}(xh,y,h^{-1}g,h^{-1}wh)\\
    &=\sum_{k}{\cal D}^{\rho}_{ik}    \big(S\big[\textstyle{w^{-1}}\mathop{\square}\limits_{\textstyle{h}}\big]\big)\,\Psi^{\rho}_{kj}(x,y,g,w)\\
    &=\sum_{k}{\cal D}^{\rho^{\ast}}_{ki}    \big(\textstyle{w^{-1}}\mathop{\square}\limits_{\textstyle{h}}\big)\,\Psi^{\rho}_{kj}(x,y,g,w).
    \end{align}
For a general graph, we introduce a spin network basis defined as follows. 
For each edge, we choose an irreducible representation $\rho_{\text{e}}$ and at each vertex a basis of intertwiners between the incoming representations of the edges attached to v to the representations $\rho_{\text{v}}$ appearing in their tensor product  
    \begin{align}
    {\cal I}^{\alpha}_{i,j_{1}\cdots j_{n}}\big(\rho^{\epsilon_{\text{v},\text{e}_{1}}}_{\text{e}_{1}}\otimes\cdots\otimes\rho^{\epsilon_{\text{v},\text{e}_{n}}}_{\text{e}_{n}}\rightarrow
    \rho_{\text{v}}\big).
    \end{align}
The irreducible representations are all incoming because if an edge e is outgoing, we replace the representation by its dual in the intertwiner, not on the edge.
We use the incidence matrix $\epsilon_{\text{v},\text{e}}=1$ (resp. $-1$) if e is incoming (resp. outgoing) and define $\rho^{1}=\rho$ and $\rho^{-1}=\rho^{\ast}$.
The edges are labelled from 1 to $n$ on the vertex starting from the corner carrying the variable $x$ and moving around the polygon in clockwise order.
Finally, the parenthesis in the tensor product is dictated by the triangulation chosen for the vertex.
An example is given in figure \ref{intertwinerfigure}, with an intertwiner obeying
    \begin{multline}
    \sum_{l_{1},\dots,l_{3}}
    \frac{\omega_{u,vw,g}\;\omega_{g,g^{-1}ug,g^{-1}vwg}}{\omega_{u,g,g^{-1}vwg}}
    \frac{\omega_{v,w,g}\;\omega_{g,g^{-1}vg,g^{-1}wg}}{\omega_{v,g,g^{-1}wg}}\;
    {\cal D}^{\rho_1}_{j_{1}l_{1}}\!\big(\textstyle{u}\mathop{\square}\limits_{\textstyle{g}}\big)
    {\cal D}^{\rho_2}_{j_{2}l_{2}}\!\big(\textstyle{v}\mathop{\square}\limits_{\textstyle{g}}\big)
    {\cal D}^{\rho_3}_{j_{3}}l_{3}\!\big(\textstyle{w}\mathop{\square}\limits_{\textstyle{g}}\big)
    {\cal I}_{i,l_{1}l_{2}l_{3}}(\rho_{1}\otimes\rho_{2}\otimes \rho_{3}\rightarrow\rho)\\
    =
    \sum_{k}{\cal D}^{\rho}_{ik}\!\big(\textstyle{uvw}\mathop{\square}\limits_{\textstyle{g}}\big)
    {\cal I}_{k,j_{1}j_{2}j_{3}}(\rho_{1}\otimes\rho_{2}\otimes \rho_{3}\rightarrow\rho).
    \end{multline}

We contract the internal indices of the spin network $s=\big\{{\cal I}^{\alpha}_{\text{v}},\rho_{\text{e}}\big\}_{\text{v}\in \text{e}, \text{v}\in V}$ and multiply it with the normalizing factor to obtain a function $F^{s}_{i_{1},\dots,i_{|V|}}(\left\{g_{\text{e}},w_{\text{e}}\right\})$, so that
    \begin{align}
    \Psi^{s}_{i_{1},\dots,i_{|V|}}(\left\{g_{\text{e}},w_{\text{e}},x_{\text{v}}\right\})
    =
    \Psi_{\text{\tiny string}}(\left\{g_{\text{e}},w_{\text{e}},x_{\text{v}}\right\})\times F^{s}_{i_{1},\dots,i_{|V|}}(\left\{g_{\text{e}},w_{\text{e}}\right\})
    \times \sqrt{\frac{\prod d_{\rho_{\text{e}}}}{|X|^{|V|}}}
    \end{align}
is an orthonormal basis of the subspace of $L^{2}(G^{2|E|}\times X^{|V|})$ projected onto by all the projectors $\Pi_{\text{v}}$.
In this basis, the action of the finite group gauge transformations are simply
    \begin{align}
    U_{\text{v},g}\,\Psi^{s}_{i_{1},\dots,i_{|V|}}(\left\{g_{\text{e}},w_{\text{e}},x_{\text{v}}\right\})
    =\sum_{j_{\text{v}}}D^{\rho_{\text{v}}}_{i_{\text{v}}j_{\text{v}}}\big(\textstyle{w_{\text{v}}}\mathop{\square}\limits_{\textstyle{g}}\big)
    \,\Psi^{s}_{i_{1},\dots,j_{\text{v}},\dots,i_{|V|}}(\left\{g_{\text{e}},w_{\text{e}},x_{\text{v}}\right\}).
    \end{align}
An invariant state corresponds to having all vertex representations trivial.

\paragraph{String insertions on faces.}

\begin{figure}[h]
\begin{center}
\begin{tikzpicture}[scale=0.8,
  midarrow/.style={
    postaction={
      decorate,
      decoration={
        markings,
        mark=at position 0.5 with {\arrow{Stealth}}}}},scale=1]

  \pgfmathsetmacro{\l}{2}
  \pgfmathsetmacro{\h}{1.5}

 \coordinate (A0) at (0,0);
 \coordinate (A0T) at (0,0+\h);
\coordinate (A1) at (-1*\l,1*\l);
\coordinate (B0) at (\l,\l);
 \coordinate (B0T) at (\l,\l+\h);
 \coordinate (B1) at (0,2*\l);
\coordinate (B2) at (-0.75*\l,2.75*\l);
\coordinate (C0) at (\l+\l,\l+\l);
\coordinate (C0T) at (\l+\l,\l+\l+\h);
\coordinate (C1) at (0+\l,2*\l+\l);
\coordinate (C2) at (0.25*\l,3.75*\l);
\coordinate (D0) at (5*\l,2*\l);
\coordinate (D0T) at (5*\l,2*\l+\h);
\coordinate (D1) at (6*\l,3*\l);
\coordinate (D2) at (6.75*\l,3.75*\l);
\coordinate (D1') at (7*\l,2*\l);
\coordinate (D2') at (7*\l,3*\l);
\coordinate (D3') at (7*\l,4*\l);
\coordinate (E0) at (6*\l,1*\l);
\coordinate (E0T) at (6*\l,1*\l+\h);
\coordinate (E1) at (7*\l,2*\l);
\coordinate (E2) at (7.75*\l,2.75*\l);
\coordinate (F0) at (3*\l,0*\l);
\coordinate (F0T) at (3*\l,0*\l+\h);
\coordinate (F1) at (5*\l,0*\l);

\coordinate (M) at (2*\l,0*\l);
\coordinate (N) at (7/2*\l,3/2*\l);
\coordinate (P) at (2*\l,3/2*\l);
\coordinate (Q) at (1/2*\l,0*\l);

\coordinate (I) at (2*\l,3/4*\l);

\fill[black!10] (F0) -- (D0) -- (D1) --  (D2) -- (E2) -- (E1) -- (E0) -- (F1);
\fill[black!10] (D0) -- (D1) -- (C1) --  (C2) -- (B2) -- (B1) -- (A1) -- (A0) -- (B0) -- (C0) -- (D0);

\fill[blue!10] (F0) -- (F0T) -- (F0T) --  (D0T) -- (D0);
\fill[blue!10] (D0) -- (D0T) -- (C0T) --  (B0T) -- (A0T) -- (A0) -- (B0) --  (C0) -- (D0);

\fill[red!10] (F1) -- (E0) -- (D1) --  (C1) -- (B1) -- (A1) -- (A0T) -- (B0T) -- (C0T) -- (D0T) -- (F0T) ;

\draw[red, below left] (I) node{f};

  \draw[midarrow,red,dashed] (M) -- (N) -- (P) -- (Q);

\draw (D0) -- (E0);

\draw[thick] (A0) -- (B0) -- (C0) -- (D0) -- (F0) ;
\draw[thick] (A1) -- (B1) -- (C1) -- (D1) -- (E1) -- (E0) -- (F1);
\draw[thick,red] (A0T) -- (B0T) -- (C0T) -- (D0T) -- (F0T) ;

\draw (A0) -- (A1) ;
        
\draw (B0) -- (B1) -- (B2);
\draw (C0) -- (C1) -- (C2);
\draw (D0) -- (D1) -- (D2);
\draw (E0) -- (E1) -- (E2) ;

\draw (F0) -- (F1) ;

\draw[red] (A0) --  (A0T) ;
\draw[red] (A0T) --  (A1);

\draw[red] (B0) --  (B0T) -- (B1) ;
\draw[red] (C0) --  (C0T) -- (C1) ;
\draw[red] (D0) --  (D0T) -- (D1) ;
\draw[red] (D0) --  (D0T) -- (E0) ;
\draw[red] (E0) -- (D1) ;

\draw[red] (F0) --  (F0T) -- (F1) ;

\draw[midarrow] (F0) -- (F1) node[midway, below left] {$w$} ;
\draw[midarrow] (F0) -- (D0) node[midway, above] {$g_{1}$} ;
\draw[midarrow] (C0) -- (D0) node[midway, above] {$g_{2}$} ;
\draw[midarrow] (B0) -- (C0) node[midway, above] {$w$} ;
\draw[midarrow] (B0) -- (A0) node[midway, above left ] {$g_{3}$} ;
\draw[midarrow] (F0) -- (A0) node[midway, below] {$h_{\text{f}}=g^{}_{1}g^{-1}_{2}w^{-1}g^{}_{3}$} ;

\draw[midarrow, red] (A0) -- (A0T) node[midway, left ] {$u$} ;
\draw[midarrow,red] (F0T) -- (A0T) node[midway, above] {$h_{\text{f}}$} ;

\draw (F0) node[below left ] {$x$} ;
\draw[red] (F0T) node[above left ] {$xu$} ;

\draw[midarrow, red] (F0) -- (F0T) node[midway, left] {$u$} ;
\draw[midarrow, red] (D0) -- (D0T) node[midway, left] {$h_{1}^{-1}uh_{1}^{}\quad$} ;
\draw[midarrow, red] (C0) -- (C0T) node[midway, right] {$h_{2}^{-1}uh_{2}^{}\quad$} ;

\draw[midarrow, red] (F0T) -- (F1) node[midway, above right ] {$u^{-1}w$} ;

 ;


\end{tikzpicture}
\end{center}
\caption{Parallel transport around a face with the lifted graph colored in red and the lateral string in blue.}
\label{holonomy:fig}
\end{figure}
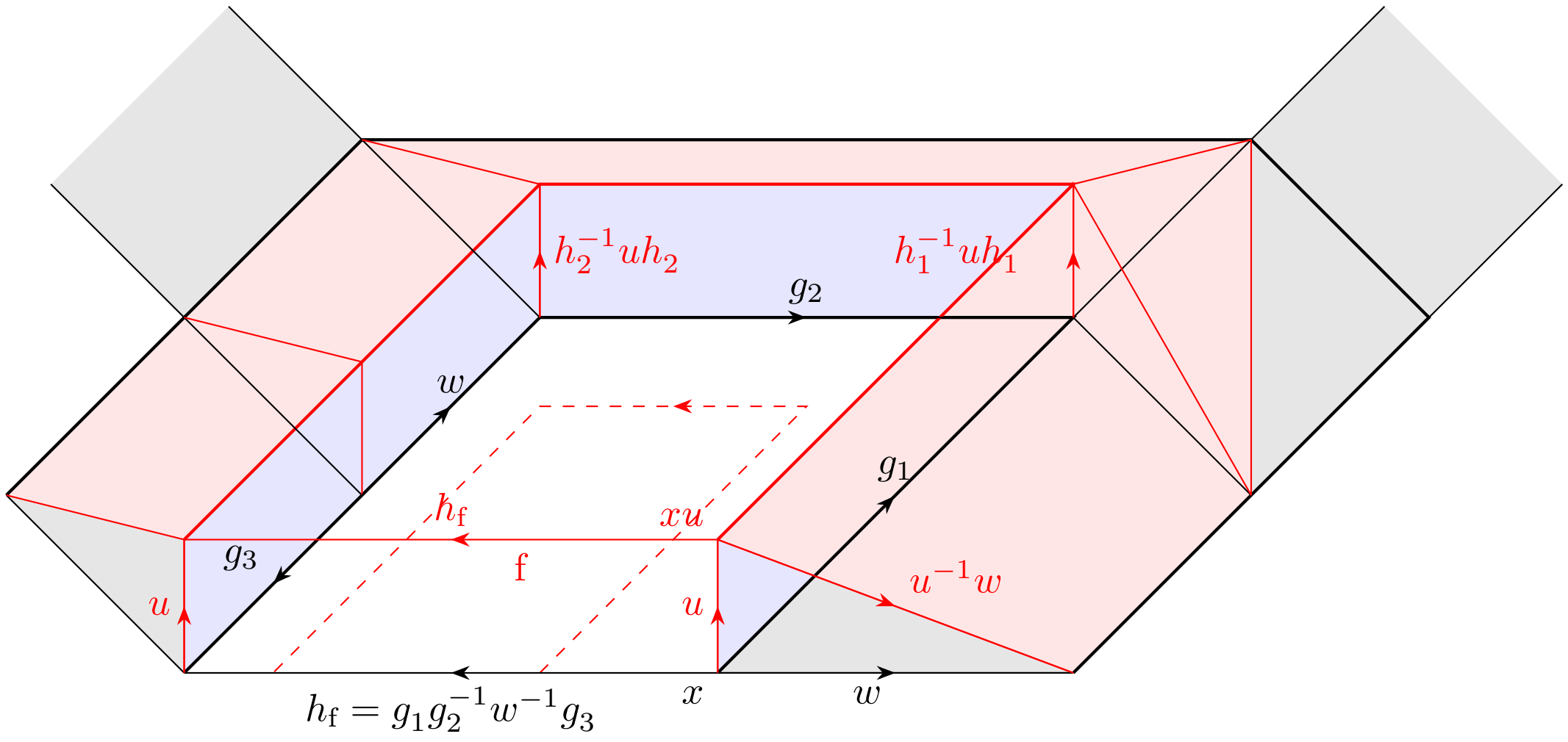

In analogy with Kitaev's model, we construct a family
of face operators $V_{\text{f},\rho}$, from which we define a third set of projectors that implements the flatness condition around the faces.
For that, fix a face $\text{f}\in F$ and an initial corner on that face, that can be on an edge or on a defect.
Traveling around the face in the orientation provided by the embedding of the graph $\Gamma$ in the surface, we multiply from  left to right the group elements encountered on the edges $g_{\text{e}}^{\pm}$ or on the vertices $w_{\text{v}}^{\pm}$ (defects), with $+$ when their orientation agree with the face and $-$ if not, see figure \ref{holonomy:fig}.
This defines a sequence $h_{1},\dots, h_{n},\dots$ of group elements and the holonomy around the face $h_{\text{f}}$ is obtained after a round trip, back to the initial corner that we chose.
Then choose $u\in G$ that commutes with $h_{\text{f}}$ and parallel transport it around the face, starting at the initial corner, by inner automorphisms,
    \begin{align}
    u_{0}=u\rightarrow u_{1}=h_{1}^{-1}u\,h_{1}\rightarrow \cdots\rightarrow 
    u_{n}=h_{n}^{-1}u\,h_{n}\rightarrow \cdots\rightarrow h_{\text{f}}^{-1}u\,h_{\text{f}}=u.
    \label{pralleltransport}
    \end{align}
Define a right action of $u$ on the windings $w_{\text{e}}$ incident to the face by translating by $u_{n}$ along the face
    \begin{align}
    u\rhd_{\text{f}}w_{\text{e}}=
    \begin{cases}
    u_{n}^{-1}w_{\text{e}}&\text{ if the side carrying $w_{\text{e}}$ is oriented away from f}\\
    w_{\text{e}}u_{n}&\text{ if the side carrying $w_{\text{e}}$ is oriented towards f}\\
    u_{n}^{-1}w_{\text{e}}u_{m}^{-1}&\text{ if the side carrying $w_{\text{e}}$ intersects f twice}\\
    w_{\text{e}}&\text{ if the side carrying $w_{\text{e}}$ does not intersect f}
    \end{cases}
    \end{align}
where $m$ and $n$ are the steps in the parallel transport   \eqref{pralleltransport} needed to reach the corresponding winding.
We also have that $u\rhd_{\text{f}}x_{\text{v}}=x_{\text{v}}u_{n}$ if v$\in$f, reached after $n$ steps, and $u\rhd_{\text{f}}x_{\text{v}}=x_{\text{v}}$ if v$\notin$f.
This defines a right action of the centralizer of $h_f$ on $G^{|E|}\times X^{|V|}$, $u\rhd_{\text{f}}(v\rhd_{\text{f}}\cdot)=vu\rhd_{\text{f}}\cdot$ and actions around different faces commute, $u\rhd_{\text{f}}(v\rhd_{\text{f}'}\cdot)=v\rhd_{\text{f}'}(u\rhd_{\text{f}}\cdot)$.

This right action on the transversal elements is supplemented by a phase to define the following operator $V_{\text{f},\rho}$, acting on wave functions in ${\cal H}$, see figure \ref{holonomy:fig}
    \begin{multline}
    V_{\text{f},\rho}\Psi\big(\left\{g_{\text{e}},w_{\text{e}},x_{\text{v}}\right\}\big)=\\
    \sum_{\substack{u\in G\\h_{\text{f}}u=uh_{\text{f}}}}
    \Psi\big(\underbrace{\left\{g_{\text{e}},u\rhd_{\text{f}}w_{\text{e}},u\rhd_{\text{f}}x_{\text{v}}\right\}}_{\text{action on variable around $\text{f}$}}\big)
    \times\underbrace{\Psi_{\substack{\text{lateral}\\\text{ string}}}\big(\left\{g_{\text{e}},w_{\text{v}},u,x_{\text{v}}\right\}\big)}_{\text{string on lateral side}}
    \times \underbrace{\prod\exp-\text{i}C}_{\text{magnetic flux on tube}}
    \times  \underbrace{\text{Tr}\prod_{n} D^{\rho}({u_{n}}\mathop{\square}\limits_{h_{n+1}})}_{\text{representation $\rho$}}
    \label{voperator}
    \end{multline}
with
    \begin{align}
    \Psi_{\substack{\text{lateral}\\\text{string}}}=\prod_{\text{\tiny edge}} \Psi_{\substack{\text{lateral}\\\text{edge}}} \prod_{\text{\tiny vertex}}\Psi_{\substack{\text{lateral}\\\text{vertex}}}.
    \end{align}
The last three factors in the definition \eqref{voperator} are:
\begin{itemize}[label=\textbullet]
\item a lateral string amplitude $\Psi_{\text{lateral string}}$ obtained by multiplying the contribution along edges given in figure \ref{lateralstringedge} and defect vertices given in figure \ref{lateralstringvertex};
 
\item a magnetic flux with 3-form $C$ on a tube triangulated by tetrahedra, see the dashed lines in figures \ref{lateralstringedge} and \ref{lateralstringvertex};

\item a trace of the product of matrices in the representation $\rho$, multiplied from left to right starting with $D^{\rho}({u}\mathop{\square}\limits_{h_{1}})$,
    \begin{align}
    \text{Tr}\Big[D^{\rho}({u}\mathop{\square}\limits_{h_{1}})D^{\rho}({u_{1}}\mathop{\square}\limits_{h_{2}})\cdots D^{\rho}({u_{n}}\mathop{\square}\limits_{h_{n+1}})\cdots\Big].
    \end{align}

\end{itemize}
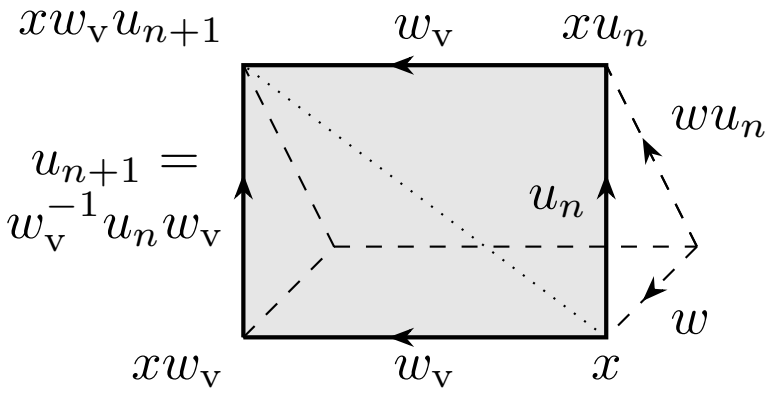
\begin{figure}[H]
\begin{center}
\begin{tikzpicture}[scale=0.60,
  midarrow/.style={
    postaction={
      decorate,
      decoration={
        markings,
        mark=at position 0.6 with {\arrow{Stealth}}}}},scale=1]
         \coordinate (A) at (-1,-1);
     \coordinate (B) at (3,-1);
    \coordinate (C) at (5,-1);
        \coordinate (C') at (5,-1+3);
  \coordinate (B') at (3,-1+3);
    \coordinate (A') at (-1,-1+3);
             \coordinate (A'') at (-1+1,-1+1);
     \coordinate (B'') at (3+1,-1+1);
    \coordinate (C'') at (5+1,-1+1);
   \fill[black!10] (A) -- (B) -- (B') -- (A');
\draw[thick] (A) -- (B) --(C) -- (C') -- (B') -- (A') --(A);
\draw[dashed] (A) -- (A'') -- (A') ;
\draw[dashed] (B) -- (B'') -- (B') ;
\draw[dashed] (A'') -- (B'') -- (C'') ;
\draw[dashed, midarrow] (C'') -- (C) node[midway, below right] {$w_{\text{e}}$};
\draw[dashed, midarrow] (C'') -- (C') node[midway, above right] {$w_{\text{e}}u_{n}$};
\draw (A) node[below left]{$y$};
\draw (A') node[above left]{$yu_{n+1}$};
\draw (B) node[below]{$xg_{\text{e}}$};
\draw (B') node[above]{$xu_{n}g_{\text{e}}\quad$};
\draw (C') node[above right]{$xu_{n}\quad$};
\draw (C) node[below right]{$x$};
\draw[midarrow] (C) -- (B) node[midway, below] {$g_{\text{e}}$};
\draw[midarrow] (C') -- (B') node[midway, above] {$g_{\text{e}}$};
\draw[midarrow] (C) -- (C') node[midway, below right] {$u_{n}$};
\draw[midarrow] (A) -- (A') node[midway, left] {$\begin{array}{c} u_{n+1}= \\ g^{-1}_{\text{e}}u_{n}g_{\text{e}} \end{array}$};
\draw[midarrow] (B) -- (B') ;
\draw[dotted] (B) -- (A') ;
\end{tikzpicture}
\end{center}

\begin{align*}
\exp\text{i}\big\{ 
A_{g_{\text{e}}}(x,xu_{n})-\Phi_{u_{n},g_{\text{e}}}(x)+\Phi_{g_{\text{e}},u_{n+1}}(x)+B(xg_{\text{e}},y,yu_{n+1})+B(yu_{n+1},xu_{n}g_{\text{e}},xg_{\text{e}})+A_{u_{n+1}}(xg_{\text{e}},y)\big\}
\end{align*}
\caption{Amplitude for a lateral string along an edge with gauge element $g_{\text{e}}$.}
\label{lateralstringedge}

\end{figure}

\begin{figure}[H]
\begin{center}
\begin{tikzpicture}[scale=0.60,
  midarrow/.style={
    postaction={
      decorate,
      decoration={
        markings,
        mark=at position 0.6 with {\arrow{Stealth}}}}},scale=1]
         \coordinate (A) at (-1,-1);
     \coordinate (B) at (3,-1);
    \coordinate (C) at (5,-1);
        \coordinate (C') at (5,-1+3);
  \coordinate (B') at (3,-1+3);
    \coordinate (A') at (-1,-1+3);
             \coordinate (A'') at (-1+1,-1+1);
     \coordinate (B'') at (3+1,-1+1);
    \coordinate (C'') at (5+1,-1+1);
   \fill[black!10] (A) -- (B) -- (B') -- (A');
\draw[thick] (A) -- (B) -- (B') -- (A') --(A);
\draw[dashed] (A) -- (A'') -- (A') ;
\draw[dashed]  (B'') -- (B') ;
\draw[dashed] (A'') -- (B'')  ;
\draw[dashed, midarrow] (B'') -- (B) node[midway, below right] {$w$};
\draw[dashed, midarrow] (B'') -- (B') node[midway, above right] {$wu_{n}$};
\draw (A) node[below left]{$xw_{\text{v}} $};
\draw (A') node[above left]{$xw_{\text{v}} u_{n+1}$};
\draw (B) node[below]{$x$};
\draw (B') node[above]{$xu_{n}$};
\draw[midarrow] (A) -- (A') node[midway, left] {${\substack{{\displaystyle u_{n+1}=}\\{\displaystyle w^{-1}_{\text{v}}u_{n}w_{\text{v}}}}}$};
\draw[midarrow] (B) -- (A) node[midway, below] {$w_{\text{v}}$};
\draw[midarrow] (B) -- (B') node[midway, left] {$u_{n}$};
\draw[midarrow] (B') -- (A') node[midway, above] {$w_{\text{v}}$};

\draw[dotted] (B) -- (A') ;
\end{tikzpicture}
\end{center}

\begin{align*}
\exp\text{i}\big\{ -\Phi_{u_{n},w_{\text{v}}}(x)+\Phi_{w_{\text{v}},u_{n+1}}(x)
+B(x,xu_{n},xw_{\text{v}}u_{n+1})+B( xw_{\text{v}}u_{n+1},xw_{\text{v}},x)\big\}
\end{align*}

\caption{Amplitude for a lateral string along a vertex  with defect element $w_{\text{v}}$.}
\label{lateralstringvertex}
\end{figure}

These operators commute with the projectors $\Pi_{\text{v}}$ 
    \begin{align}
    V_{\text{f},\rho}\Pi_{\text{v}}=\Pi_{\text{v}}V_{\text{f},\rho}\label{commutationVPi}
    \end{align}
and the gauge transformations $U_{\text{v}}$, 
    \begin{align}
    V_{\text{f},\rho}U_{\text{v}}=U_{\text{v}}V_{\text{f},\rho}\label{commutationVU}
    \end{align}
so that they also commute with the projectors $P_{\text{v}}$ since the latter is a sum of gauge transformations.
These commutation relations are clear if the vertex v does not belong to the face f.
If v$\in$f, first consider \eqref{commutationVPi}.
Since $\Pi_{\text{v}}$ only acts on the variables in $X$, the part containing the trace of the representation is unaffected by the action of $V_{\text{f},\rho}$.
Then,  $\Pi_{\text{v}}$ inserts a series of tetrahedra and a string propagation on the defect, see \eqref{definitionPiv} and comparing the tetrahedra, string propagation and string lateral amplitude on both sides yields the result.
Similarly, to establish \eqref{commutationVU}, we first notice that the magnetic tube amplitude is unaffected by the action of $U_{\text{v}}$ because $C$ is invariant.
Then, the lift of the  string lateral amplitude yields a pair of opposite string translations on the defect vertices and an extra phase, made of products of the 3-cocycle $\omega$, which is precisely cancelled by an extra phase in the trace of the representation matrices.

\medskip

It turns out that the composition of the face  operators $V_{\text{f},\rho}$ obeys the multiplication law
    \begin{align}
    V_{\text{f},\rho}V_{\text{f},\rho'}=\sum_{\rho''}N_{\rho,\rho'}^{\rho''}V_{\text{f},\rho''}\label{compositionV}
    \end{align}
where $N_{\rho,\rho'}^{\rho''}$ are the fusion coefficients of the representations of the quasi-Hopf algebra $\mathcal{D}^{\omega}[G]$, obtained by decomposing the tensor product of the representations $\rho$ and $\rho'$ into irreducible ones
    \begin{align}
    \rho\otimes\rho'\;\sim\;\bigoplus_{\rho''}N_{\rho,\rho'}^{\rho''}\rho''.
    \end{align}
These are integers that obey commutativity and associativity conditions, see for instance \cite{Bullivant:2019fmk}.
To establish \eqref{compositionV}, note that
    \begin{multline}
    V_{\text{f},\rho}V_{\text{f},\rho'}\Psi\big(\left\{g_{\text{e}},w_{\text{e}},x_{\text{v}}\right\}\big)=
    \sum_{\substack{u\in G\\h_{\text{f}}u=uh_{\text{f}}}}\sum_{\substack{v\in G\\h_{\text{f}}v=vh_{\text{f}}}}
    \Psi\big(\left\{g_{\text{e}},(uv)\rhd_{\text{f}}w_{\text{e}},(uv)\rhd_{\text{f}}x_{\text{v}}\right\}\big)\\    \Psi_{\substack{\text{lateral}\\\text{string}}}\big(\left\{g_{\text{e}},w_{\text{v}},u,x_{\text{v}}\right\}\big)
    \times \prod\exp-\text{i}C\big(\left\{g_{\text{e}},w_{\text{v}},u,x_{\text{v}}\right\}\big)
    \times \text{Tr}\prod_{n} D^{\rho}({u}\mathop{\square}\limits_{h_{n}})
    \\
    \Psi_{\substack{\text{lateral}\\\text{string}}}\big(\left\{g_{\text{e}},u\rhd_{\text{f}}w_{\text{v}},v,u\rhd_{\text{f}}x_{\text{v}}\right\}\big)
    \times \prod\exp-\text{i}C\big(\left\{g_{\text{e}},u\rhd_{\text{f}}w_{\text{v}},v, u\rhd_{\text{f}}x_{\text{v}}\right\}\big)
    \times \text{Tr}\prod_{n} D^{\rho'}({v}\mathop{\square}\limits_{h_{n}}).
    \end{multline}
Then, the lateral string amplitude and the magnetic flux tube conspire to give
    \begin{multline}
    \Psi_{\substack{\text{lateral}\\\text{string}}}\big(\left\{g_{\text{e}},w_{\text{v}},u,x_{\text{v}}\right\}\big)
    \times \prod\exp-\text{i}C\big(\left\{g_{\text{e}},w_{\text{v}},u,x_{\text{v}}\right\}\big)
    \\
    \times\Psi_{\substack{\text{lateral}\\\text{string}}}\big(\left\{g_{\text{e}},u\rhd_{\text{f}}w_{\text{v}},v,u\rhd_{\text{f}}x_{\text{v}}\right\}\big)
    \times \prod\exp-\text{i}C\big(\left\{g_{\text{e}},u\rhd_{\text{f}}w_{\text{v}},v, u\rhd_{\text{f}}x_{\text{v}}\right\}\big)\\
    =    \Psi_{\substack{\text{lateral}\\\text{string}}}\big(\left\{g_{\text{e}},w_{\text{v}},uv,x_{\text{v}}\right\}\big)
    \times \prod\exp-\text{i}C\big(\left\{g_{\text{e}},w_{\text{v}},uv,x_{\text{v}}\right\}\big)\times
    \prod_{n}\frac{\omega_{u_{n},v_{n},h_{n}}\;\omega_{h_{n},h_{n}^{-1}u_{n}h^{}_{n},h_{n}^{-1}v_{n}h_{n}^{}}}{\omega_{u_{n},h_{n},h_{n}^{-1}v_{n}h_{n}^{}}}.
    \end{multline}
The last phase is the coproduct phase that is needed to construct the tensor product representation, see \eqref{tensorrepresentations}
    \begin{multline}
    \sum_{u,v\in G}
    \text{Tr}\Big[D^{\rho}({u}\mathop{\square}\limits_{h_{1}})D^{\rho}({u_{1}}\mathop{\square}\limits_{h_{2}})\cdots D^{\rho}({u_{n}}\mathop{\square}\limits_{h_{n+1}})\cdots\Big]
    \text{Tr}\Big[D^{\rho'}({v}\mathop{\square}\limits_{h_{1}})D^{\rho'}({v_{1}}\mathop{\square}\limits_{h_{2}})\cdots D^{\rho'}({v_{n}}\mathop{\square}\limits_{h_{n+1}})\cdots\Big]\\
    \times\prod_{n}\frac{\omega_{u_{n},v_{n},h_{n}}\;\omega_{h_{n},h_{n}^{-1}u_{n}h^{}_{n},h_{n}^{-1}v_{n}h_{n}^{}}}{\omega_{u_{n},h_{n},h_{n}^{-1}v_{n}h_{n}^{}}}
    \quad =\quad 
    \sum_{w\in G}
    \text{Tr}\Big[D^{\rho\otimes\rho'}({w}\mathop{\square}\limits_{h_{1}}) D^{\rho\otimes \rho'}({w_{1}}\mathop{\square}\limits_{h_{2}})\cdots D^{\rho\otimes\rho'}({w_{n}}\mathop{\square}\limits_{h_{n+1}})\cdots\Big].
    \label{definitionQf}
    \end{multline}
Thus, we obtain $V_{\text{f},\rho}V_{\text{f},\rho'}=V_{\text{f},\rho\times\rho'}$, and the result follows from the decomposition of the tensor product into irreducible representations.

Then, let us define the face operator as 
    \begin{align}
    Q_{\text{f}}=\frac{1}{|G|^{2}}\sum_{\rho } d_{\rho}V_{\text{f},\rho}.
    \end{align}
First, notice that the dependence on the representations is entirely contained in the trace of the product of matrix elements, that reads, up to a phase factor  $(\cdots)$ see \eqref{groupflatnes} for a more explicit expression, independent of $\rho$
    \begin{align}
    \sum_{\rho} d_{\rho}\text{Tr}\Big[\prod_{n} D^{\rho}({u_{n}}\mathop{\square}\limits_{h_n}\big)\Big]
    =(\cdots) \sum_{\rho} d_{\rho}\chi_{\rho}({u}\mathop{\square}\limits_{h_{\text{f}}}\big)
    \end{align}
where $h_n$ denotes the sequence $g_{\text{e}}^{\pm}$ or $w_\text{v}^{\pm}$ around the face, and $h_{\text{f}}$ is the total holonomy around the face and $\chi_{\rho}=\text{Tr}\big(D_{\rho}\big)$  the characters of the irreducible representations of $\mathcal{D}^{\omega}[G]$.
Thanks to the Peter--Weyl theorem \eqref{PeterWeyl}, the sum of characters provides a representation of the Kronecker symbol on the group $G$
    \begin{align}
    \sum_{\rho} d_{\rho}\chi_{\rho}({w}\mathop{\square}\limits_{g}\big)=|G|\delta_{g,e},\quad\forall w\in G.
    \label{Fourierdelta}
    \end{align}
Consequently, the operator  $Q_{\text{f}}$ implements the flatness condition around the face,
    \begin{align}
    Q_{\text{f}}\Psi\big(\big\{g_{\text{e}},w_{\text{e}},x_{\text{v}}\big\}\big)=0
    \qquad\text{if}\quad h_{\text{f}}\neq e.
    \label{groupflatnes}
    \end{align}
Second, let us show that $Q_{\text{f}}$ is a projector.
Its square is given by
    \begin{align}
    Q_{\text{f}}^{2}=\frac{1}{|G|^{4}}\sum_{\rho,\rho' } d_{\rho}d_{\rho'}V_{\text{f},\rho}V_{\text{f},\rho'}
    =\frac{1}{|G|^{4}}\sum_{\rho,\rho',\rho''} d_{\rho}d_{\rho'}
    N_{\rho,\rho'}^{\rho''}
    V_{\text{f},\rho''}.
    \label{Qsquare}
    \end{align}
In order to evaluate the RHS, let us express the fusion coefficients in terms of characters.
Taking the trace in a tensor product of representations, one has
    \begin{align}
    \sum_{uv=w}
    \frac{\omega_{u,v,g}\omega_{gg^{-1}ug,g^{-1}vg}}{\omega_{u,g,g^{-1}vg}}\,
    \chi_{\rho}\big(u\mathop{\square}\limits_{g}\big) \chi_{\rho'}\big(v\mathop{\square}\limits_{g}\big)
    =\sum_{\rho''}N^{\rho''}_{\rho\rho'}\,\chi_{\rho''}\big(w\mathop{\square}\limits_{g}\big).
    \end{align}
Then, as a consequence of the Peter--Weyl orthogonality relations \eqref{PeterWeyl}, the characters are themselves orthogonal
    \begin{align}
    \frac{1}{|G|}\sum_{g,w}\chi_{\rho}\big(w\mathop{\square}\limits_{g}\big)
    \;\overline{\chi}_{\rho'}\big(w\mathop{\square}\limits_{g}\big)
    =\delta_{\rho\rho'}.
    \end{align}
This leads to the following expression of the fusion coefficients
    \begin{align}
    N^{\rho''}_{\rho\rho'}=\frac{1}{|G|}\sum_{g,u,v}
    \frac{\omega_{u,v,g}\,\omega_{g, g^{-1}ug,g^{-1}vg}}{\omega_{u,g,g^{-1}vg}}\,
    \chi_{\rho}\big(u\mathop{\square}\limits_{g}\big) \chi_{\rho'}\big(v\mathop{\square}\limits_{g}\big)
    \overline{\chi}_{\rho''}\big(uv\mathop{\square}\limits_{g}\big).
    \end{align}
Multiplying by $d_{\rho}d_{\rho'}$ and summing over $\rho$ and $\rho'$ leads to   
    \begin{align}
    \sum_{\rho,\rho'}d_{\rho}d_{\rho'}
    N^{\rho''}_{\rho\rho'}=|G|^{2}d_{\rho''}
    \end{align}
where we used the Fourier representation of the Kronecker symbol \eqref{Fourierdelta}. 
Finally, replacing in \eqref{Qsquare}
    \begin{align}
    Q_{\text{f}}^{2}
    =\frac{1}{|G|^{4}}\sum_{\rho,\rho',\rho''} d_{\rho}d_{\rho'}
    N_{\rho,\rho'}^{\rho''}
    V_{\text{f},\rho''}
    =\frac{1}{|G|^{2}}\sum_{\rho''}d_{\rho''}V_{\text{f},\rho''}=Q_{\text{f}}
    \end{align}
so that $Q_{\text{f}}$ is indeed a projector.

\paragraph{Hamiltonian and symmetries.}

To summarize, we have a Hamiltonian which is a sum of mutually commuting projectors
    \begin{align}
    H=\sum_{\text{v}}\lambda_{\text{v}}\big(1-\Pi_{\text{v}}\big)+\sum_{\text{v}}\mu_{\text{v}}\big(1-P_{\text{v}}\big)
    +\sum_{\text{f}}\nu_{\text{f}}\big(1-Q_{\text{f}}\big)
    \label{Hamiltonian}
    \end{align}
with $\lambda_{\text{v}},\mu_{\text{v}},\nu_{\text{f}}$ positive coupling constants, $\Pi_{\text{v}}$ projects onto magnetic amplitudes, see \eqref{PivString}, $P_{\text{v}}$ a projector onto gauge invariant states
    \begin{align}
    P_{\text{v}}=\frac{\delta_{w_{{\text v}},e}}{|G|}\sum_{g\in G}U_{\text{v},g}
    \end{align}
and $Q_{\text{f}}$ implementing the flatness condition together with the translation of the transverse variables
    \begin{align}
    Q_{\text{f}}=\frac{1}{|G|^{2}}\sum_{\rho } d_{\rho}V_{\text{f},\rho}.
    \end{align}
The Hamiltonian commutes with the unitary symmetry $U_{\text{v},g}$ and the non invertible symmetry $V_{\text{f},\rho}$, obeying the algebra 
    \begin{align}
    U_{\text{v},g}U_{\text{v},h}&=\frac{\omega_{w_{\text{v}},g,h}\;\omega_{g,h,(gh)^{-1}w_{\text{v}}gh}}{\omega_{g,g^{-1}w_{\text{v}}g,h}}\;U_{\text{v},gh}\\
    U_{\text{v},g}U_{\text{v}',h}&=U_{\text{v}',h}U_{\text{v},g}\quad\text{for v$\neq$v'}\\
    V_{\text{f},\rho}V_{\text{f},\rho'}&=\sum_{\rho''}N_{\rho,\rho'}^{\rho''}V_{\text{f},\rho''}\\
    V_{\text{f},\rho}V_{\text{f}',\rho'}&=V_{\text{f}',\rho'}V_{\text{f},\rho}\quad\text{for f$\neq$f'}\\
    U_{\text{v},g}V_{\text{f},\rho}&=V_{\text{f},\rho}U_{\text{v},g}
    \end{align}
with $N_{\rho,\rho'}^{\rho''}$ the fusion coefficients of the representations of $\mathcal{D}^{\omega}[G]$.

\paragraph{Factorization of string amplitudes.}

When acting on the subspace of string magnetic amplitudes $\prod_{\text{v}}\Pi_{\text{v}}{\cal H}$ consisting of wave functions of the type $\Psi_{\text{\tiny string}}(\left\{g_{\text{e}},w_{\text{e}},x_{\text{v}}\right\})
\times F(\left\{g_{\text{e}},w_{\text{e}}\right\})$ with $F$ an arbitrary function of group elements, there is an expression of the action of $V_{\text{f},\rho}$ similar to that of $U_{\text{v},g}$ given in \eqref{groupgauge}.
This follows from the commutation of $\Pi_{\text{v}}$ with $V_{\text{f},\rho}$ which forces the resulting wave function to depend on the $x_{\text{v}}$ variables only through the string magnetic amplitude.
This leads to the expression
    \begin{multline}
    V_{\text{f},\rho}\Psi_{\text{\tiny string}}(\left\{g_{\text{e}},w_{\text{e}},x_{\text{v}}\right\})
    \times F(\left\{g_{\text{e}},w_{\text{e}}\right\})
    =\\ \Psi_{\text{\tiny string}}(\left\{g_{\text{e}},w_{\text{e}},x_{\text{v}}\right\})\times
    \sum_{u\in G}
    F(\left\{g_{\text{e}},u\rhd_{\text{f}}w_{\text{e}}\right\})
    \times 
    \Xi_{\text{f},u}(\left\{g_{\text{e}},w_{\text{e}}\right\})
    \times
    \chi_{\rho}\big(u\mathop{\square}\limits_{h_{\text{f}}}\big)
    \label{groupface}
    \end{multline}
where we recall that $h_{\text{f}}$ is the total holonomy around the face, including the defects $w_{\text{v}}$ and gauge fields $g_{\text{e}}$. $\Xi_{\text{f},u}(\left\{g_{\text{e}},w_{\text{e}}\right\})$ is a product of 3-cocycles $\omega$ obtained by multiplying the matrices in the trace in the definition \eqref{definitionQf}, together with a contribution from the prisms on the graph. 
Geometrically, it interpolates between the bottom and top surfaces in figure \ref{holonomy:fig}.

\paragraph{Ground state.}

The ground states of the Hamiltonian \eqref{Hamiltonian} are those states for which the three type of constraints are fulfilled,
\begin{align}
H\Psi_{0}=0\qquad\Leftrightarrow\qquad 
\begin{cases}
\Pi_{\text{v}}\Psi_{0}=0&\forall\text{v}\in V\\
P_{\text{v}}\Psi_{0}=0&\forall\text{v}\in V\\
Q_{\text{f}}\Psi_{0}=0&\forall\text{f}\in F\\
\end{cases}
\end{align}
The first set of constraints $\Pi_{\text{v}}\Psi_{0}=0$ imposes that the dependence of the wave function on coordinates in $X$ is completely fixed to be the string magnetic amplitude, see \eqref{PivString}. Then, the only remaining degrees of freedom are the group elements $g_{\text{e}}$ and  $w_{\text{e}}$ on the edges. 

The remaining projectors impose both some flatness conditions and a translationnal invariance
\begin{itemize}
\item $P_{\text{v}}$ imposes that absence of defect at vertex $\text{v}$ and a gauge transformation of the edge variables
    \begin{align}
    w_{\text{v}}=\prod_{\substack{\text{e} \text{ such that}\\\text{v}\in\partial\text{e} }} w_{\text{e}}^{\pm}=e,\qquad  
    \big\{g_{\text{e}},w_{\text{e}}\big\}\rightarrow
    \big\{g\rhd_{\text{v}}g_{\text{e}},g\rhd_{\text{v}}w_{\text{e}}\big\}.
    \end{align}
\item $Q_{\text{f}}$ imposes the flatness of the face $\text{f}$ (taking into account both defects and gauge fields) and a transformation of the windings
    \begin{align}
    h_{\text{f}}=\prod_{\text{e,v} \in\partial\text{f} } \big(w_{\text{v}}^{\pm},g_{\text{e}}^{\pm} \big)
    =e,\qquad   
    \big\{g_{\text{e}},w_{\text{e}}\big\}\rightarrow
    \big\{g_{\text{e}},u\rhd_{\text{f}}w_{\text{e}}\big\}.
    \end{align}
\end{itemize} 
Considering a state $\Psi_{0}$ that fulfills these constraints. The first one imposes $w_{\text{v}}=e$, so that only the gauge fields $g_{\text{e}}$ appear in the constraint $h_{\text{f}}=e$ imposed by   $Q_{\text{f}}$, so that the gauge fields are also flat. Then, on a genus zero surface,
$g_{\text{e}}$ and $w_{\text{e}}$ can be written as pure gauges
    \begin{align}
    g_{\text{e}}=g^{-1}_{\text{v}}g^{}_{\text{v'}}\qquad
    w_{\text{e}}=u^{-1}_{\text{f}}u^{}_{\text{f'}}
    \end{align} 
for a suitable set of vertex variables $g_{\text{v}}$ and face variables $u^{}_{\text{f}}$. Then, the second part of the projectors implement the unicity of the wave function under this expression.

\begin{figure}[h]
\begin{align}
\parbox{2cm}{
\begin{tikzpicture}[scale=1.2,
  midarrow/.style={
    postaction={
      decorate,
      decoration={
        markings,
        mark=at position 0.5 with {\arrow{Stealth}}}}},scale=0.3]
\coordinate (A) at (0,0);
\coordinate (B) at (6,0);
\coordinate (C) at (4,2);
\coordinate (D) at (3,-3);
\draw[midarrow, thick] (B) -- (C) node[midway, above right] {$h$};
\draw[midarrow, thick] (C) -- (A) node[midway, above left] {$k$};
\draw[thick] (A) -- (D) ;
\draw[midarrow, thick] (D) -- (B) node[midway, below right] {$g$};
\draw[thick] (D) -- (C) ;
\draw[thick,dotted] (A) -- (B) ;
\end{tikzpicture}}
\qquad \rightarrow \qquad \omega_{g,h,k}
\end{align}
\caption{Tetrahedron in the Dijkgraaf--Witten finite group Chern--Simon theory.}
\label{DWtetrahedron}
\end{figure}
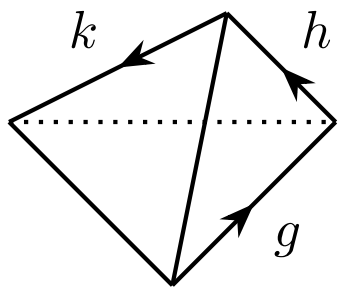

Moreover, once the two flatness conditions are imposed, the resulting phases are products of the 3-cocycle $\omega$ that correspond to the ratio of the partition function of the Dijkgraaf--Witten finite group Chern--Simons theory \cite{DW_1990} on triangulations of ball ${\cal B}$ that bounds the surface $\Sigma$ on which the graph is drawn,
    \begin{align*}    \Psi_{0}\big(\big\{x_{v},g_{\text{e}},w_{\text{e}}\big\}\big)=\Psi_{\text{string}}\big(\big\{x_{v},g_{\text{e}},w_{\text{v}}\big\}\big)
    \times \frac{\Psi_{\text{DW}}\big(\big\{g_{\text{e}},w_{\text{e}}=e\big\}\big)}{\Psi_{\text{DW}}\big(\big\{g_{\text{e}},w_{\text{e}}\big\}\big)}.
    \end{align*}
Recall that the partition function of the Dijkgraaf--Witten model is constructing by triangulating 3-manifolds and associating to any tetrahedra a 3-cocycle, assuming that all faces are flat, see figure \ref{DWtetrahedron}.

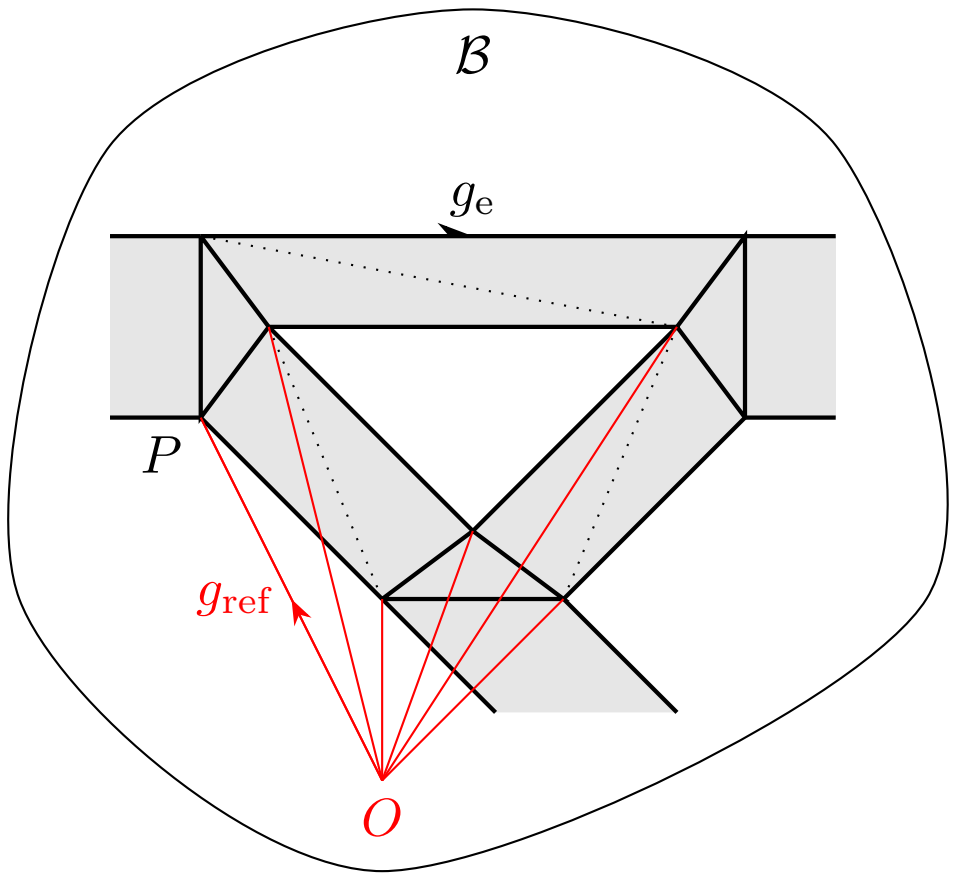
\begin{figure}[h]
\begin{center}

\begin{tikzpicture}[scale=1,
  midarrow/.style={
    postaction={
      decorate,
      decoration={
        markings,
        mark=at position 0.5 with {\arrow{Stealth}}}}},scale=0.3]

\coordinate (A1) at (-2,0);
\coordinate (B1) at (2,0);
\coordinate (C1) at (0,3/2);

\coordinate (I) at (-10,0);
\coordinate (J) at (0,13);
\coordinate (K) at (10,0);
\coordinate (L) at (0,-13);

 \pgfmathsetmacro{\a}{6}
 \pgfmathsetmacro{\b}{6}

\coordinate (A2) at (0+\a,-2+\b);
\coordinate (B2) at (0+\a,2+\b);
\coordinate (C2) at (-3/2+\a,0+\b);

\coordinate (A3) at (0-\a,0+2+\b);
\coordinate (B3) at (0-\a,0-2+\b);
\coordinate (C3) at (3/2-\a,0+\b);

\coordinate (A1') at (-2+2.5,0-2.5);
\coordinate (B1') at (2+2.5,0-2.5);

\coordinate (A2') at (0+\a+2,-2+\b);
\coordinate (B2') at (0+\a+2,2+\b);

\coordinate (A3') at (0-\a-2,0+2+\b);
\coordinate (B3') at (0-\a-2,0-2+\b);

 \pgfmathsetmacro{\Ox}{-2}
 \pgfmathsetmacro{\Oy}{-4}

\coordinate (O) at (\Ox,\Oy);

\draw plot [smooth cycle] coordinates {(10,0) (8,10)  (0,13) (-8,10) (-10,0) (-2,-6) };

\draw (0,12) node {${\cal B}$};
\draw[midarrow, thick] (B3) -- (A3) node[midway, left] {$w_{\text{e}}$};

\draw[midarrow, thick] (B3) -- (A3) node[midway, left] {$w_{\text{e}}$};
\draw[midarrow, thick] (A3) -- (B2) node[midway, above] {$g_{\text{e}}$};
\draw[midarrow,red] (O) -- (B3) node[midway, left] {$\textcolor{red}{g_{\text{ref}}}$};

\draw[red] (O) node[below] {$O$} ;
\draw (B3) node[below left ] {$P$} ;

\fill[black!10] (A1) -- (B1) -- (C1) -- (A1);
\fill[black!10] (B1) -- (A2) -- (C2) -- (C1);
\fill[black!10] (A3) -- (C3) -- (C2) -- (B2);
\fill[black!10] (A1) -- (C1) -- (C3) -- (B3);

\fill[black!10] (A1) -- (B1) -- (B1') -- (A1');
\fill[black!10] (A2) -- (B2) -- (B2') -- (A2');
\fill[black!10] (A3) -- (B3) -- (B3') -- (A3');

\fill[black!10] (A2) -- (B2) -- (C2) -- (A2);

\fill[black!10] (A3) -- (B3) -- (C3) -- (A3);

\draw[thick] (A1) -- (B1) -- (C1) -- (A1);

\draw[thick] (A2) -- (B2) -- (C2) -- (A2);

\draw[thick] (A3) -- (B3) -- (C3) -- (A3);

\draw[thick] (B1) -- (A2);
\draw[thick] (C1) -- (C2);
\draw[dotted] (B1) -- (C2);

\draw[thick] (C1) -- (C3);
\draw[thick] (A1) -- (B3);
\draw[dotted] (A1) -- (C3);

\draw[thick] (C2) -- (C3);
\draw[thick] (A3) -- (B2);
\draw[dotted] (A3) -- (C2);

\draw[thick] (A1) -- (A1');
\draw[thick] (B1) -- (B1');

\draw[thick] (A2) -- (A2');
\draw[thick] (B2) -- (B2');

\draw[thick] (A3) -- (A3');
\draw[thick] (B3) -- (B3');

\draw[red] (O) -- (A1);
\draw[red] (O) -- (B1);
\draw[red] (O) -- (C1);
\draw[red] (O) -- (C2);
\draw[red] (O) -- (C3);
\draw[red] (O) -- (B3);

\end{tikzpicture}
\end{center}

\caption{Construction of the triangulation}
\label{Chern-Simons}
\end{figure}

In the construction of $\Psi_{\text{DW}}\big(\big\{g_{\text{e}},w_{\text{e}}\big\}\big)$, the ribbon graph is also triangulated as part of the surface, in addition to its faces, see figure \ref{Chern-Simons}. The triangulation of the ball ${\cal B}$ such that  $\partial{\cal B}=\Sigma$ is constructed by adding an extra point $O$, choosing an arbitrary point $P$ in the triangulation of $\sigma$ and assigning it a arbitray group element $g_{\text{ref}}$.  Then all other group elements $g'_{\text{ref}}, g''_{\text{ref}},\dots$ joining $O$ to the other points of the triangulation of $\Sigma$ are determined by the flatness condition.  Then, $\Psi_{\text{DW}}\big(\big\{g_{\text{e}},w_{\text{e}}\big\}\big)$ is the product of all 3-cocycles associated to this triangulation. The topological nature of Dijkgraaf--Witten theory ensures that it does not depend on the choice of   $g_{\text{ref}}$.

Acting with the projector $P_{\text{v}}$ on this state generates translates the group elements and generates a phase $\Upsilon_{\text{v},g}(\left\{g_{\text{e}},w_{\text{e}}\right\})$ (see equation \eqref{groupgauge}), which is precisely cancelled by the ration of Chern--Simons partition functions
    \begin{align}
    \frac{
    \Psi_{\text{DW}}\big(\big\{g\rhd_{\text{v}}g_{\text{e}},g\rhd_{\text{v}}w_{\text{e}}\big\}\big)}
    {\Psi_{\text{DW}}\big(\big\{g\rhd_{\text{v}}g_{\text{e}},g\rhd_{\text{v}}(w_{\text{e}}=e)\big\}\big)}=
    \Upsilon_{\text{v},g}(\left\{g_{\text{e}},w_{\text{e}}\right\}).
    \end{align}
Similarly, the phase $\Xi_{\text{f},u}(\left\{g_{\text{e}},w_{\text{e}}\right\})$ involved in the projector $V_{\text{f},\rho}$ (see equation \eqref{groupface}) is cancelled by the action on the ratios of Chern--Simons partition functions
    \begin{align}
    \frac{
    \Psi_{\text{DW}}\big(\big\{g_{\text{e}},u\rhd_{\text{f}}w_{\text{e}}\big\}\big)}
    {\Psi_{\text{DW}}\big(\big\{g_{\text{e}},u\rhd_{\text{v}}(w_{\text{e}}=e)\big\}\big)}=
    \Xi_{\text{f},u}(\left\{g_{\text{e}},w_{\text{e}}\right\}).
    \end{align}

\paragraph{Elementary ribbon operators.}

Having sketched the general structure of the ground state on a genus 0 graph, let us consider some excitations.
As for Kitaev's quantum double model, the simplest excitations correspond, on a genus 0 surface, to the breakdown of two constraints of the same nature, either $P_{\text{v}}\Psi_{0}=0$ and $P_{\text{v}'}\Psi_{0}=0$, or $Q_{\text{f}}\Psi_{0}=0$ and $Q_{\text{f}'}\Psi_{0}=0$.
The latter constraints on the face tie the gauge fields along ${\text{g}_{e}}$ to the total defect at the vertices $w_{\text{v}}$.
Interpreting the defect as a magnetic charge, this is similar to the constraint appearing in Chern--Simons theory, relating the spatial curvature of the gauge field to the magnetic charge.
As such, we do not break it, so that the interpretation as a model of magnetic charge coupled to a finite Chern--Simons still holds.

Therefore, we focus our interest on breaking the constraints  $P_{\text{v}}\Psi_{0}=0$ and $P_{\text{v}'}\Psi_{0}=0$ for a pair of distant vertices.
This is done using ribbon operators which are very similar to the ones used in Kitaev's model.
We choose a path $\ell$ connecting two sites where the defect is inserted.
Starting at a corner, let us choose a group element $u\in G$. Then, we parallel transport it along the edges $g_{\text v}$ and the defect vertices  $w_{\text v}$, see the parallel transport equation \eqref{pralleltransport}.
However, when we change face and cross an edge $\text{e}$ transversally, we do not use the winding $w_{\text{e}}$ to parallel transport $u$, the latter remains un changed, $u_{n}\rightarrow u_{n+1}=u_{n}$ when crossing transversally the edge.
Consequently, one to modify $w_{\text{e}}\rightarrow u_{n}^{\pm}w_{\text{e}}  u_{n}^{\mp}$ in order to maintain the absence of defect at the vertex and $g_{\text{e}}\rightarrow u_{n}^{\pm}f_{\text{e}}$ to maintain the gauge constraint around the face, with $\pm$ being chosen according to the relative orientation of sides of the edge with respect to the path, see figure \ref{ribbon:fig}.
Then, we evaluate the different group elements in a product of matrices in a fixed representation $\mathcal{D}^{\omega}[G]$, insert a magnetic tube and a lateral string amplitude,  as for the construction of $V_{\rho,\text{f}}$ \eqref{voperator}, with the notable exception that the path does not close.
Therefore the defects are created on the boundary of the ribbon. They cancel on a closed ribbon, since the latter is nothing but the operator $V_{\rho,\text{f}}$.

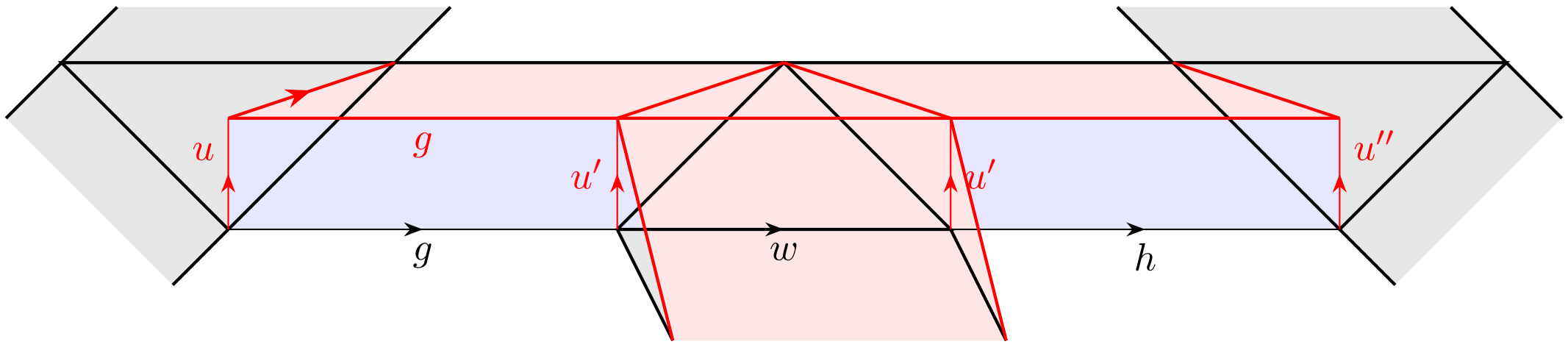
\begin{figure}[h]
\begin{center}
\begin{tikzpicture}[scale=1,
  midarrow/.style={
    postaction={
      decorate,
      decoration={
        markings,
        mark=at position 0.5 with {\arrow{Stealth}}}}},scale=2]

  \pgfmathsetmacro{\a}{2.5}
  \pgfmathsetmacro{\b}{0.75}
  \pgfmathsetmacro{\c}{0.25}
  \pgfmathsetmacro{\d}{0.5}

 \coordinate (A1) at (-\a,0);

 \coordinate (B1) at (-\a+\b,+\b);
 \coordinate (C1) at (-\a-\b,+\b);
 
 \coordinate (A1') at (-\a-\c,0-\c);
 \coordinate (B1') at (-\a+\b+\c,+\b+\c);
 \coordinate (C1') at (-\a-\b+\c,+\b+\c);
\coordinate (C1'') at (-\a-\b-\c,+\b-\c);

 \coordinate (A1'') at (-\a,0+\d);

  \coordinate (A2) at (-\b,0);
 \coordinate (B2) at (+\b,0);
 \coordinate (C2) at (0,+\b);
  \coordinate (B2'') at (+\b,0+\d);
 \coordinate (A2'') at (-\b,0+\d);
 
   \coordinate (A2') at (-\b+\c,0-2*\c);
 \coordinate (B2') at (+\b+\c,0-2*\c);
 
 \coordinate (A3) at (+\a,0);
 \coordinate (B3) at (+\a+\b,+\b);
 \coordinate (C3) at (+\a-\b,+\b);

 \coordinate (A3') at (+\a+\c,0-\c);
 \coordinate (B3') at (+\a+\b+\c,+\b-\c);
  \coordinate (B3'') at (+\a+\b-\c,+\b+\c);
\coordinate (C3') at (+\a-\b-\c,+\b+\c);
 
  \coordinate (A3'') at (+\a,0+\d);

\fill[black!10] (A1') -- (A1) -- (A2) -- (A2') -- (B2') -- (B2) -- (A3) -- (A3') -- (B3') -- (B3) -- (B3'') -- (C3') -- (C3) -- (C2) -- (B1) -- (B1') --(C1') -- (C1) -- (C1'') -- (A1') ;

\fill[blue!10] (A1) -- (A1'') -- (A3'') -- (A3) -- (A1) ;

\fill[red!10] (A1'') -- (A2'') -- (A2') -- (B2') -- (B2'') -- (A3'') -- (C3) -- (C2) -- (B1) -- (A1'');


  \draw[thick] (A1) -- (B1) -- (C1) -- (A1);
  \draw[thick] (A1) -- (A1');
  \draw[thick] (B1) -- (B1');
  \draw[thick] (C1) -- (C1');
  \draw[thick] (C1) -- (C1'');

  \draw[thick] (A2) -- (B2) -- (C2) -- (A2);
    \draw[thick] (A2) -- (A2');
\draw[thick] (B2) -- (B2');
  
  \draw[thick] (A3) -- (B3) -- (C3) -- (A3);
  \draw[thick] (A3) -- (A3');
  \draw[thick] (B3) -- (B3');
  \draw[thick] (B3) -- (B3'');
  \draw[thick] (C3) -- (C3');

\draw[midarrow] (A1) -- (A2) node[midway, below] {$g$};
\draw[midarrow] (A2) -- (B2) node[midway, below] {$w$};
\draw[midarrow] (B2) -- (A3) node[midway, below] {$h$};
  \draw[thick] (B1) -- (C2) -- (C3);

  \draw[thick,red] (A1'') -- (A2'') node[midway, below] {$g$};
    \draw[thick,red] (A2'') -- (B2'') -- (A3'');


\draw[midarrow,red] (A1) -- (A1'') node[midway, above left] {$u$};
\draw[midarrow,red] (A2) -- (A2'') node[midway, left] {$u'$};
\draw[midarrow,red] (B2) -- (B2'') node[midway,  right] {$u'$};
\draw[midarrow,red] (A3) -- (A3'') node[midway, above right] {$u''$};

  \draw[thick,red,midarrow] (A1'') -- (B1);
  \draw[thick,red] (A2'') -- (C2);
  \draw[thick,red] (A2'') -- (A2');
 \draw[thick,red] (B2'') -- (C2);
  \draw[thick,red] (B2'') -- (B2');
  \draw[thick,red] (A3'') -- (C3);


 ;


\end{tikzpicture}
\end{center}
\caption{Ribbon operator with lifted graph in red and lateral string in blue.}
\end{figure}

In analogy with the Kitaev model, the ribbon operators can be superposed and concatenated.
However, some extra phase appear, in particular when doing multiple tensor products.
Similarly, they are homotopically invariant, under deformation of the path, with fixed endpoints, if no defect is encountered by moving the ribbon.

Besides their potential use in quantum computing, ribbon operators also provide a realization of quasi-Hopf symmetry, in the sense of \cite{Mack:1991tg}.

\section*{Conclusion}

With the aim of extending Kitaev's quantum double model to strings coupled to a 3-form background, we have built a theoretical framework centered on unitary gauge transformations $U_{\text{v},g}$ and non-invertible face operators $V_{\text{f},\rho}$ implementing the action of a finite group $G$ and the twisted sector structures on the graph.
Our main results can be summarized in three points.

Kitaev's quantum double model and its symmetries are the starting point of our construction: from this framework, we extend the description of particles in a magnetic field to strings coupled to a field strength $C = \mathrm{d}B$.
In order to preserve the geometric and cohomological structure of the continuous case, we then construct a discrete formulation for a finite group $G$.
This formulation allows us to define the gauge transformations and the magnetic translations, which act on the twisted sectors.

Following the construction of the magnetic translation operators $T_g^w$ acting on the twisted sectors and the derivation of their composition law, we have established their compatibility with string propagation.
Considering the fusion of strings, we naturally recover the coproduct of the twisted Drinfel'd double introduced by R. Dijkgraaf, V. Pasquier and P. Roche in \cite{DPR_1990}, thus allowing us to derive a quasi-triangular quasi-Hopf structure from the string operators, characterized in particular by its associator, $R$-matrix and antipode.
This highlights the central role played by the 3-cocycle $\omega$ in this formalism, encoding the magnetic contributions and twisting the algebraic structure of the Drinfel'd double.

This meticulous algebraic construction led us to establish a Kitaev-type Hamiltonian from three families of compatible local projectors characterizing the constraints associated with magnetic amplitudes, gauge invariance and dynamics on the faces of a given graph.
Its study then highlighted, on the one hand, the non-invertible symmetries carried by the face operators, whose fusion algebra provides an enriched algebraic description of the symmetries of our model.
On the other hand, analyzing this Hamiltonian has enabled us to relate its properties to the representation structure of the twisted Drinfel'd double and to the description of the excitations by means of ribbon operators.

By establishing a link between Kitaev-type topological quantum computing models and string dynamics coupled with a higher-form magnetic field, the construction raises the crucial question of universality \cite{Chen_2025, lo2025universalquantumcomputations3}, which has remained a central issue throughout this work, but is left open at this stage.
Moreover, this construction suggests several directions for future developments, including the study of the algebra of ribbon operators, and the possible enlargement of the formalism, such as a continuum formulation, a generalization to a 2-group or an extension to $3{+}1$ dimensions involving the translation of a membrane in a 4-form background.

Let us briefly sketch how they can be achieved.
\begin{itemize}[label=\textbullet]
\item In the continuum formulation, the finite set $X$ is replaced by a manifold ${\cal M}$, on which the finite group $G$ acts, preserving a closed and integral 3-form $C$.
The latter is the curvature of a gerbe and this gerbe serves to construct the holonomy of the twisted sectors.
The projector $\Pi_{\text{v}}$ is replaced by an equivalence class of homotopic surfaces and the resulting degrees are solely encoded in the group elements decorating the edges. 

\item More in general, the finite group $G$ can be replaced by a finite 2-group, specified by the data of a crossed module $(G,H,t,\alpha)$. 
Here $G$ and $H$ are finite groups, $t:\, H\rightarrow G$ a group morphism and $\alpha:\,(g,u)\in G\times H\mapsto \alpha_{g}(u)\in H$ a right action\footnote{
We can compose it with inversion to get the more standard left action.
} of $G$ on $H$ by group automorphisms.
This data is further assumed to satisfy the Pfeiffer identity $\alpha_{t(u)}(v)=u^{-1}vu$ for any $u,v\in H$ and equivariance condition $t(\alpha_{g}(u)\big)=g^{-1}t(u)g$ for any $u\in H$ and $g\in G$.   Furthermore, let $\omega$ by a 3-cocycle on $G$ and define an algebra $\mathcal{D}^{\omega}[G,H]$ with generators obeying
    \begin{align}    w\mathop{\square}\limits_{g}\;v\mathop{\square}\limits_{h}=\delta_{\alpha_{g}(w),v}\;\frac{\omega_{t(w),g,h}\;\omega_{g,h,\alpha_{gh}[t(w)]}}{\omega_{g,\alpha_{g}[t(w)],h}}\;w\mathop{\square}\limits_{gh}
    \qquad\text{for any $g,h\in G$ and $w\in H$.}
    \end{align}
A similar formula holds for the coproduct
    \begin{align}
    \Delta\!\left(w\mathop{\square}\limits_{g}\right)
    =\sum_{u,v\in H}\delta_{uv,w}\;
    \frac{\omega_{t(u),t(v),g}\;\omega_{g,\alpha_{g}[t(u)],\alpha_{g}[t(v)]}}{\omega_{u,g,\alpha_{g}[t(v)]}}
    \; u\mathop{\square}\limits_{g}\,\otimes\,w\mathop{\square}\limits_{g}
    \end{align} 
and the braiding
    \begin{align}
    R=\sum_{u,v\in H} u\mathop{\square}\limits_{e}\otimes  v\mathop{\square}\limits_{u}.
    \end{align}
$\mathcal{D}^{\omega}[G,H]$ turns out to be a quasi-Hopf algebra that can serve as an input to define a Kitaev model, where edges are decorated by elements of the 2-group, i.e., pairs $(g,u)$, with a Hamiltonian constructed as in the model presented here.
Since the twisted sectors and the gauge elements are labelled by different groups, this allows to enlarge the scope of applicability of the construction.
\item The construction of the  $3{+}1$ dimensional analogue is based  on the magnetic translations of a membrane in a 4-form background.
The analogue of the twisted sectors are given by $2n$ group elements $g_{1},\dots,g_{2n}$, obeying the constraints $g^{}_{1}g^{}_{2}g^{-1}_{1}g^{-1}_{2}\cdots g^{}_{2n-1}g^{}_{2n}g^{-1}_{2n-1}g^{-1}_{2n}$ that define a representation of the fundamental group of a genus $n$ surface. 
It corresponds to a $4n$-gon, with pairwise identifications of the edges.
For $n=1$, the magnetic translation $T_{g}^{u,v}$ generate the quantum triple introduced by Bullivant and Delcamp \cite{Bullivant:2019fmk}. 

\end{itemize}


\printbibliography

\end{document}